\documentclass[a4paper,11pt]{article}
\pdfoutput=1 
\usepackage{xspace}
\usepackage{jcappub} 

\usepackage[T1]{fontenc} 
\usepackage{subfig}
\usepackage{aas_macros} 
\usepackage{tabularx}  
\usepackage{longtable}  %
\usepackage{booktabs} %
\usepackage{units} %
\usepackage{comment}
\usepackage{xspace}
\usepackage{lineno}
\usepackage{multirow}
\usepackage{siunitx}

\newcommand{\gr}{$\gamma$-ray\xspace}
\newcommand{\Gr}{$\gamma$~ray\xspace}
\newcommand{\Grs}{$\gamma$~rays\xspace}
\newcommand{\gammapy}{\textsc{gammapy}\xspace}
\newcommand{\ctools}{\textsc{ctools}\xspace}
\newcommand{\non}{n_\mathrm{On}}
\newcommand{\noff}{n_\mathrm{Off}}
\newcommand{\nexcess}{n_\mathrm{excess}}
\renewcommand{\vec}{\mathbf}
\newcommand{\params}{\boldsymbol{\pi}}
\newcommand{\nuisparams}{\boldsymbol{\theta}}
\newcommand{\ts}{\mathrm{TS}}
\def\LIV{\ifmmode {\mathrm{LIV}}\else{\scshape LIV}\fi\xspace}
\def\LI{\ifmmode {\mathrm{LI}}\else{\scshape LI}\fi\xspace}

\title{Sensitivity of the Cherenkov Telescope Array for probing cosmology and fundamental physics with gamma-ray propagation}

\author{H.~Abdalla,$^{1}$}
\author{H.~Abe,$^{2}$}
\author{F.~Acero,$^{3}$}
\author{A.~Acharyya,$^{4}$}
\author{R.~Adam,$^{5}$}
\author{I.~Agudo,$^{6}$}
\author{A.~Aguirre-Santaella,$^{7}$}
\author{R.~Alfaro,$^{8}$}
\author{J.~Alfaro,$^{9}$}
\author{C.~Alispach,$^{10}$}
% ...

 \bigskip 
 \bigskip 
\affiliation{$^{1} \ $Centre for Space Research, North-West University, Potchefstroom, 2520, South Africa}

\affiliation{$^{2} \ $Institute for Cosmic Ray Research, University of Tokyo, 5-1-5, Kashiwa-no-ha, Kashiwa, Chiba 277-8582, Japan}

\affiliation{$^{3} \ $AIM, CEA, CNRS, Université Paris-Saclay, Université Paris Diderot, Sorbonne Paris Cité, CEA Paris-Saclay, IRFU/DAp, Bat 709, Orme des Merisiers, 91191 Gif-sur-Yvette, France}

\affiliation{$^{4} \ $Centre for Advanced Instrumentation, Dept. of Physics, Durham University, South Road, Durham DH1 3LE, United Kingdom}

\affiliation{$^{5} \ $Laboratoire Leprince-Ringuet, École Polytechnique (UMR 7638, CNRS/IN2P3, Institut Polytechnique de Paris), 91128 Palaiseau, France}

\affiliation{$^{6} \ $Instituto de Astrofísica de Andalucía-CSIC, Glorieta de la Astronomía s/n, 18008, Granada, Spain}

\affiliation{$^{7} \ $Instituto de Física Teórica UAM/CSIC and Departamento de Física Teórica, Universidad Autónoma de Madrid, c/ Nicolás Cabrera 13-15, Campus de Cantoblanco UAM, 28049 Madrid, Spain}

\affiliation{$^{8} \ $Universidad Nacional Autónoma de México, Delegación Coyoacán, 04510 Ciudad de México, Mexico}

\affiliation{$^{9} \ $Pontificia Universidad Católica de Chile, Av. Libertador Bernardo O'Higgins 340, Santiago, Chile}

\affiliation{$^{10} \ $University of Geneva - Département de physique nucléaire et corpusculaire, 24 rue du Général-Dufour, 1211 Genève 4, Switzerland}

% ...

\abstract{
The Cherenkov Telescope Array (CTA), the new-generation ground-based observatory for \gr astronomy, provides unique capabilities to address significant open questions in astrophysics, cosmology, and fundamental physics. We study some of the salient areas of \gr cosmology that can be explored as part of the Key Science Projects of CTA, through simulated observations of active galactic nuclei (AGN) and of their relativistic jets. Observations of AGN with CTA will enable a measurement of \gr absorption on the extragalactic background light with a statistical uncertainty below 15\% up to a redshift $z=2$ and to constrain or detect \gr halos up to intergalactic-magnetic-field strengths of at least 0.3\,pG. Extragalactic observations with CTA also show promising potential to probe physics beyond the Standard Model. The best limits on Lorentz invariance violation from \gr astronomy will be improved by a factor of at least two to three. CTA will also probe the parameter space in which axion-like particles could constitute a significant fraction, if not all, of dark matter. We conclude on the synergies between CTA and other upcoming facilities that will foster the growth of \gr cosmology.
}

\begin{document}
\maketitle
\flushbottom
\newpage

\section{Gamma-ray propagation on cosmic scales}\label{sec:intro}

Over the past decade, the study of \gr propagation over cosmological distances has emerged as a successful branch of ground-based \gr astronomy. This new field, sometimes called \gr cosmology, exploits bright and distant very-high energy (VHE, $E > 30\,$GeV) emitters as beacons to probe the electromagnetic content and fabric of the Universe.

Gamma~rays from extragalactic sources such as blazars, which are active galactic nuclei (AGN) with jets viewed at small angles \cite{1995PASP..107..803U}, can interact en route to the observer through processes in the Standard Model of particle physics and beyond. The main effect impacting VHE \gr propagation is the production of electron-positron pairs on near-UV to far-infrared photon fields \cite{REF::NIKISHOV::JETP1962,1967PhRv..155.1404G, 1967PhRv..155.1408G}. This process results in a horizon {\cite{1970Natur.226..135F}}, located around a redshift $z \sim 1.2$ ($z\sim 0.03$) for \Grs with an energy of $100\,$GeV ($10\,$TeV), beyond which the Universe is increasingly opaque to higher-energy emission and more-distant \gr sources ({e.g.}, Refs.~\cite{1992ApJ...390L..49S, 2008A&A...487..837F, 2011MNRAS.410.2556D, 2012MNRAS.422.3189G, 2016ApJ...827....6S}). This effect also provides a probe of a photon field that populates large voids: the extragalactic background light (EBL)~\cite{1992ApJ...390L..49S}. The EBL is composed of the light emitted by stars, through nucleosynthesis, and by AGN, through accretion, since the epoch of reionization. About half of this light is absorbed by dust grains and reprocessed to mid- and far-infrared wavelengths, while the rest populates the near-UV to near-infrared range (see, {e.g.}, Ref.~\cite{2018MNRAS.474..898A}). The EBL is thus a tracer of the integral cosmic star-formation history. While the specific intensity of the EBL remains uncertain due to the difficulties of foreground subtraction in direct observations, current-generation \gr observatories (in particular imaging atmospheric Cherenkov telescopes, IACTs: H.E.S.S.~\cite{2004NewAR..48..331H}, MAGIC~\cite{2008ApJ...674.1037A}, VERITAS~\cite{2008AIPC.1085..657H}) show agreement with expectations from galaxy counts at the $\sim$\,30\,\% level for EBL wavelengths up to a few tens of $\mu$m \cite{2015ApJ...812...60B, 2017A&A...606A..59H, 2019MNRAS.486.4233A, 2019ApJ...885..150A, 2019ApJ...874L...7D}. On the other hand, the redshift evolution of the EBL, partly probed by observations with the {\it Fermi} Large Area Telescope (LAT, \cite{2009ApJ...697.1071A}) up to hundreds of GeV \cite{2012Sci...338.1190A,2018Sci...362.1031F}, remains poorly constrained by ground-based observatories due to the limited number of \gr sources detected beyond $z\sim0.5$.

The electron-positron pairs produced by the interaction of \Grs with target EBL photons are sensitive, due to their charged nature, to the intergalactic magnetic field (IGMF), whose strength and coherence length remain poorly constrained
\cite{2013A&ARv..21...62D}. An IGMF seed is often invoked for dynamo amplification to explain the ${\sim}\,\mu$G fields observed in galaxies and clusters of galaxies, but the IGMF origin remains disputed~(see, {e.g.}, Refs.~\cite{2013A&ARv..21...62D,2019Galax...7...47S}). It could be either of astrophysical origin, produced with the formation of large-scale structures, or it could be produced in first-order phase transitions prior to recombination. The reprocessing of the energy of the pairs through Comptonization of photons of the cosmic microwave background (CMB) is expected to produce a lower-energy \gr signal. For an IGMF strength $\lesssim 10^{-17}\,$G and for primary \gr energies up to ${\sim}\,10\,$TeV, the \gr signal could be observed up to few hundreds of GeV as a beamed component with an extension $\lesssim 0.1^\circ$~(see, {e.g.}, Ref.~\cite{2018ApJS..237...32A}),
and as an extended emission for higher field strengths \cite{1994ApJ...423L...5A}. The non-detection of such spectral and spatial features by current VHE observatories has constrained the IGMF strength to lie outside a range from $3\times 10^{-16}$ up to $7 \times 10^{-14}\,$G with at least 95\,\% confidence for a coherence length larger than $1\,$Mpc and blazar duty cycles $\gtrsim 10^5$\,years \cite{2014A&A...562A.145H, 2017ApJ...835..288A}. A combination of VHE observations and \textit{Fermi}-LAT data discards configurations of the IGMF with smaller strength   \cite{2018ApJS..237...32A}. These constraints hold if electron-positron pairs lose their energy predominantly through inverse-Compton scattering with the CMB. The pairs could also lose their energy to the intergalactic plasma by cooling through plasma instabilities. The relative strength of plasma-instability and Compton cooling is under active theoretical debate ({e.g.}, Refs.~\cite{2012ApJ...752...22B,2018ApJ...857...43V,2019MNRAS.489.3836A}).

Besides the classical processes discussed above,  propagation could be altered in non-standard scenarios beyond the Standard Model of particle physics. Such an alteration occurs for a coupling of \Grs to sub-eV particles, often referred to as weakly interacting slim particles (WISPs), such as axion-like particles (ALPs) inside magnetic fields (see, {e.g.}, Refs.~\cite{deangelis2007,mirizzi2007,deangelis2011}). Oscillations between \Grs and ALPs would in particular result in a spectral signature that has been searched for in the \gr spectra of AGN lying in clusters with known magnetic fields \cite{hess2013:alps, ajello2016,zhang2018}. Furthermore, Lorentz invariance violation (LIV) either for photons alone or for both photons and leptons at energies up to the Planck scale and {above} could result in a modification of their dispersion relation that could {lead to} an increase of the pair-production threshold, reducing the opacity of the Universe to \Grs with energies larger than tens of TeV \cite{1999ApJ...518L..21K, 2008PhRvD..78l4010J}. Current-generation ground-based instruments have already placed bounds on this process above the Planck scale for first-order modifications of the pair-production threshold \cite{2015ApJ...812...60B,2019ApJ...870...93A}.

The advent of the Cherenkov Telescope Array (CTA, \cite{2011ExA....32..193A}), with a sensitivity improvement with respect to current-generation instruments by a factor of five to twenty depending on energy, with a lower energy threshold enabling spectral reconstruction down to $30\,$GeV, and with improved angular and energy resolutions, will open the way to characterizing VHE blazars with unprecedented accuracy. These observations will trigger tremendous growth of the young field of \gr cosmology. CTA will be based at two sites, one in each hemisphere, and will thus be able to observe any part of the sky. CTA is conceived as an observatory with about $50\,$\% of its time open to observing proposals from the scientific community. A large fraction of the remaining observing time is dedicated to Key Science Projects (KSPs, \cite[][]{2019scta.book.....C}), focused on populations of \gr sources and deep-field observations that would be difficult to pursue with proposals from single observers. 

In this paper, we assess how the AGN KSP \cite{2019scta.book..231Z} and the Cluster of Galaxies KSP \cite{2019scta.book..273Z} can be exploited, beyond the study of astrophysical processes at play at the sources, as observation programs probing \gr cosmology. After a brief description in Sec.~\ref{sec:cta-sim} of the tools used for simulation and analysis of CTA data, we propose a list of foremost targets among currently VHE-detected AGN to measure \gr absorption and we determine the ensuing sensitivity to the EBL imprint in Sec.~\ref{sec:source_and_ebl}. We assess CTA capabilities to constrain or detect the IGMF (Sec.~\ref{sect::IGMF}), ALPs (Sec.~\ref{sec:axions}), and LIV (Sec.~\ref{sec:LIV}) with deep targeted spectral and morphological observations of AGN. Finally, we discuss in Sec.~\ref{sec:Ccl} the multi-wavelength and multi-messenger context of the measurements and the synergies between the exploration of \gr cosmology with CTA and upcoming observatories.

\section{CTA simulations and data analysis}\label{sec:cta-sim}

We assess the potential of CTA to probe \gr-cosmology based on the optimized baseline layouts discussed in Ref.~\cite{2019APh...111...35A}. As a cost-effective solution for improved performance with respect to current-generation IACTs, the baseline layouts of the Northern- and Southern-hemisphere sites of CTA feature telescopes of different sizes. Four large-sized telescopes (LSTs, 23\,m diameter reflector) on each site will enable the trigger of the arrays down to \gr energies of 20\,GeV, yielding an analysis threshold close to 30\,GeV. The CTA-North and -South sites will be equipped with up to 15 and 25 medium-sized telescopes (MSTs, 12\,m diameter reflector), respectively, to scrutinize the \gr sky at hundreds of GeV and up to several TeV. Finally, the baseline configuration of CTA-South features 70 small-sized telescopes (SSTs, 4.3\,m diameter reflector) to observe \Grs up to 300\,TeV, which are only expected to be detected from Galactic \gr sources, as they are not affected by attenuation induced by propagation on extragalactic scales.

The results presented in this work are based on the layouts of the baseline arrays, as well as on the live time proposed for the KSPs of CTA. Improved performance of the arrays with respect to the baseline layout could result in a reduction of the proposed live time for equal scientific return. Conversely, a configuration of the arrays featuring fewer telescopes of each type could, to some extent, be mitigated by an increase in the amount of live time allocated to the targets discussed in this paper. As such, the scientific return of CTA on \gr cosmology presented in this work should be viewed as a realistic goal, whose full achievement or exceeding will depend on the implementation of the arrays and on the allocation of observation time to the variety of science cases covered by CTA. The goals set in this work are aimed to provide one of the important scientific cornerstones for the detailed assessment of the deployment and time allocation of CTA.

\subsection{Data simulation and analysis}\label{sec:cta-sim_subsec}

We use Monte Carlo simulations to derive the instrument response functions (IRFs, version \texttt{prod3b-v1} for this work) and the expected background from cosmic-ray induced air showers~\cite{bernlohr2013,2019APh...111...35A}.\footnote{The latest versions of the IRFs and background Monte Carlo simulations are available at \url{https://www.cta-observatory.org/science/cta-performance/}.} For this study, we generate \Grs and background events based on these IRFs and background rates. The southern site in Paranal, Chile is considered for AGN with negative declinations; the northern array at La Palma, Spain is used otherwise. Furthermore, we adopt the following choices for the simulations and analyses throughout this paper if not stated otherwise:
\begin{itemize}
\item We set the minimum threshold energy for analyses to 30\,GeV \cite{2019APh...111...35A}. 
\item We use a ratio of exposures between signal (``On'') and background (``Off'') regions of $\alpha_\mathrm{exp} = 0.2$. This value corresponds to five background regions of the same size as the signal region.
\item We analyze energy bins for which the binned flux of the AGN is detected (a) with a significance $S >2\,\sigma$,\footnote{$\sigma$ denotes a Gaussian standard deviation, with variations of ${\pm}\,1\,\sigma$ around the median yielding the 84\% and 16\% quantiles of a 1D Gaussian distribution, respectively.} where $S$ is the Li \& Ma estimate given by Eq.~(17) in Ref.~\cite{lima1983}, (b) with excess events, $\nexcess = \non{} - \alpha_\mathrm{exp}\times\noff{}$, such that $\nexcess / \noff{} > 5\,\%$, and (c) with $\nexcess > 10$. Criterion (b) in particular ensures control of the cosmic-ray background at low energies. 
\item We use ten energy bins per energy decade. This value corresponds to a bin width $\Delta \ln E = 0.1\times\ln 10$ that encompasses events within ${\pm}\,1\,\sigma(E)$ around true energies above ${\sim}\,200\,$GeV, where $\sigma(E)/E$ is the energy resolution of CTA. Although the energy binning is a factor of about two smaller than the energy resolution at $30\,$GeV, the convolution of the true underlying spectrum with the IRFs ensures a full treatment of energy dispersion at all energies.
\end{itemize}

Given a model for the \gr differential energy spectrum at Earth, $\phi$ in units of $\mathrm{TeV}^{-1}\,\mathrm{cm}^{-2}\,\mathrm{s}^{-1}$ 
(or $\mathrm{TeV}^{-1}\,\mathrm{cm}^{-2}\,\mathrm{s}^{-1}\,\mathrm{sr}^{-1}$ for extended sources discussed in Sec.~\ref{sect::IGMF}),
we use \gammapy and \ctools~\cite{Deil2017,ctools}\footnote{See \url{http://docs.gammapy.org/} and \url{http://cta.irap.omp.eu/ctools/}.} to compute the expected number of signal counts, $\mu_i$, in the $i$-th bin of reconstructed energy integrated over the energy range $\Delta E^\prime_i$ and solid angle $\Omega$ of reconstructed arrival direction $\vec{p}^\prime$
(reconstructed values are denoted in this section with a prime, while true values are not), obtained in an observation of duration $\mathrm{T}_\mathrm{obs}$,
\begin{equation}
    \label{eq:npred}
    \mu_i(\params, \nuisparams) = \mathrm{T}_\mathrm{obs} \int\limits_{\Delta E_i^\prime} \mathrm{d}E^\prime     \int\limits_{E-\textrm{range}}\mathrm{d}E 
    \int\limits_{\Omega}\mathrm{d}\vec{p}^\prime
    \int\limits_{\vec{p}-\textrm{range}}
    \mathrm{d}\vec{p}\ R(E,E^\prime, \vec{p}, \vec{p}^\prime)\ \phi(E, \vec{p},\params, \nuisparams).
\end{equation}

In the above equation, the source model, $\phi$, is a function of energy, of the parameters of interest for the particular science case, $\params$, and of additional nuisance parameters, $\nuisparams$. For example, when looking for \gr absorption, $\params$ includes the parameters of the EBL model, whereas $\nuisparams$ encodes the parameters of the intrinsic spectrum. The IRF, $R$, includes the effective area, point spread function (PSF) and energy dispersion. 
For point sources, the model $\phi$ contains a delta function in $\mathbf{p}$.
The solid angle $\Omega$ is either bound by the angular region centered on the source position in a classical On/Off analysis of point sources or by the chosen size of the spatial pixels in a template-based analysis of extended sources, such as used in Sec.~\ref{sect::IGMF}.
The number of expected background events, $b$, is obtained from Monte Carlo simulations. 

Given $\mu$ and $b$, we generate CTA simulations by sampling the number of observed events in the signal and background regions from Poisson distributions: $P(\non | \mu + b)$ and $P(\noff |b/\alpha_\mathrm{exp})$, respectively. We derive best-fit parameters by maximizing the associated log-likelihood summed over all energy bins which pass the selection criteria, as discussed in App.~\ref{sect::data_analysis}. The significance of the sought-after effect is determined through a log-likelihood ratio test with respect to a null hypothesis under which the effect is absent. Using the example of \gr absorption parametrized with an EBL normalization, this corresponds to the case when the normalization is zero, {i.e.}\ no absorption on the EBL.

\subsection{Systematic uncertainties}
\label{sec:cta-sim_subsec_sys_unc}

We distinguish between two classes of systematic uncertainties: those of instrumental origin and those arising from the underlying physical model.
We opt whenever possible for treating them separately to assess their relative magnitude with respect to statistical uncertainties.

Systematic uncertainties arising from the modeling are identified on a case-by-case basis and their impact is explored using bracketing assumptions that are deemed reasonable based on the current knowledge of astrophysical conditions at play in the environments of \gr emitters. A general approach addressing this source of systematic uncertainties is difficult to formulate, as the underlying model may be affected in a non-continuous manner by variations of the parameters of interest. Such variations occur, {e.g.},  for random variations of a vector field in multiple domains along the path of particles propagating on cosmic scales.

Systematic uncertainties arising from mismatches between Monte-Carlo simulations and the true instrument response are similarly treated using bracketing IRFs. This approach follows the procedure developed by the \emph{Fermi}-LAT Collaboration~\cite{Ackermann:2012kna}.\footnote{\url{https://fermi.gsfc.nasa.gov/ssc/data/analysis/scitools/Aeff_Systematics.html}} The reconstruction of spectral features imprinted along the line of sight is primarily affected by uncertainties on the effective area and on the energy dispersion. The uncertainty on the background rate is sub-dominant at the energies of interest for signal-dominated \gr sources and is thus neglected. The uncertainty on the angular dispersion primarily affects the normalization of the flux for a point-like source, a nuisance parameter which is profiled over in the analyses presented in this work. We model variations of the effective area and energy scale as a fractional shift of ${\pm}\,5\%$ and ${\pm}\,6\%$, respectively, corresponding to ${\pm}\,1\,\sigma$ estimates matching the projected systematic uncertainties for CTA. Except close to the energy threshold, a constant shift of the effective area as a function of energy only affects the normalization of the flux of the AGN (nuisance parameter). Instead of a constant shift, we consider discrete step-wise variations of the effective area (smoothed at the energy-resolution scale) with an amplitude of ${\pm}\,5\%$, at energies where one sub-system of telescopes starts to dominate the point-source sensitivity; the LSTs of CTA are assumed to dominate the sensitivity up to ${\sim}\,150\,$GeV while MSTs take over up to a few TeV. In the southern site, SSTs dominate the sensitivity above ${\sim}\,5\,$TeV. The variations of the energy scale are modeled here as a constant shift by ${\pm}\,6\%$ at all energies. For most science cases (Sec.~\ref{sec:source_and_ebl} and Sec.~\ref{sec:LIV}), a simple quadratic summation of systematic uncertainties (see App.~\ref{sect::systematics}) is sufficient. Some science cases (see Sec.~\ref{sect::IGMF}) require discussion of the impact of bracketing hypotheses. 

While adapted to the search for phenomena impacting a broad energy or angular range, bracketing IRFs are not well suited to the search for narrow spectral features, such as expected from ALPs (see Sec.~\ref{sec:axions}). For the latter, we directly incorporate systematic uncertainties in Eq.~\eqref{eq:npred} as additional nuisance parameters.

The simple models adopted for systematic uncertainties of instrumental origin will be refined in the future with the measured instrumental response, extracted from CTA data.

\section{Interaction of \Grs with the extragalactic background light}\label{sec:source_and_ebl}

Interactions of \Grs with the EBL result in an effective opacity of the Universe that depends on the distance of the \gr source and on the \gr energy. The interaction process involved is pair production of electrons and positrons. The threshold of this process is related to the mass of the particles, $m_{\rm e}$, following
\begin{align}
    &E'_{\gamma}\epsilon' \geq \frac{2\left(m_{\rm e}c^2\right)^2}{1-\cos\theta'},\nonumber\\
    {\rm or\quad} &E'_{\gamma} \geq \unit[210]{GeV}\times\left(\frac{\lambda'}{\unit[1]{\mu m}}\right) \text{\quad for\ head-on\ collisions},\label{Eq:threshold}
\end{align}
where $E'_{\gamma}$ and $\epsilon'$ ($\lambda'$) are the energies (wavelength) of the \Gr and low-energy photon, respectively, in the comoving cosmological frame at which the interaction occurs, and where $\theta'$ is the angle between the momenta of the two photons.

The effective absorption of \Grs of observed energy $E_{\gamma}$ from a \gr source at redshift $z_0$ is quantified by the optical depth ({e.g.}, Ref.~\cite{2005ApJ...618..657D}),
\begin{equation}
    \label{Eq:optInteg}
    \small
    \tau (E_{\gamma},z_0) = \int_{0}^{z_0} \mathrm{d} z \frac {\partial D(z)}{\partial z} \int_{0}^{\infty} \mathrm{d} \epsilon' \frac{\partial n (\epsilon', z)}{\partial \epsilon'} \int_{-1}^{1} \mathrm{d}\cos\theta' \frac{1-\cos\theta'}{2} \sigma_{\gamma\gamma}\big((1+z)E_\gamma,\epsilon',\cos\theta'\big) \mathcal{H}(\epsilon'-\epsilon'_{\rm th}).
\end{equation}

The integration in Eq.~\eqref{Eq:optInteg} is performed over (a) the line of sight, $\partial D(z)/\partial z$ being the distance element in $\Lambda$CDM cosmology, (b) the energy of the target photons, $\partial n/\partial \epsilon'$ being the differential number density of EBL photons, and (c) the angle, $\theta'$, between the \gr and the target photons, which are assumed to be isotropically distributed in the comoving frame. The integrand, including the Breit-Wheeler differential cross section for photo-production of pairs, $\sigma_{\gamma\gamma}$, is non-zero when Eq.~\eqref{Eq:threshold} is satisfied, {i.e.}\ $\epsilon' \geq \epsilon'_{\rm th}$, as encoded in the argument of the Heaviside function, $\mathcal{H}$.

The differential pair-production cross section can be integrated analytically over $\cos\theta'$, as discussed, {e.g.}, in Ref.~\cite{1988ApJ...335..786Z} (Eqs. B11--12) or Ref.~\cite{2015ApJ...812...60B} (Eq. 7). When further integrated over the line of sight, the so called ``EBL kernel'' in Ref.~\cite{2015ApJ...812...60B} is maximal in the observer's frame at a \gr energy about a factor of two larger than that imposed by kinematics in Eq.~\eqref{Eq:threshold}. Neglecting redshift dependence to first order, EBL photons at wavelengths $\unit[0.5]{\mu m}$ (optical), $\unit[5]{\mu m}$ (mid-infrared), and $\unit[50]{\mu m}$ (far-infrared) are thus primarily responsible for absorption of \Grs with energies ${\sim}\,\unit[200]{GeV}$, ${\sim}\,\unit[2]{TeV}$, and ${\sim}\,\unit[20]{TeV}$, respectively. Gamma-ray astronomy thus probes in an indirect manner both the cosmic optical background (COB, $\unit[0.1-8]{\mu m}$) and part of the cosmic infrared background (CIB, $\unit[8-1000]{\mu m}$), the two components of the EBL. {Gamma-ray observations at PeV energies are limited to Galactic distance scales by interactions with the CMB.}

The EBL photon density, $\partial n / \partial \epsilon$, can be parametrized (with parameters $\pi_{\rm EBL}$) to estimate the  optical depth to \Grs, $\tau(E_\gamma,z_0)$, from \gr data. Best-fit EBL parameters are obtained by profiling over the ``intrinsic'' parameters, $\theta_{\rm int}$, when modeling the \gr source spectrum, 
\begin{equation}
    \phi(E_\gamma,z_0) = \phi_{\rm int}(E_\gamma; \widehat{\theta}_{\rm int}(\pi_{\rm EBL}))\times \exp\left(-\tau(E_\gamma,z_0; \pi_{\rm EBL})\right).
\end{equation}

The intrinsic spectrum, $\phi_{\rm int}(E_\gamma)$, corresponds to the flux expected if all \Grs escaped absorption on the EBL. It is often modeled as a smooth function of energy, increasingly steep as energy increases (concave function), {e.g.}, a log-parabola with exponential cut-off 
\begin{equation}
    \label{eq:lpe}
    \phi_{\rm int}(E_\gamma) = \phi_{0} (E_\gamma/E_0)^{-\Gamma - \beta \ln(E_\gamma/E_0)} \exp{(-E_\gamma/E_\mathrm{cut})},
\end{equation}
where $\phi_0$ is the flux normalization at a fixed reference energy $E_0$. $\Gamma$ is the power-law index at energy $E_0$ or at any energy when the curvature parameter $\beta$ of the log-parabola is null, and $E_\mathrm{cut}$ describes a cut-off at the high-energy end of the spectrum. Different intrinsic parameters, $\theta_{\rm int} = \{\phi_0, \Gamma, \beta, E_\mathrm{cut}\}$, are assigned to each spectrum and likelihoods from multiple spectra can readily be combined to jointly fit the EBL parameters, $\pi_{\rm EBL}$.

The specific intensity of the EBL has been constrained with current VHE data in two manners: EBL-model dependent approaches ({e.g.}, Refs.~\cite{2013A&A...550A...4H,2015ApJ...815L..22A, 2016A&A...590A..24A, 2019MNRAS.486.4233A}) and approaches where the spectral energy distribution is modeled independently from any prescriptions (only the redshift evolution is tuned to follow that of EBL models, see, {e.g.}, Refs.~\cite{2015ApJ...812...60B, 2017A&A...606A..59H, 2019ApJ...885..150A}). Model-dependent approaches have thus far mostly relied on a simple scaling by a factor $\alpha \geqslant 0$ of the photon density from an existing EBL model, which results in $\tau(E_\gamma,z_0; \pi_{\rm EBL}) = \alpha\times\tau(E_\gamma,z_0)$. The deviation of the EBL parameter from $\alpha=0$ quantifies the detection of the EBL signal, while its deviation from $\alpha=1$ quantifies the departure from the model, {be it in normalization or dependence on source redshift and \gr energy}. This approach, adapted from the study of distant $\gamma$-ray bursts (GRBs) and blazars \cite{2010ApJ...723.1082A}, enabled the {\it Fermi}-LAT and H.E.S.S. Collaborations to disentangle the imprint of the EBL from any intrinsically curved or cut-off intrinsic spectral shapes for the first time in Refs.~\citep{2012Sci...338.1190A, 2013A&A...550A...4H}. Model-independent approaches usually proceed from disentangling the redshift evolution of the photon field from its local spectrum at $z=0$, assuming that the photon density dependences can be factorized as $\displaystyle {\partial n(\epsilon', z)/\partial \epsilon' = \partial n(\epsilon, z=0)/\partial \epsilon \times f(z)}$, where $\epsilon$ is the EBL photon energy in the observer's frame and $f(z)$ models the redshift evolution of the field. The latter approach is well justified in the local Universe up to $z\sim 0.6-0.8$, where the redshift evolution of the emissivity of the EBL sources as a function of wavelength is not too strong \cite{2015ApJ...812...60B}. Nonetheless, the approach fails to reproduce a realistic evolution of the EBL at higher redshifts, where CTA will have  sensitivity to spectral features imprinted by propagation.

Gamma-ray constraints on the EBL enable the study of numerous science cases. For example, they can be used to study the specific intensity of the EBL at $z=0$, which probes the proportion of direct starlight to dust re-emission, the UV background, as well as near-infrared signatures of hydrocarbons. Alternatively, the evolution of the constraints with redshift provides information on the cosmic star-formation history, reionization, and cosmological parameters (see, e.g., Refs.~\cite{2018MNRAS.474..898A,2018Sci...362.1031F}). In this work, we adopt a simple parametrization of the EBL density through a scale factor, $\alpha$, applied to the optical depth in order to illustrate the overall performance of CTA with respect to current-generation instruments. We aim at assessing the capability of CTA to constrain the optical-depth normalization as a function of the redshift of the \gr sources. The scale factor encompasses information both on the evolution of the EBL as well as on the specific intensity of the EBL, integrated over a wavelength window bound at its high-end by kinematics (Eq.~\eqref{Eq:threshold}).

\subsection{Source selection from AGN KSP}\label{subsect::ebl_ksp_sel}

As noted in the introduction, the CTA Consortium has prepared a program of KSPs, which are intended to cover ${\sim}\,\unit[40]{\%}$ of the available observing time in the first ten years of CTA operations. The KSP dedicated to AGN involves three different observing programs: the long-term monitoring program, the search for and follow-up of AGN flares, and a program devoted to establishing high-quality spectra of selected AGN, with a systematic coverage of redshift and AGN type. These observations will provide a rich database of VHE spectra that can be used to study the physics at play in the \gr sources and the effects of \gr absorption. While a detailed evaluation of the live time allocated to each KSP remains to be established upon commissioning of CTA, we propose in the following sub-sections (Secs.~\ref{sec:longterm}--\ref{sec:flare}) a list of candidates that we recommend for observation to constrain \gr absorption on cosmic scales. Readers who may want to skip the detailed discussion of source and dataset selections are directed to the source summary provided in Sec.~\ref{sec:srcsummary}.

\subsubsection{Long-term monitoring sample}\label{sec:longterm}

The AGN long-term monitoring program is proposed to provide long-term light curves of about 15 \gr sources representing the sub-categories of AGN ({e.g.}, blazars and radio galaxies) currently detected by ground-based instruments. These light curves will be obtained through weekly 30-minute observations during the period of observability of the AGN, resulting in less than 12\,hours per year per target. A list of potential targets is presented in chapter 12 of Ref.~\cite{2019scta.book.....C}. The total foreseen observing time accumulated over the duration of the program is about 1100\,hours from the northern CTA site and 400\,hours from the southern CTA site. 

The long-term average spectra of AGN may be an aggregate of multiple emission states, including possible spectral variability that could limit the accuracy of the reconstruction of the absorption.\footnote{This limitation could nonetheless be mitigated by analyzing single AGN observations grouped by spectral state, as done in Ref.~\cite{2013A&A...550A...4H}.} The duty cycle for \gr elevated states of AGN has been estimated to about 50\% on monthly time-scales based on long-term observations of blazars with \textit{Fermi}~LAT \cite{2020A&A...634A..80R}. While such a duty cycle remains uncertain in the VHE band, we assume based on \textit{Fermi}-LAT observations that half of the long-term monitoring program of CTA will yield spectra with observed flux levels that are representative of the average emission state of the AGN.

For our purpose, we exclude M87 and NGC\,1275 from the list of 15 targets suggested in Ref.~\cite{2019scta.book.....C}. The proximity of these AGN (17 and 75\,Mpc, that is $\tau=1$ at ${\sim}\,30\,$TeV and ${\sim}\,15\,$TeV, respectively) would require the assumption of emission well beyond 10\,TeV to constrain the EBL. We also exclude the FSRQs\footnote{Blazars are classified into BL Lac objects (BLLs) and flat spectrum radio quasars (FSRQs), showing weak and strong emission lines, respectively~\cite{1995PASP..107..803U}.} PKS\,1510$-$089 and PKS\,1222+216, which may not have been observed in their average state by current-generation instruments. The remaining 11 blazars, which have been regularly monitored, are all BL Lac objects. Two are low-synchrotron peaked blazars (LSP), two are intermediate-synchrotron peaked (ISP), five are high-synchrotron peaked (HSP), and two are extremely-high-synchrotron peaked (EHSP), with classifications detailed in Refs.~\cite{2010ApJ...715..429A, 2020NatAs...4..124B}. We extracted the intrinsic parameters of all 11 blazars from published VHE observations that are representative of an average state of each blazar, following the methodology of Ref.~\cite{2015ApJ...812...60B} and taking as benchmark the EBL model of Ref.~\cite{2011MNRAS.410.2556D}. This benchmark EBL model is consistent with current observational constraints and is used throughout this paper. The intrinsic properties of each AGN are given in App.~\ref{sec:intr-models}. These AGN have firm redshift determinations based on published spectroscopic observations with at least two well-detected and identified spectral features. The only exception among the 11 AGN in this sample is 3C\,66A. Lyman $\alpha$ absorption systems have been detected in the HST/COS spectrum~\cite{2012ApJ...744...60G} of this blazar up to $z=0.34$, thus setting a firm lower limit on its redshift~\cite{fur13a}. Further support to the use of this value as the true redshift of 3C\,66A has been provided by the possible hosting of this ISP by a cluster of galaxies at $z=0.34$ \cite{tor18}. 

We simulate CTA observations of 50-hour duration for each of the 11 blazars using \gammapy, as described in Sec.~\ref{sec:cta-sim_subsec}. Gamma-ray absorption is modeled following our benchmark EBL model and, to ensure a conservative modeling at the highest energies, an ad hoc exponential cut-off is incorporated in the spectral model at an energy $E_{\rm cut}'$ in the cosmological comoving frame (that is $E_{\rm cut} = E_{\rm cut}'/(1+z)$ for the observer). The value of the energy cut-off is set as accounting for the observed correlation between the synchrotron and gamma-ray peak locations ({e.g.}, Ref.~\cite{2017MNRAS.469..255G}). We assume $E_{\rm cut}'=100\,$GeV for LSP and ISP, $E_{\rm cut}'=1\,$TeV for HSP, and $E_{\rm cut}'=10\,$TeV for EHSP AGN.

To illustrate the reconstruction of the \gr absorption, we retain AGN detected at least at the  $5\,\sigma$ confidence level above the energy, $E(\tau=1)$, at which the optical depth, $\tau$, is equal to 1. This optical-depth value corresponds to the so-called \gr horizon. The selection on detection significance at energies beyond the cosmic \gr horizon effectively removes six blazars from the sample. This selection thus yields a list of five blazars, for a total observing time of 250\,hours.

\subsubsection{High-quality spectral sample}
\label{sec:hq}

The high-quality spectra program is proposed to cover deep observations of AGN of different classes and at different distances. About 350\,hours (200\,hours for the northern CTA site and 150\,hours for the southern CTA site) will be devoted to this program, spread over 10\,years.  A list of possible targets is given in Ref.~\cite{2019scta.book.....C}, on the basis of the extrapolation of blazar spectra from the 1FHL catalog of the \emph{Fermi}-LAT Collaboration \cite{2013ApJS..209...34A}. Here, we study \gr sources of interest for EBL studies from the 3FHL catalogue \cite{2017ApJS..232...18A}. 

Starting from the 3FHL, we select blazars with a constrained redshift and exclude those already considered in the long-term monitoring program. We simulate 20\,hours of observation of each blazar, incorporating a class-dependent comoving cut-off at $E_{\rm cut}$ (see Sec.~\ref{sec:longterm}). The most-promising blazars are selected in an intermediate redshift range, $0.05 \lesssim z \lesssim 0.6$, based on their detection significance above $E(\tau=1)$ (see App.~\ref{sec:intr-models}). 
This selection results in a list of 12 blazars with firm redshift, among which 10 have been detected by current-generation ground-based instruments. Three other selected blazars, without firm redshift, lie at $z>0.3$; two of which have already been detected by IACTs. The lower limits on the distances of these three blazars, based on spectroscopic absorption lines, are considered here as putative true redshift values. Firm redshift determination will presumably be available by the time of the analysis of the CTA observations.\footnote{One of the three above-mentioned blazars, PKS\,1424+40, exemplifies the potential of future spectroscopic observations of AGN, with the distance of this \gr source being estimated in Ref.~\cite{pai17} from two emission lines at the edge of the minimum detectable equivalent width.}

\subsubsection{Flare sample}
\label{sec:flare}

The AGN-flare program has been devised to search for and follow-up VHE flares from AGN, triggered either by external facilities or internally by the monitoring program performed with CTA. On the basis of the results of present-day facilities (mainly {\it Fermi} and {\it Swift}), about 25 alerts per year are expected, $10-15$ of which would be followed with CTA.  The time budget proposed for the follow-up program of AGN flares is about 700\,hours for the northern CTA site and 500\,hours for the southern CTA site. 

We base the simulations on two different samples of observed elevated-flux states. The first sample, consisting of 14 AGN, exploits observations during flaring states by VHE ground-based instruments, as listed in Ref.~\cite{2015ApJ...812...60B}. Intrinsic spectra are derived, as for the long-term monitoring sample, by fitting a spectral model accounting for absorption on the EBL to the VHE data \cite{2011MNRAS.410.2556D}. The only AGN out of 14 that do not have a firm redshift are 3C\,66A (see above) and S5\,0716+714. A lower limit from Lyman $\alpha$ absorption systems places S5\,0716+714 at least at $z=0.2315$ \cite{dan13}, which we use as putative true redshift. Four of the 14 AGN are FSRQs, 9 are BL\,Lacs (one LSP, three ISPs, four HSPs, one EHSP), and the remaining one, IC\,310, is classified as a radio galaxy (RDG). We note that the classification of this \gr source related to its jet viewing angle remains debated \cite{2012A&A...538L...1K}. Following Ref.~\cite{2020NatAs...4..124B}, IC\,310 is classified during elevated states as an extreme-TeV object (ETEV) based on its observed \gr spectrum. Similarly, during flaring states, Mrk\,501 is classified as EHSP. The properties of each AGN are given in App.~\ref{sec:intr-models}.

The second sample is selected from the {\it Fermi}-LAT monitored source list\footnote{\url{https://fermi.gsfc.nasa.gov/ssc/data/access/lat/msl_lc/}.} with redshift in SIMBAD\footnote{\url{http://simbad.u-strasbg.fr/simbad/}.} greater than 0.05. We retain AGN showing a flux above 1\,GeV averaged over one day greater than $\unit[10^{-7}]{cm^{-2}\ s^{-1}}$, detected with a test statistic ${\rm TS}\ge 100$. We exclude from this preliminary sample nine AGN that have an uncertain redshift (see App.~\ref{uncertain_z}). We also excluded nine AGN that are already included in the TeV-flare sample. Finally, only the brightest single-day observation of each \gr source is retained, resulting in 63 flares from 63 different AGN. Fifty-two of these \gr sources are classified as FSRQs, ten as BL\,Lacs and one is an AGN of uncertain class (PKS\,0521$-$36, see Ref.~\cite{2015MNRAS.450.3975D}). It should be noted that three of the BL\,Lacs at high redshift (PKS\,0537$-$441, A0\,235+164, and PKS\,0426$-$380) can be classified as FSRQs based on their spectral-energy-distribution (SED) shapes and on the luminosity of the broad-line regions in units of the Eddington luminosity~\cite{2011MNRAS.414.2674G}. All of these four blazars with uncertain classification are LSPs, resulting in a firm assignment of their cut-off at $E_{\rm cut}'=100\,$GeV.

The spectrum of each AGN during its one-day flare is determined from a dedicated \emph{Fermi}-LAT analysis. Events passing the \texttt{P8R2\_SOURCE} selection are analyzed in an energy range spanning $1-500\,$GeV and in a region of interest (ROI) covering an area of $10^\circ\times10^\circ$ centered on the \gr source position. 
A spatial binning of $0.1^\circ$ per pixel is used along with eight energy bins per decade.
We model the ROI by including all sources listed in the 3FGL catalog \cite{2015ApJS..218...23A} up to $15^\circ$ away from the central source. All the spectral parameters of \gr sources within $5^\circ$ of the center are left free to vary, while only the flux normalizations are free for \gr sources  $5^\circ-10^\circ$ away from the source of interest. Standard templates are used for the isotropic diffuse emission and Galactic diffuse emission.\footnote{
We use \texttt{gll\_iem\_v06.fits} for the Galactic and \texttt{iso\_P8R2\_SOURCE\_V6\_v06.txt} for the isotropic diffuse emission, see 
\url{https://fermi.gsfc.nasa.gov/ssc/data/access/lat/BackgroundModels.html}.} After an initial optimization using \textsc{fermipy} v.0.16.0+188~\cite{2017ICRC...35..824W} and the \emph{Fermi} Science tools v.11-07-00,\footnote{See \url{http://fermipy.readthedocs.io/} and \url{https://fermi.gsfc.nasa.gov/ssc/data/analysis/documentation/}.} we fix all the parameters of \gr sources detected with $\mathrm{TS} < 10$. 
The flux normalizations of \gr sources with $\mathrm{TS} > 50$ are then freed. As shown in App.~\ref{sec:intr-models}, most of the spectra are well reproduced by a power-law model, except in one case (3FGL\,J2025.6$-$0736, {i.e.}\ PKS\,2023$-$077) for which a log parabola is preferred. Since we are analyzing only one day of {\it Fermi}-LAT data for each \gr source, the small event counts at the highest energies do not result in a measurable contribution of \gr absorption to the hardness of the {\it Fermi}-LAT spectra and we thus consider the spectral parameters obtained in the $1-500\,$GeV range as the intrinsic ones.

A live time of 10\,hours, corresponding to two to three nights of ground-based observations in an elevated state is considered for each AGN in the ``flare'' program. Assuming a class-based spectral cut-off removes 9 and 40 AGN from the list of flares detected with VHE ground-based instruments (TeV-flare sample) and with \textit{Fermi}-LAT (GeV-flare sample), respectively.  In total, 27 flaring AGN are expected to be detected beyond $5\,\sigma$ above $E(\tau=1)$. The \gr source IC\,310 at $z=0.019$, with a $4.4\,\sigma$ detection above $E(\tau=1)$, is kept in the sample used for low-redshift constraints on the EBL. These 28 AGN amount to 280\,hours of simulated observations. The TeV-flare sample spans a range up to $z\sim1$, illustrative of the cosmic volume covered by current-generation IACTs, while the GeV-flare sample extends to $z\sim2$, demonstrating the tremendous increase in the range accessible to CTA.

\subsubsection{Sources selected for EBL studies: a summary}
\label{sec:srcsummary}

\begin{figure}[t]
  \center
  \includegraphics[width=\columnwidth]{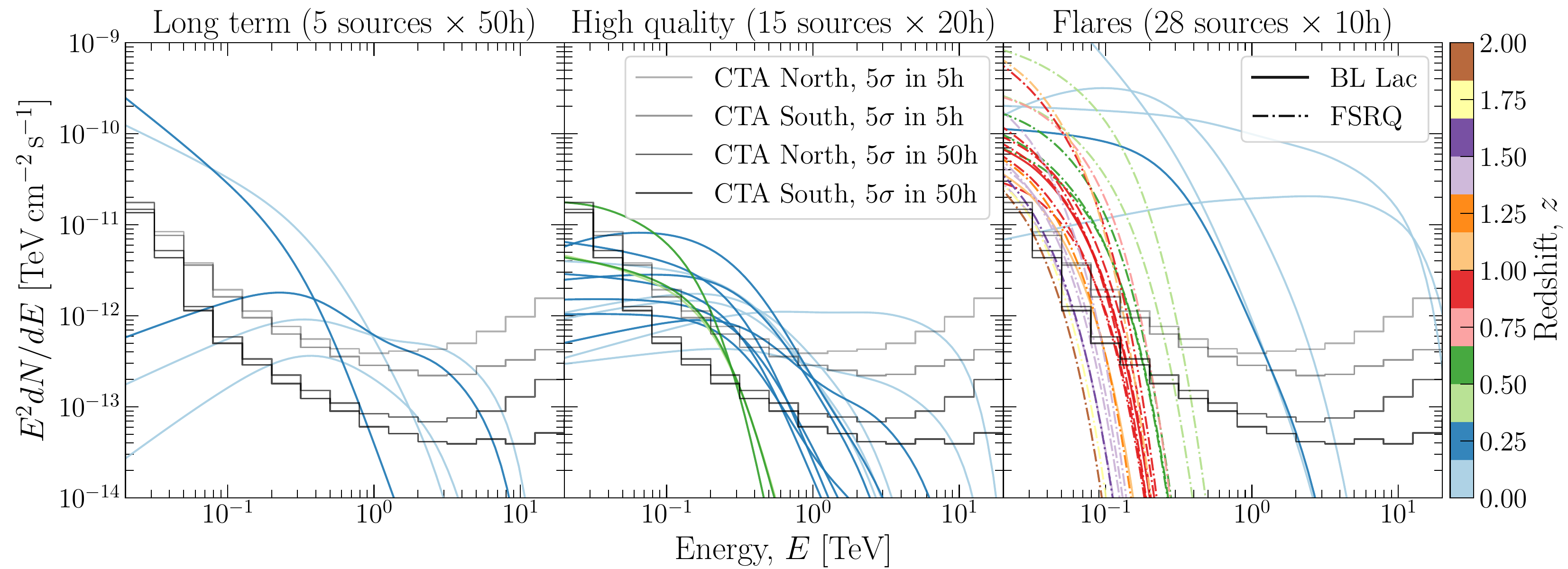}
  \caption{
Simulated spectral models of AGN retained for the reconstruction of the EBL scale factor with CTA. 
The differential spectrum of each AGN multiplied by the square of energy is displayed as a function of \gr energy as continuous (BL\,Lac objects) or dashed lines (FSRQs). The color scale indicates the redshifts of the AGN. {\it Left}: Long-term monitoring program with 50\,hours per field. 
{\it Center}: High-quality spectra program with 20\,hours per field. 
{\it Right}: Flare program with 10\,hours per field.
An optical-depth normalization $\alpha=1$ is adopted, using the model in Ref.~\cite{2011MNRAS.410.2556D}. The sensitivities of CTA North and South in 5\,hours and 50\,hours for a pointing zenith angle of $20^\circ$ are shown as grey line segments.   
  }
  \label{fig::ebl_spectra}
\end{figure}

As illustrated in Fig.~\ref{fig::ebl_spectra} and tabulated in App.~\ref{sec:intr-models}, we consider a total of 5 long-term monitored blazars with 50\,hours dedicated to each field. High-quality spectral observations are suggested for 15 selected blazars, assuming 20\,hours per field. All the blazars in these two samples are BL Lacs. For the flare program, we consider a total of 28 AGN with 10\,hours of observation each. BL Lacs and IC\,310 represent 7 out of the 28 AGN and span a redshift range of $0.019-1.11$. FSRQs correspond to 21 of the 28 AGN and cover a redshift range of $0.36-1.84$. This distribution is in line with that observed in the GeV range by {\it Fermi}~LAT, but it should not necessarily be considered as a realistic prediction of the distribution of flares at TeV energies across redshift and AGN classes, which is precisely one of the key questions to be addressed by the extragalactic observing programs of CTA, including the extragalactic survey of one quarter of the sky \cite{2019scta.book..143M}.

The list of 48 simulated spectra described above is obtained with a class-dependent spectral cut-off. We alternatively considered comoving spectral cut-off values fixed for all AGN at 100\,GeV, 1\,TeV and 10\,TeV, respectively. The lowest energy cut-off removes all the AGN below redshift $z<0.35$, for which the \gr horizon is located at energies greater than 300\,GeV. None of the AGN from the long-term monitoring or high-quality spectra programs would be detected beyond the \gr horizon for $E_{\rm cut}' = 100\,$GeV. With a total of 23 detections beyond $E(\tau=1)$ out of 103 simulated spectra, this scenario should be considered pessimistic as the presence of a cut-off at 100\,GeV in all extragalactic \gr sources would result in only a handful being detectable by current-generation instruments, compared to ${>}\,80$ VHE AGN being detected by ground-based instruments to date. 
The most optimistic scenario, on the other hand, with a comoving spectral cut-off at 10\,TeV, results in the detection of significant signal beyond $E(\tau=1)$ for 93 simulated spectra, compared to 82 for a 1\,TeV cut-off. While a high-energy cut-off at 10\,TeV is not expected for every nearby AGN over a 50-hour time span (therefore the label ``optimistic''), past observations have revealed long-term extreme emission, {e.g.},  up to at least 20\,TeV during HEGRA observations of Mrk\,501 in 1997 \cite{1999A&A...349...11A} and H.E.S.S. observations in 2014 \cite{2019ApJ...870...93A}. The detection of such extreme, likely rare states will be invaluable for constraints of the EBL at the longest wavelengths.

In total, we simulate $830\,$hours of CTA observations of AGN expected to be detected at energies beyond the cosmic \gr horizon with a class-dependent cut-off. This observation budget represents about a quarter of the AGN KSP. Cumulative constraints on \gr propagation could also be expected from the remaining three quarters of the total live time, although likely in a subdominant manner. These latter observations will be of major interest for variability and spectral studies, even possibly morphology studies for the nearest AGN, which will constrain the acceleration and radiative conditions at play in astrophysical jets.

\subsection{Determination of the optical depth}

We use the simulated signal and background counts for AGN selected above to reconstruct the scale factor of the benchmark EBL model, while profiling over the intrinsic spectral parameters. The fit is performed using both the Levenberg-Marquardt and Monte-Carlo minimizations of \textsc{Sherpa}\footnote{\url{http://cxc.cfa.harvard.edu/contrib/sherpa/}} for cross checks. We use the same intrinsic spectral models for the fit as used in the simulations, {i.e.}\ a power law with an exponential cut-off in most cases, leaving all the parameters free to vary. The selection of the intrinsic model, and more particularly the number of degrees of freedom allowed in the fit, constitutes a source of uncertainty in the reconstruction of the EBL (see, {e.g.}, Ref.~\cite[][]{2019A&A...627A.110B}). We do not address here this source of uncertainty, noting that future joint works of the GeV and TeV communities on this methodological aspect are highly desirable.

We group the selected AGN ranging from $z=0.019$ to $z=1.84$ in several redshift intervals. We consider at first a bin size of $\Delta z=0.05$ and then merge consecutive bins until a minimum number of five AGN is reached in each interval. A lower minimum of four AGN is considered for the first redshift bin to illustrate the low-redshift performance of CTA. We simulate 1000 realizations of the spectra for each bin, reconstruct the optical-depth normalization for each realization through the profile likelihood method, and store the 16\%, 50\%, and 84\% quantiles of the distribution in each bin to estimate the median reconstructed normalization of the optical depth, as well as the associated $1\,\sigma$ uncertainties. This approach is repeated with bracketing IRFs to estimate the amplitude of the systematic uncertainties induced by the instrument, as indicated in Sec.~\ref{sec:cta-sim}. 

\begin{figure}[t]
  \center
  \includegraphics[width=\columnwidth]{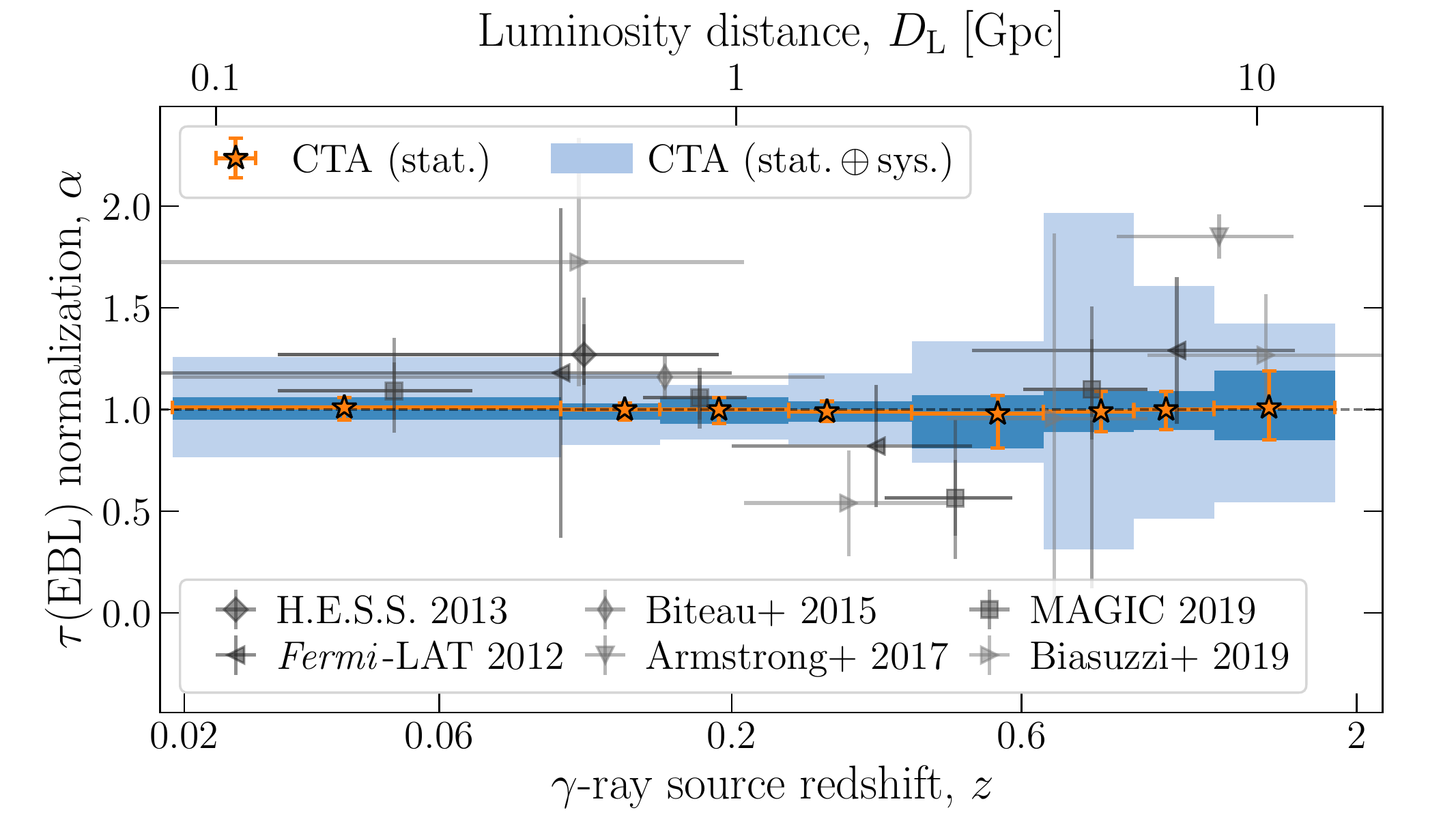}
  \caption{
Projected CTA constraints on the EBL scale factor as a function of redshift of the \gr sources. The median reconstructed scale factor and the 16--84\% quantiles of the distribution in each redshift bin are shown as orange stars and error bars (also dark blue shaded regions), respectively. The accumulated effect of statistical uncertainties and systematic uncertainties resulting from changes in the energy scale and effective area are illustrated with the light blue shaded regions (16--84\% quantiles). A class-dependent cut-off energy is considered for AGN in the joint-fit. The most up-to-date constraints available in the literature, shown in grey, are extracted from Refs.~\cite[][]{2012Sci...338.1190A, 2013A&A...550A...4H, 2015ApJ...812...60B, 2017MNRAS.470.4089A, 2019MNRAS.486.4233A, 2019A&A...627A.110B}. 
The upper row in the bottom legend indicates constraints from ground-based instruments while the lower row shows constraints from \textit{Fermi}-LAT data. 
  }
  \label{fig::ebl_constrains1}
\end{figure}

The results are summarized in Fig.~\ref{fig::ebl_constrains1} (see also App.~\ref{sec:ebl-extra-results}). Assuming a class-dependent comoving cut-off, the normalization of the optical depth is reconstructed with an average statistical uncertainty of $5\%$ in the first four redshift bins ($z < 0.4$) and of $10-15\%$ at higher redshifts ($0.4 < z < 1.85$). In comparison, measurements with current-generation IACTs access a redshift range limited to $z<1$, with statistical uncertainties of $10-15\%$ for $z<0.4$ and $20-25\%$ for $0.4<z<1$. The systematic uncertainties of instrumental origin are expected to be below $25\%$ up to a redshift $z < 0.65$, with a minimum of $12\%$ around $z\sim 0.2$, and to increase at large redshifts with an average value of $50\%$ for $0.65 < z < 1.85$, in line with expectations from current-generation instruments. Variations on the EBL scale factor resulting from varying IRFs up to $z=0.65$ are comparable to those obtained from state-of-the-art models of the EBL ({e.g.}, Refs.~\cite[][]{2011MNRAS.410.2556D,2010ApJ...712..238F}). These models converge on comparable spectra up to $8\,\mu$m (cosmic optical background) and on similar evolutions up to $z \sim 1$.

For the most distant \gr sources, the presence of a cut-off in the intrinsic spectral model at $E_{\rm cut}' = 100\,$GeV is hard to disentangle from absorption on the EBL, resulting in large systematic uncertainties, particularly marked for the redshift bin $z=0.65-0.9$. For such redshifts, the cosmic \gr horizon is located in the transition region between the LSTs and the MSTs, around $150\,$GeV, and therefore the systematic uncertainties of instrumental origin distort the observed spectra while the observed energy range only offers a limited handle on the intrinsic spectra of AGN. 

The spectral fit quality decreases for an instrument response increasingly deviating from the nominal one, which alludes to a likely overestimation of the impact of systematic uncertainties of instrumental origin. To tackle this limitation, a marginalization over the corresponding nuisance parameters, {e.g.}, within the framework of Bayesian hierarchical models, could be a promising approach to explore. Similarly, the EBL density is only parametrized in this work through a scale factor, which is a normalization of the optical depth. This approach is motivated by the intertwined dependences on redshift and wavelength of the EBL photon field. More advanced parametrizations of the population of UV-to-far-infrared sources contributing to the EBL will enable an assessment of the detailed constraints within the reach of CTA.

Nonetheless, the simplified approach adopted in this work provides a useful first glimpse at the territory to be covered by CTA. The low-redshift constraints from AGN with a comoving cut-off at 10\,TeV will crucially affect the capability of CTA to probe the still under-constrained CIB component up to 100$\,\mu$m. The highest sensitivity is expected from CTA for \gr sources in the redshift range $0.1-0.4$. Such intermediate-redshift observations will probe the core wavelength range of the COB, and in particular test for possible excesses, such as suggested by observations from the CIBER rocket experiment \cite{2017ApJ...839....7M}. CTA low-energy observations of \gr sources beyond $z\sim 0.5$, even with a cut-off at $100\,$GeV, will play an important role in constraining \gr interactions with UV photons down to 0.1$\,\mu$m. The capabilities of CTA at low energies will be crucial to constrain the cosmic star-formation history, particularly up to its peak located at $z\sim 1.5-2.5$, as probed by \textit{Fermi}~LAT \cite{2018Sci...362.1031F}. The combination of high-precision measurements from CTA with the large sample of \gr sources detected with \textit{Fermi}~LAT beyond $z=1$ holds a formidable potential not only to probe the EBL spectrum at $z=0$ over four decades in wavelength, in the range $0.1-100\,\mu$m,\footnote{{The UV component of the EBL remains underconstrained to a large extent (see Ref.~\cite{2016RSOS....350555C}). The near-UV region offers an interesting window of opportunity for combined analyses of \textit{Fermi}-LAT and CTA data.}}   but also its evolution over cosmic ages, including contributions from UV sources beyond $z \sim 2$ to which CTA will be sensitive by means of the integral nature of the EBL.

Observations from the AGN KSP of CTA will be essential to constrain the spectrum and evolution of the EBL. Complementary constraints of cosmological parameters, such as $H_0$ and $\Omega_M$, can also be expected from this observation program. 
The \gr optical depth is to first order proportional to the EBL density and inversely proportional to $H_0$ \cite{1994ApJ...423L...1S, 2005APh....23..598B, 2013ApJ...771L..34D, 2015ApJ...812...60B, 2019ApJ...885..137D}. As a result, for an EBL spectrum fixed to the level expected from galaxy counts, the uncertainty on $H_0$ is at least as great as the uncertainty on the scale factor $\alpha$. Dedicated studies will enable the determination of the full potential of CTA observations to constrain cosmological parameters.

\section{Deflections of electron-positron pairs in the intergalactic magnetic field }\label{sect::IGMF}

The IGMF is another cosmological entity that can be studied with \gr data~\citep{Plaga95, Ichiki08, NeronovSemikoz09}. The presence of an IGMF modifies the development of electromagnetic cascades initiated by \gr absorption. The lower-energy cascade emission thus provides a handle on the properties of the intervening IGMF. Various aspects of the IGMF influence can be probed: time delay of the cascade emission~\citep{2004ApJ...613.1072R, Murase08, Ichiki08, NeronovSemikoz09}, presence of broad spectral features due to the cascade contribution~\citep{Neronov10, Taylor11, Tavecchio11}, and extended, halo-like emission around point-like primary \gr sources~\citep{HEGRA_IGMF, 2010A&A...524A..77A, 2014A&A...562A.145H, 2017ApJ...835..288A}. CTA observations promise to address all of these effects at once.

Magnetically induced time delays~\citep{Murase08, Ichiki08, NeronovSemikoz09} originate from the difference in paths between the primary emission, propagating straight from the \gr source, and the cascade flux produced by electrons and positrons, which are subject to deflections. The value of the delay depends on the observed \gr energy and on the IGMF strength, $B_\mathrm{IGMF}$. For an observed \Gr with energy $E_\gamma \sim 100$\,GeV, the mean delay is expected to be on the order of $t_{\rm delay} \sim 3 \,{\rm years} \times (B_\mathrm{IGMF}/10^{-16}\,\mathrm{G})^2$ for a \gr source located at $z \sim 0.1$~\citep{Taylor11} and an injected \gr energy of 100\,TeV. The IGMF influence can be differentiated from the intrinsic source behavior through the energy dependence of the delay, which scales with measured \gr  energy as $t_{\rm delay} \propto E_\gamma^{-2}$. A detection of such coherent time delays at different energies, repeating from flare to flare, could thus be used to measure the IGMF strength.
Additionally, absence of multi-wavelength flaring activity, e.g., at X-ray energies, could also indicate that flares are not source intrinsic but reflect the delayed emission. 

A non-negligible IGMF not only delays but also spreads the cascade contribution in time, effectively decreasing its instantaneous flux. For short flares, delayed emission would only be detectable if the spread is sufficiently small for the cascade flux not to be suppressed below the sensitivity of CTA. The magnitude of the flux suppression is on the order of $t_\mathrm{flare} / (t_\mathrm{delay} + t_\mathrm{flare})$, where $t_\mathrm{flare}$ is the flare duration. Assuming a suppression by a factor of ten, a flare duration of one day translates into a maximum IGMF strength $B_\mathrm{IGMF} \lesssim 10^{-17}$\,G. With regular observations planned throughout the AGN long-term monitoring, CTA would be able to detect delayed emission from relatively bright primary flares, during which the \gr flux increases by a factor five to ten with respect to the median emission. 

In what follows, we focus the study on IGMF strengths higher than those probed with time delays, which could result in spectral and morphological signatures in \gr observations. In particular, we perform simulations of the prototypical extreme blazar 1ES\,0229+200 ($z=0.14$). Due to its hard intrinsic \gr spectrum with index $\Gamma < 2$, which extends to ${\sim}\,10\,$TeV, and the lack of strong \gr variability, 1ES\,0229+200 proves to be among the best-suited \gr sources for searching for cascade signatures. {Readers who may want to skip the discussion of the simulation setup are directed to Sec.~\ref{Sec:CombIGMF}.}

\subsection{Simulation of the cascade flux}\label{sect::IGMF_simulation_setup}

We simulate the development of electromagnetic cascades with the \textsc{CRPropa} code~\citep{CRPropa,2018APh...102...39H}, by injecting ${\gtrsim}\,10^6$ \Grs in the energy range 100\,GeV -- 100\,TeV, at a redshift of $z=0.14$ (${\sim}\,660$\,Mpc), within a conical jet with a $\theta_\mathrm{j} = 10^\circ$ opening angle. The \textsc{CRPropa} simulation includes pair production on the EBL and CMB, inverse Compton scattering on these two backgrounds, as well as adiabatic energy losses due to propagation in an expanding Universe.  
Despite considerable efforts on magnetohydrodynamic simulations (see {e.g.}\ Refs.~\cite{2014MNRAS.445.3706V,2016MNRAS.462..448G}), 
the coherence length of the IGMF is unknown. Here, we consider an IGMF composed of uniform cells of either 1\,Mpc or 0.01\,Mpc size, with random orientations of the magnetic field and a strength fixed to a value of $B_\mathrm{IGMF}$. 
The development of the cascade is probed over a volume occupied by $100\times100\times100$ such cells, which is periodically repeated in the simulations as needed to fill the relevant space. We do not track particles whose trajectory length exceeds 4\,Gpc to speed up the simulations. 
Secondary and primary \Grs falling on a sphere of radius $R=10$\,Mpc around the observer are retained and their arrival directions on the sphere are recorded. In order to simulate any spectral shape at the source, we record the observed spectrum as a function of injected \gr energy. We can then generate observed spectra with arbitrary spectral shapes by simply re-weighting the injected spectrum without the need to re-run the \textsc{CRPropa} simulation (similarly to Ref.~\cite{2018ApJS..237...32A}). 

We assume that the intrinsic spectrum of 1ES\,0229+200 follows a power law, with spectral parameters provided in App.~\ref{sec:intr-models}, incorporating an exponential cut-off at $E_\mathrm{cut}' = 10$\,TeV, which is in line with the lower limit set by VHE observations of this object~\citep{HESS_1ES0229, VERITAS_1ES0229}. We then use \ctools to simulate an exposure of CTA North, {i.e.}\ one Monte-Carlo realization, of 50\,hours at a zenith angle of $20^\circ$ and to compute the likelihood for a given set of spectral parameters ($N_{0}, \Gamma, E_\mathrm{cut}$) for each tested IGMF setup. We simulate CTA data above an energy threshold of 50\,GeV for this science case. Given the steep increase of the effective area at lower energies, systematic uncertainties in the energy scale could lead to a distortion of the reconstructed spectrum affecting IGMF searches at lower energies. 

\subsection{Gamma-ray observables of the IGMF at high field strengths}\label{sec:IGMFillustration}

\begin{figure}[t]
  \center
  \includegraphics[width=0.8\columnwidth]{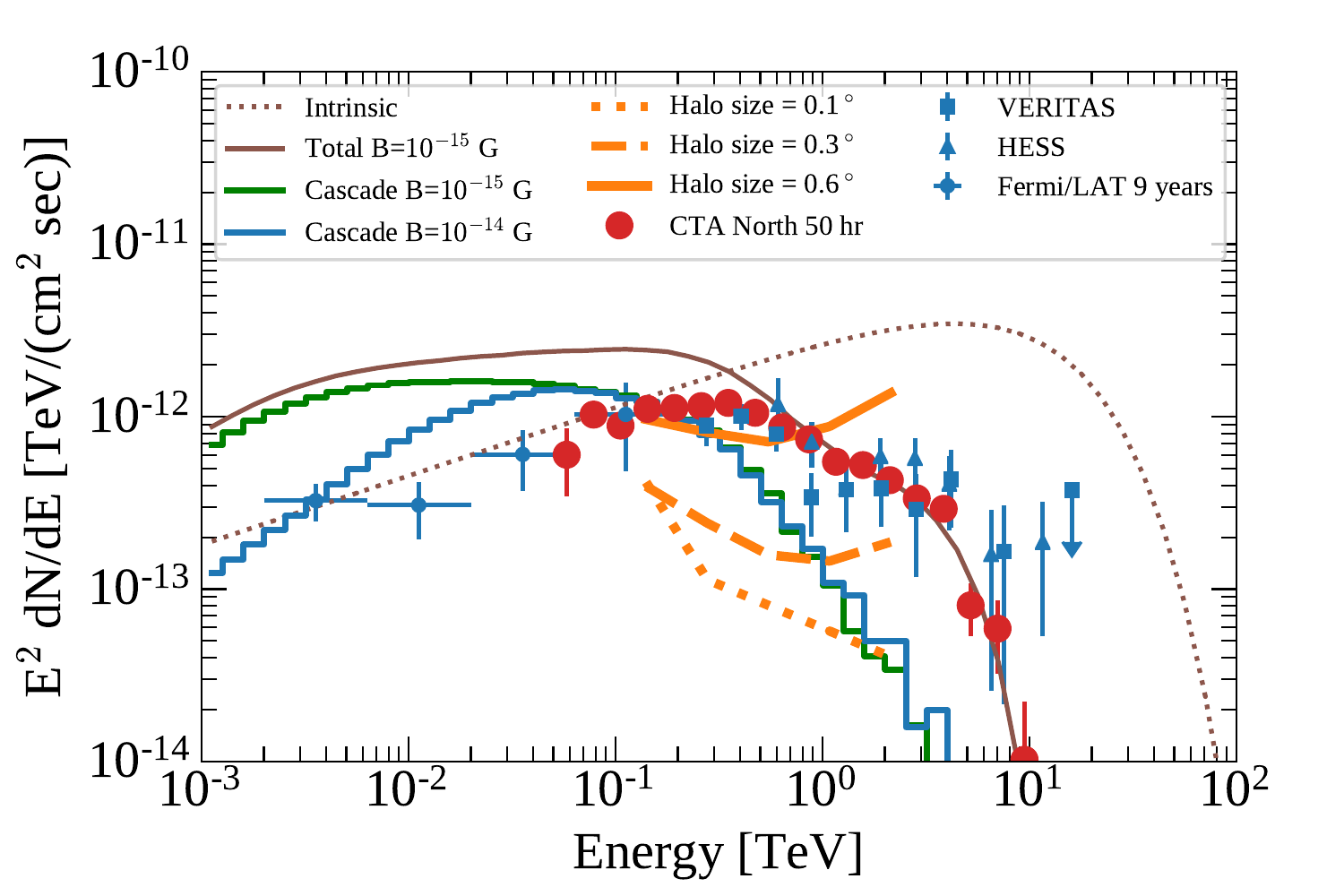}
  \caption{
Simulated spectrum of the blazar 1ES\,0229+200 and sensitivity of CTA to an IGMF-induced halo.
The differential photon spectrum of 1ES\,0229+200 multiplied by the square of energy is displayed as a function of \gr energy as red points. For comparison, spectra from \textit{Fermi}~LAT (custom Pass~8 analysis of the 9-year long data set), H.E.S.S.~\cite[][]{HESS_1ES0229} and VERITAS~\cite[][]{VERITAS_1ES0229} are shown as blue dots, blue triangles and blue squares, respectively. The  3\,$\sigma$ sensitivities of CTA~North to the cascade emission from a single \gr source are shown for fixed extensions of $0.1^\circ$, $0.3^\circ$ and $0.6^\circ$ as dotted, dashed and continuous orange lines, respectively. Fifty hours of observations with CTA~North are simulated assuming an intrinsic comoving cut-off at 10\,TeV and no cascade component. The brown dotted line illustrates the intrinsic spectrum. The corresponding cascade fluxes for IGMF strengths of $10^{-15}\,$G and $10^{-14}\,$G are shown as green and blue line segments, respectively.  The total flux, resulting from the sum of the direct and cascade emissions, is shown as a brown solid line for an IGMF strength of $10^{-15}\,$G.   
  }
  \label{fig::igfm_halo_uls}
\end{figure}

For IGMF strengths above $10^{-17}\,$G, two main observable effects could be probed with CTA in \gr observations of blazars. The first observable would be the presence of a low-energy spectral component on top of the observed point-source spectra. The all-sky spectra from such cascade components are shown as solid lines in Fig.~\ref{fig::igfm_halo_uls} for $B_\mathrm{IMGF} = 10^{-15}$\,G and $B_\mathrm{IGMF} = 10^{-14}$\,G, assuming that the blazar 1ES\,0229+200 has been emitting \Grs for  $10^7$\,years.\footnote{This value is used as an estimate of the maximum AGN activity timescale \cite{2002NewAR..46..313P}. Shorter activity times are discussed in Sec.~\ref{Sec:CombIGMF}, as they can suppress the cascade emission~\cite{dermer11, finke15, meyer16,2018ApJS..237...32A}.}

The second, inarguable observable of the IGMF would be the presence of extended \gr halos around distant blazars. These halos would originate from the deflections of the electron-positron pairs in the presence of a sufficiently strong IGMF and have already been searched for with existing IACTs, currently offering the best angular resolution above several tens of GeV~\citep{HEGRA_IGMF, 2010A&A...524A..77A, 2014A&A...562A.145H, 2017ApJ...835..288A}. First studies showed that such halos could be within the reach of CTA \cite{2013APh....43..215S, meyer16, 2018ApJ...869...43F}. 
A characteristic angular spread of the cascade caused by the intervening IGMF is $\theta \simeq 0.5^\circ \times (B_\mathrm{IGMF}/10^{-14}\,\mathrm{G})$ at 100\,GeV~\citep{NeronovSemikoz09}. 
The angular resolution of \gr instruments sets the minimum observable angular spread induced by the IGMF. 
With a foreseen angular resolution of  ${\sim}\,0.13^\circ$ at 100\,GeV, CTA could search for small halos corresponding to IGMF strengths down to $B_\mathrm{IGMF} \gtrsim 3 \times 10^{-15}\,\mathrm{G}$.\footnote{The indicated minimum IGMF strength is a conservative estimate as sub-PSF-scale structures can be resolved by the instrument provided a sufficiently large number of signal events.}
The maximum observable strength of the IGMF is set by the halo surface brightness, whose detection depends both on the cascade suppression due to isotropization and on the sensitivity of the instrument. Improved sensitivities enable searches for larger halos with lower surface brightness, corresponding to higher magnetic fields. 

 Because of a larger angle between primary and secondary photons, the cascade component is increasingly suppressed with increasing IGMF strength. The limiting case for cascade suppression corresponds to an IGMF stronger than $B_\mathrm{IGMF} \gtrsim 10^{-13}\mathrm{-}10^{-12}$\,G, for which the suppression factor, ${\sim}\,(D_\gamma / D_\mathrm{src})^2$, becomes independent of $B_\mathrm{IGMF}$. In the case of 1ES\,0229+200, where $D_\mathrm{src} \simeq 660$\,Mpc is the distance of the blazar and $D_\gamma \simeq 80$\,Mpc is the mean free path of a 10\,TeV \Gr, the secondary radiation around 100\,GeV is maximally suppressed by a factor $(D_\gamma / D_\mathrm{src})^2 \sim 1/70$.

We estimate the sensitivity of CTA to such halos by simulating a 50-hour observation of 1ES\,0229+200. 
As the halo brightness distribution depends on several unknown parameters, such as the IGMF coherence length, jet orientation, and AGN activity evolution~\citep{Neronov:2010, Neronov:2013}, we use in Fig.~\ref{fig::igfm_halo_uls} {the} simplified assumption of a disk-like halo brightness profile. Despite its simplicity, this approach enables a first evaluation of the effect of the cascade spread induced by the IGMF. We fit the sum of a point source and of an extended halo component to the angular \gr distribution. The sensitivity limit is computed as the minimal halo flux of fixed extension that results in a $3\,\sigma$ detection of the extended component. The outcome of this simulation, shown in Fig.~\ref{fig::igfm_halo_uls} as orange lines, demonstrates that 50\,hours of data would be sufficient to detect the putative halo around the blazar.

\subsection{Combined CTA sensitivity to IGMF}
\label{Sec:CombIGMF}

The sections above provide an intuition on the CTA sensitivity to IGMF-induced effects. We now quantify the sensitivity by combining the spectral and morphological approaches. This combination is enabled by updating the halo model at each step of the fit consistently with the point source spectrum tested against the data, using the pre-computed halo library described in Sec.~\ref{sect::IGMF_simulation_setup}. In this way, the fit takes into account the halo spectrum and the angular spread in a self-consistent way.  It should be noted, though, that an accurate simulation of the intergalactic cascade appearance relies on knowledge of the blazar-specific parameters like jet opening angle, orientation, and temporal flux evolution in the past ${\sim}\,10-10^7$\,years. These properties are often poorly constrained, which would suggest marginalizing over such nuisance parameters to constrain the IGMF. However, an accurate account of these uncertainties requires coverage over a large parameter space, making both simulations and analyses challenging in terms of computing time. 

For this reason, a simplified procedure is employed, where it is assumed that the jet of 1ES\,0229+200 has a simple conical shape with a $10^\circ$ opening angle and is either aligned with the line of sight or tilted by an angle of $5^\circ$. We checked that reducing the opening angle by a factor of two does not impact the results in a measurable way in the energy range covered by CTA, in agreement with the results presented in Ref.~\cite{2017MNRAS.466.3472F}.\footnote{
It should be noted though that opening angles smaller than $1^\circ$ would suppress the observed cascade emission in the energy range accessible to CTA~\cite{2017MNRAS.466.3472F}.
}
Two limiting cases are further considered: on the one hand, the blazar is assumed to have been active at the current-day average flux for $10^7$\,years and, on the other hand, that it has been active only in the last 10\,years (the period over which this blazar has been observed by \gr instruments). The underlying assumption, as in other related studies, is that the current TeV emission of the blazar traces its past emission averaged over timescales comparable to that of the cascade development (on the order of a Myr for 100\,GeV secondary photons and an IGMF strength of $10^{-14}\,$G \cite{dermer11}). Such an assumption could be either pessimistic or optimistic if the average long-term emission of 1ES\,0229+200 was larger or smaller, respectively, than the currently observed one. Finally, two possible values of the IGMF coherence scale are considered, $\lambda_\mathrm{B}=1$\,Mpc and $\lambda_\mathrm{B}=0.01$\,Mpc.

\begin{figure}[t]
  \center
  \includegraphics[width=0.8\columnwidth]{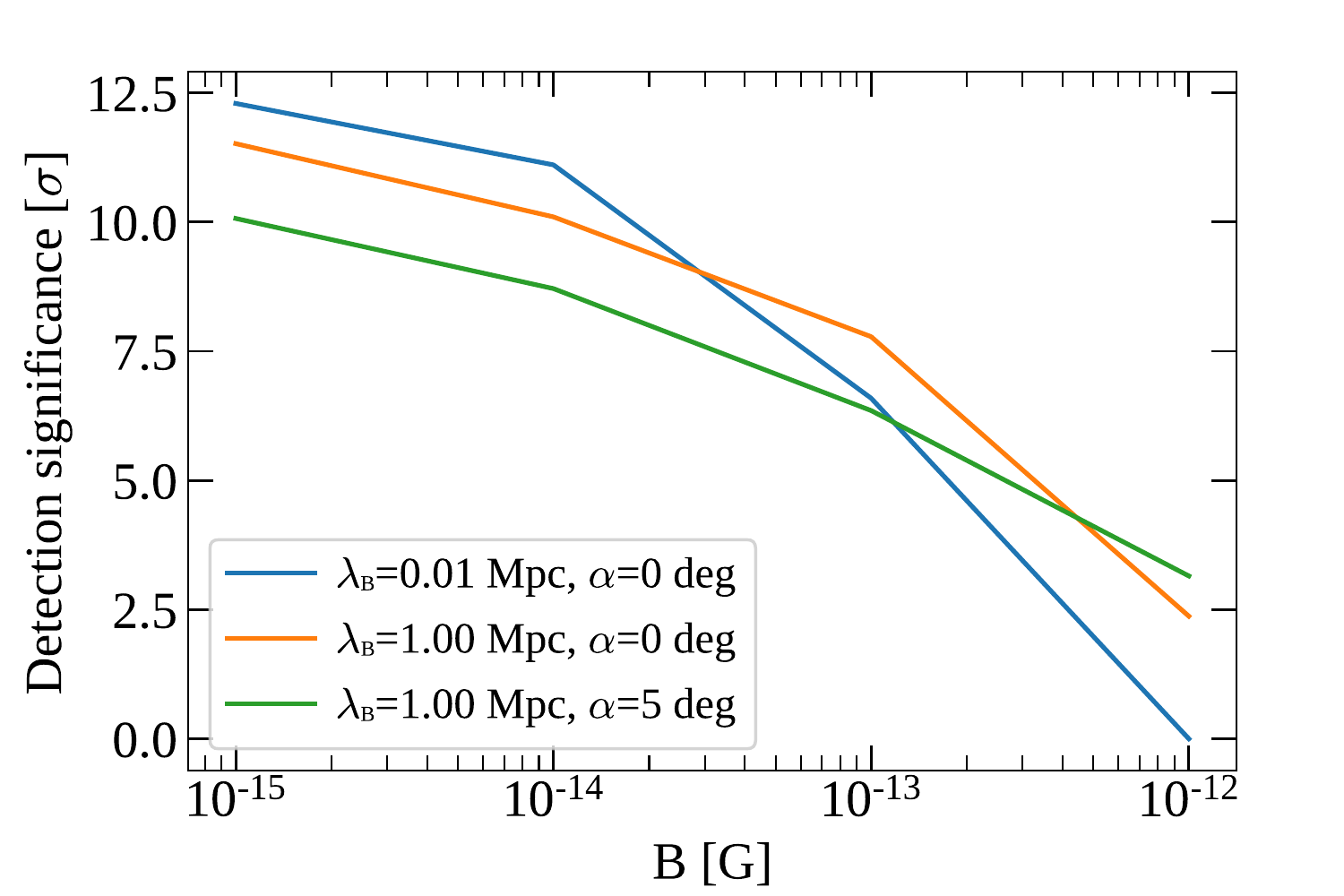}
  \caption{
Detection significance of the secondary cascade component in the spectrum of 1ES\,0229+200 as a function of IGMF strength, $B$. The configuration corresponding to a coherence scale $\lambda_{B} = 1\,$Mpc and a jet orientation $\alpha = 0^\circ$ with respect to the line of sight is shown as the orange line. The blue and green lines illustrate the impact of a change in the coherence scale ($\lambda_{B} = 0.01\,$Mpc, blue line) or jet orientation ($\alpha = 5^\circ$, green line). The estimates are obtained under the assumption of average emission over a $10^7$-year time scale.
  }
  \label{fig::igfm_halo_significance}
\end{figure}

We compute the detection significance of the cascade component accounting for both its spectral and morphological signatures. The cascade normalization scale, $s$, is fixed relative to the prediction of the \textsc{CRPropa} simulation and the log-likelihood $\ln\mathcal{L}(s)$ is computed by profiling over the intrinsic source spectral parameters. Two values of $s$ are used to compute the cascade detection significance as $\sqrt{-2(\ln\mathcal{L}(s=0) - \ln\mathcal{L}(s=1))}$ which is always smaller or equal to the log-likelihood ratio $\sqrt{-2(\ln\mathcal{L}(s=0) - \ln\mathcal{L}(\hat{s}))}$, where $\hat{s}$ is the best-fit value. 
This latter ratio follows a $\chi^2$ distribution with one degree of freedom. 

The results of this calculation are shown in Fig.~\ref{fig::igfm_halo_significance}. In all cases, CTA will be able to detect a cascade emission for IGMF strengths smaller than ${\sim}\,10^{-13}$\,G. Limits from current-generation instruments have disfavored IGMF strengths up to a few times $10^{-15}\,$G~\citep{HEGRA_IGMF, 2010A&A...524A..77A, 2014A&A...562A.145H}, with the most stringent ones disfavoring strengths up to $7 \times 10^{-14}\,$G at the 95\% confidence level \cite{2017ApJ...835..288A}. Ample room for discovery remains, with IGMF strengths up to an order of magnitude larger yielding a signal detectable by CTA, depending on the configuration of the jet and coherence length of the field. 

\begin{figure}[t]
  \center
  \includegraphics[width=0.92\columnwidth]{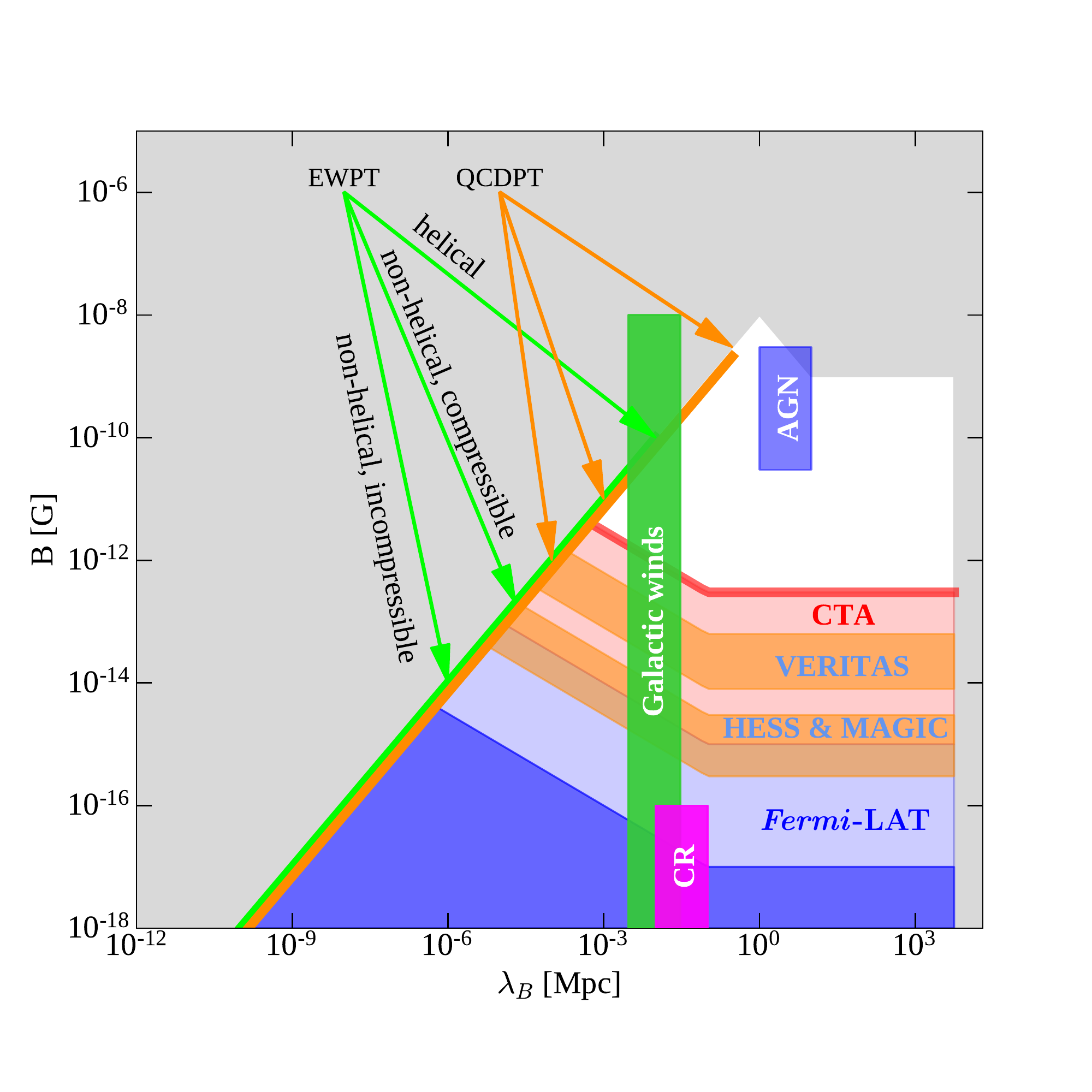}
  \caption{
Projected CTA sensitivity to IGMF signatures as a function of the IGMF strength, $B$, and coherence scale, $\lambda_{B}$. The red line marks the maximal IGMF strength that would be detectable at $5\,\sigma$ level in a 50-hour long CTA observation of 1ES\,0229+200, assuming that the blazar was active for ${\sim}\,10^7$~yr with a $10^\circ$-wide jet inclined by $5^\circ$ with respect to the line of sight. The white region is beyond the sensitivity of the instruments discussed here. Exclusion regions at the 95\% to 99\% confidence level from H.E.S.S., MAGIC (overlapping,~\cite[][]{2014A&A...562A.145H, 2010A&A...524A..77A}) and VERITAS~\cite{2017ApJ...835..288A} are shown in orange. The dark and light blue regions denote the \emph{Fermi}-LAT lower limit on IGMF strength from Ref.~\citep{Taylor11} and Ref.~\citep{2018ApJS..237...32A}, within the validity range of their simulations. Grey regions are disfavored by direct probes and theoretical considerations. 
The orange and green lines represent theoretically favored regions for the generation of the IGMF during either electro-weak (EWPT) or QCD (QCDPT) phase transitions (see Ref.~\citep[][]{2013A&ARv..21...62D} and references therein). Filled vertical boxes show favored regions of models where the IGMF is generated by a frozen-in magnetic field, originating from AGN outflows~\cite{Furlanetto:2001}, galactic winds~\citep{Bertone:2006} or induced by cosmic-ray streaming~\citep{Miniati:2011}, as labelled in the figure. Figure adapted from~\citep{2013A&ARv..21...62D}.  
  }
  \label{fig::igfm_exclusion_plot}
\end{figure}

The sensitivity region where the IGMF could be detected with CTA at the 5\,$\sigma$ confidence level is shown in Fig.~\ref{fig::igfm_exclusion_plot} under the assumption of a 10$^\circ$ jet opening angle, a 5$^\circ$ inclination of the jet axis to the line of sight, and a blazar lifetime of $10^7$\,yr. The parameter space that can be probed with CTA extends up to $3\times 10^{-13}$\,G for this set of parameters. CTA measurements will thus almost close the gap between the existing IGMF constraints and the maximal IGMF strength consistent with models of galaxy formation~\cite{dolag05, donnert09}.

An improvement by two orders of magnitude is expected with respect to constraints from \emph{Fermi}-LAT observations combined with spectra from current-generation IACTs \cite{2018ApJS..237...32A}. 
The factor limiting the constraints in Ref.~\cite{2018ApJS..237...32A} lies in the use of the small-angle approximation to compute the cascade, which becomes unreliable for large values of the IGMF strength. The validity region of their limits is shown in Fig.~\ref{fig::igfm_exclusion_plot} as a blue shaded region. 

Assuming shorter maximum blazar duty cycles would substantially reduce CTA sensitivity to IGMF signatures. As shown in the recent combined \emph{Fermi}-LAT and IACT search, the limits on the IGMF degrade by 1.5 orders of magnitude each when the activity time scale is reduced from $10^7$ to $10^4$ and from $10^4$ to $10$ years \cite{2018ApJS..237...32A}, in line with analytical expectations~\cite{dermer11}. A similar degradation should also apply here. 
On the other hand, a combination of CTA and \emph{Fermi}-LAT data could further broaden the probed parameter space, provided contemporaneous observations that are crucially needed for variable \gr sources. Even assuming discontinued observations at GeV energies during the CTA observation program, a combination of GeV--TeV spectral and morphological data from steady \gr sources could be expected to strengthen the constraints on the IGMF.

We have assumed a cut-off energy around ${\sim}\,10$\,TeV for 1ES\,0229+200, close to the lowest value compatible with current observations \cite{HESS_1ES0229, VERITAS_1ES0229}. A higher value of the cut-off energy would boost the expected cascade flux, putting CTA in a more favorable position to detect the signature of the IGMF. Although CTA will provide a significantly improved sensitivity above 10\,TeV, a direct measurement of the high-energy cut-off in the spectrum of 1ES\,0229+200 may remain out of reach. This lack of sensitivity to a high-energy cut-off is related to the rapidly increasing \gr optical depth, which reaches $\tau \approx 7$ at 10\,TeV and $\tau \approx 25$ at 20\,TeV. 

We have not addressed the systematic uncertainties introduced by a mis-modelling of the instrumental response.
Constraints of the halo could be affected in particular by an erroneous effective area (leading to wrong estimates of the intrinsic blazar spectrum and hence the cascade component) and PSF. 
A too broad (too narrow) PSF could lead to an under- (over-) estimate of the significance of the additional halo component. 
The reason {for} not {including} this uncertainty here is because of the dominating systematic uncertainty in the blazar duty cycle mentioned above. A shorter duty cycle could easily degrade the sensitivity by orders of magnitude.  
We therefore leave a re-analysis of the simulations with bracketing IRFs for future work.

In conclusion, the detection of secondary emission would provide valuable information on the IGMF strength. Our estimates suggest that CTA will provide sensitivity to field strengths up to $3\times 10^{-13}$\,G.
This region includes the IGMF parameter space where {indications have} been found for a helical IGMF in arrival directions of diffuse \Grs~\cite{2015MNRAS.450.3371C}.
More generally, if CTA detects \gr halos around blazars, the spatial shape of these halos could also be used to constrain the IGMF helicity~\cite{2016PhRvD..94h3005A}. 
CTA observations will probe a range where the IGMF could arise through different mechanisms. For example,  plasma outflows of supernova-induced winds might have ``polluted'' the intergalactic medium with a magnetic field, and thus an IGMF detection could provide possible clues on the feedback of such winds on galaxy formation and evolution. 

Alternatively, primordial phase transitions, provided they are of first order, could generate an IGMF.
These do not occur in the Standard Model, however, QCD axions of $\sim$meV mass (which can make up the dark matter) can also seed the formation of an IGMF at the QCD ``crossover'' in the early universe \cite{2018PhRvL.121b1301M}.
This process provides an independent channel for CTA to probe dark-matter WISPs, complementary to the searches that we discuss in Sec.~\ref{sec:axions}.
Finally, even in case of no signal, the absence of cascade emission in CTA observations could indicate that the electron-positron pairs predominantly lose their energy in plasma instabilities, thereby heating the intergalactic medium~\cite{2012ApJ...752...22B,2018ApJ...868...87B}.

\section{Coupling of \Grs to axion-like particles}
\label{sec:axions}

The propagation of \Grs could be affected by interactions with yet undiscovered particles beyond the Standard Model of particle physics. In particular, \Grs could oscillate into ALPs in ambient magnetic fields. ALPs are spin-zero particles, either as pseudo Nambu-Goldstone bosons that arise when additional fundamental gauge symmetries are broken or as Kaluza-Klein zero modes of compactified string theories (see, {e.g.}, Ref.~\cite[][]{2018PrPNP.102...89I} for a review). As their name suggests, they are closely related to axions that have been proposed to solve the so-called strong CP problem of the strong interaction~\cite{pq1977,weinberg1978,wilczek1978}. In contrast to axions, the ALP mass, $m_a$, and coupling to photons, $g_{a\gamma}$, are independent parameters. When sufficiently light, axions and ALPs can be dark matter candidates if produced non-thermally in the early Universe~\cite{abbott1983,preskill1983,dine1983,arias2012, 2016Natur.539...69B}.

In the presence of a homogeneous external magnetic field of strength $B$ that is transverse to the photon propagation, the photon state parallel to the field mixes with ALPs~\cite{raffelt1988}. Above a critical energy marking the transition to the so-called strong mixing regime, $E_\mathrm{crit}$, the conversion probability in a homogeneous field is given by $P_{a\gamma} \sim  \sin^2 (g_{a \gamma} \, B l/2)$, where $l$ is the distance traveled within the field. This further simplifies to $P_{a\gamma} \sim (g_{a \gamma} \, B l/2)^2$ if the photon-ALP oscillation length is much larger than the distance travelled in the field (see, e.g., Refs.~\cite{raffelt1988,2009JCAP...12..004M,deangelis2011,2015PhRvD..91h3003D,2018PhRvD..98d3018G} for detailed studies of the full expressions of $P_{a\gamma}$). 
Oscillatory features are expected to occur around the critical energy, which depends on the magnetic field as~({e.g.}, Ref.~\cite[][]{2007PhRvL..99w1102H})
\begin{equation}
    E_\mathrm{crit} \sim 2.5\,\mathrm{GeV}\left(\frac{|m_a - \omega_\mathrm{pl}|}{1\,\mathrm{neV}}\right)^2\left(\frac{B}{1\,\mu\mathrm{G}}\right)^{-1}\left(\frac{g_{a\gamma}}{10^{-11}\,\mathrm{GeV}^{-1}}\right)^{-1},
\end{equation}
where $\omega_\mathrm{pl}$ is the plasma frequency of the medium. The strong mixing regime terminates at an energy $E_\mathrm{max}$. For the specific AGN selected for our study, $E_\mathrm{max}$ is beyond the CTA sensitivity reach for the ALP parameters and the magnetic fields considered here (see below and {e.g.},  Ref.~\cite[][]{meyer2014}).

Photon-ALP conversions can lead to distinctive signatures in AGN spectra. On the one hand, ALPs evade pair production with photon fields such as the EBL and the photon-ALP oscillation can thus significantly reduce the effective optical depth. Indications for such a reduction could have been found in blazar observations both with ground-based instruments and \emph{Fermi}~LAT~\cite{deangelis2007,deangelis2011,horns2012,essey2012,rubtsov2014,kohri2017} and were interpreted as evidence for ALPs~\cite{mirizzi2007,deangelis2007,sanchezconde2009,deangelis2011,dominguez2011alps,meyer2013}. However, recent analyses found that \gr spectra are in general compatible with predictions from attenuation on the EBL~\cite{sanchez2013,dominguez2015,2015ApJ...812...60B}. Such effects will be actively probed with CTA~\cite{2013JCAP...11..023M, meyer2014cta,wouters2014,2015PhLB..744..375T,2018JHEAp..20....1G}.

On the other hand, at energies around $E_\mathrm{crit}$, oscillatory patterns are expected in AGN spectra that depend on the morphology of the traversed magnetic fields~\cite[][]{ostman2005,wouters2012}. The search for such features at \gr energies has resulted in strong bounds on the photon-ALP coupling; in particular H.E.S.S. and \emph{Fermi}-LAT observations of PKS\,2155$-$304~\cite{hess2013:alps,zhang2018} as well as \emph{Fermi}-LAT observations of the radio galaxy NGC\,1275~\cite{ajello2016} resulted in the strongest bounds on the photon-ALP coupling to date for $4\,\mathrm{neV} \lesssim m_a\lesssim 100\,\mathrm{neV}$. In what follows, we focus on the CTA sensitivity to these spectral features using simulated observations of the radio galaxy NGC\,1275 {(see {e.g.}\ Ref.~\cite{2019MNRAS.487..123G} for ALP searches with BL\,Lacs). Readers who may want to skip the discussion of the simulations and analysis are directed to Sec.~\ref{sec:alp-results}.}

\subsection{Model for photon-ALP oscillations for \gr observations of NGC\,1275}

NGC\,1275 is the central galaxy of the Perseus cluster, at a distance of ${\sim}\,75\,$Mpc ($z = 0.01756$). The cool-core Perseus cluster harbors a strong magnetic field, as large as $25\,\mu$G in the cluster center based on Faraday rotation measurements~\cite{taylor2006}, in which \Grs and ALPs can couple. 

We closely follow Ref.~\cite{ajello2016} for the computation of the photon-ALP conversion probability. The magnetic field of the Milky Way, with a strength of the order of $\mu$G, is modeled according to Ref.~\cite{jannson2012}. We include the magnetic field of the Perseus cluster, whose central magnetic-field strength is conservatively assumed to be $10\,\mu$G. The morphology of the cluster magnetic field is modeled as a random field with Gaussian turbulence~\cite{meyer2014}. The turbulence spectrum is assumed to follow a power law of index $q_B = 2.8 \pm 1.3$, between distance scales $\Lambda_\mathrm{min} = (0.7\pm0.1)$\,kpc and $\Lambda_\mathrm{max} = (35\pm 23)$\,kpc. These values are taken from observations of the cool-core cluster A\,2199~\cite{vacca2012}, as they are not known for Perseus. The extent of the cluster field is assumed to be 500\,kpc and the electron density, $n_\mathrm{el}(r)$, is taken from Ref.~\cite{churazov2003}. The magnetic field is assumed to follow the electron distribution as $B \propto (n_\mathrm{el})^\eta$, with $\eta = 0.5$ (also in agreement with the observations of A\,2199). Since NGC\,1275 is a radio galaxy observed under large viewing angles with respect to the line of sight \cite{2017MNRAS.465L..94F}, we neglect photon-ALP mixing in the magnetic field of AGN jets.

We generate 100 random realizations of cluster magnetic-field configurations and numerically calculate, using the \textsc{gammaALPs} code,\footnote{\url{https://github.com/me-manu/gammaALPs}} the probability to observe at Earth a \Gr of either polarization for an initially unpolarized photon beam, {i.e.}, the so-called photon survival probability, $P_{\gamma\gamma}$. The code computes the solution of the equations of motion of the photon-ALP system based on the transfer-matrix formalism and incorporates all relevant terms in the photon-ALP mixing matrix, including the dispersion terms of QED effects and the CMB~\cite{kartavtsev2017},\footnote{For the considered magnetic-field model, the dispersion terms of QED effects and the CMB effects become important for energies $\gtrsim1\,$TeV.} as well as \gr absorption on the EBL.

\subsection{Simulation and analysis of CTA observations of NGC\,1275}
\label{Sec:ALPsim}

A total of 300\,hours within the first five years of operations is planned to be dedicated to the Cluster of Galaxies KSP of CTA. We use these observations of the Perseus cluster and NGC\,1275 to assess the sensitivity of CTA to ALP-induced oscillations assuming (i) a 300-hour exposure and an intrinsic spectrum equal to the average spectrum observed in over 250\,hours with MAGIC in a quiescent flux state~\cite{magic-preseus2016} and (ii) a 10-hour exposure with the spectrum obtained with MAGIC during a \gr flare~\cite{2018A&A...617A..91M}, also observed with VERITAS~\cite{benbow2017}. The quiescent (flare) photon spectrum is well described by a power law (power law with exponential cut-off) with a normalization $N_0 = 2.1\times10^{-11}\,\mathrm{TeV}^{-1}\,\mathrm{cm}^{-2}\,\mathrm{s}^{-1}$ ($N_0 = 1.54\times10^{-9}\,\mathrm{TeV}^{-1}\,\mathrm{cm}^{-2}\,\mathrm{s}^{-1}$) at energy $E_0 = 200\,\mathrm{GeV}$ ($E_0 = 300\,\mathrm{GeV}$) and power-law index $\Gamma =3.6 $ ($\Gamma = 2.11$ and exponential cut-off energy $E_\mathrm{cut} = 560\,$GeV). 

As discussed in Sec.~\ref{sec:cta-sim_subsec_sys_unc}, we include systematic uncertainties connected to the instrument response directly in our model by introducing two additional nuisance parameters. 
These two parameters, $s$ and $\delta$, describe a potential shift in the energy scale and an additional smearing of the CTA energy dispersion, respectively. 
The full model is then given by
\begin{equation}
    \phi(E) = \frac{1}{\mathcal{N}} \int\limits_0^\infty \mathrm{d}E' \exp\left(-\frac{(E-E')^2}{2(\delta E')^2}\right)
    \phi_\mathrm{int}\left((1-s)E'\right) P_{\gamma\gamma}\left((1-s)E', g_{a\gamma}, m_a, \mathbf{B} \right),
\end{equation}
where
\begin{equation}
\mathcal{N} = \int\limits_0^\infty \mathrm{d}E' \exp\left(-\frac{(E-E')^2}{2(\delta E')^2}\right)    
\end{equation}
is a normalization factor and $\mathbf{B}$ is the considered magnetic field realization.

\begin{figure}[t]
    \centering
    \includegraphics[width = 0.49\linewidth]{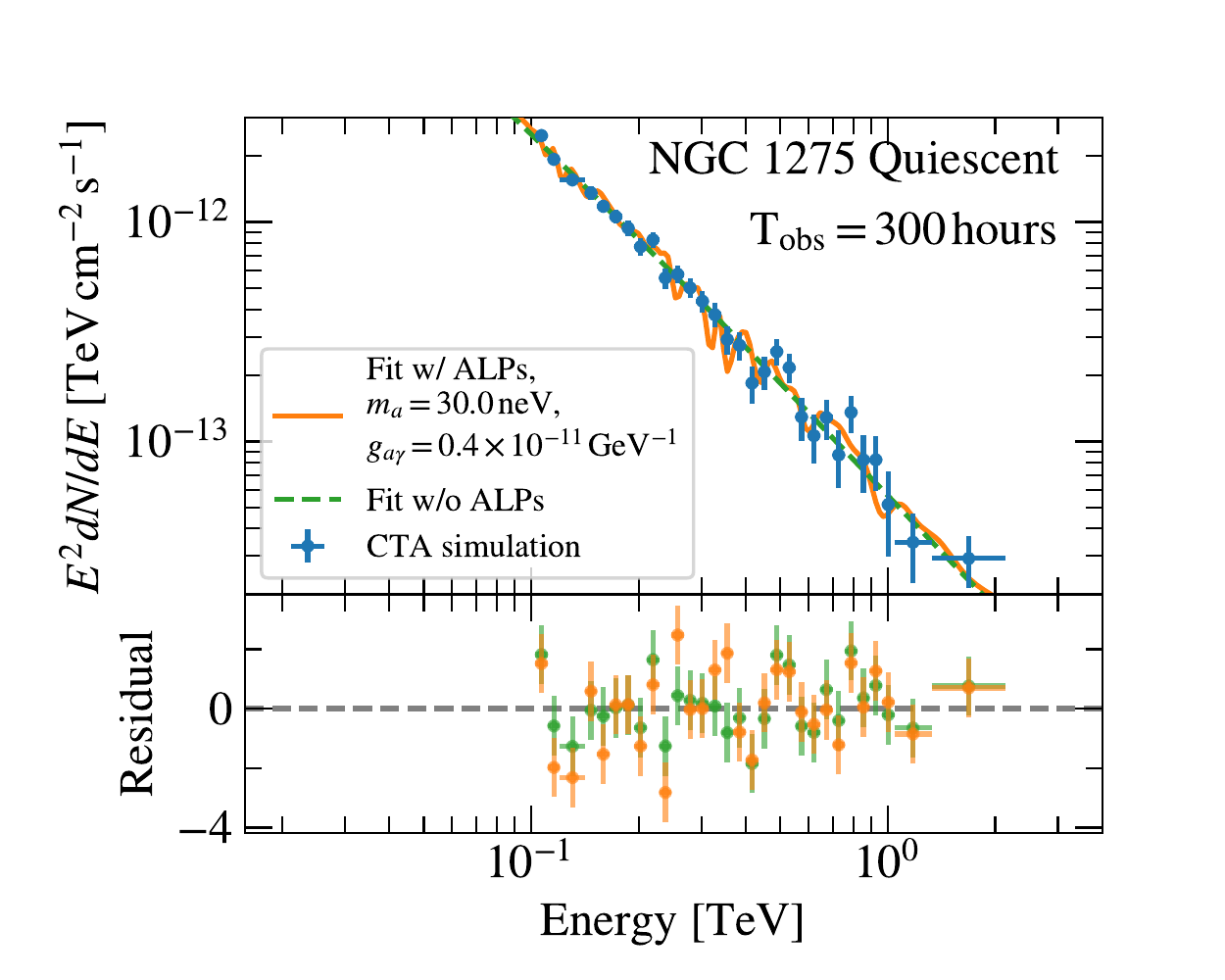}\hfill
    \includegraphics[width = 0.49\linewidth]{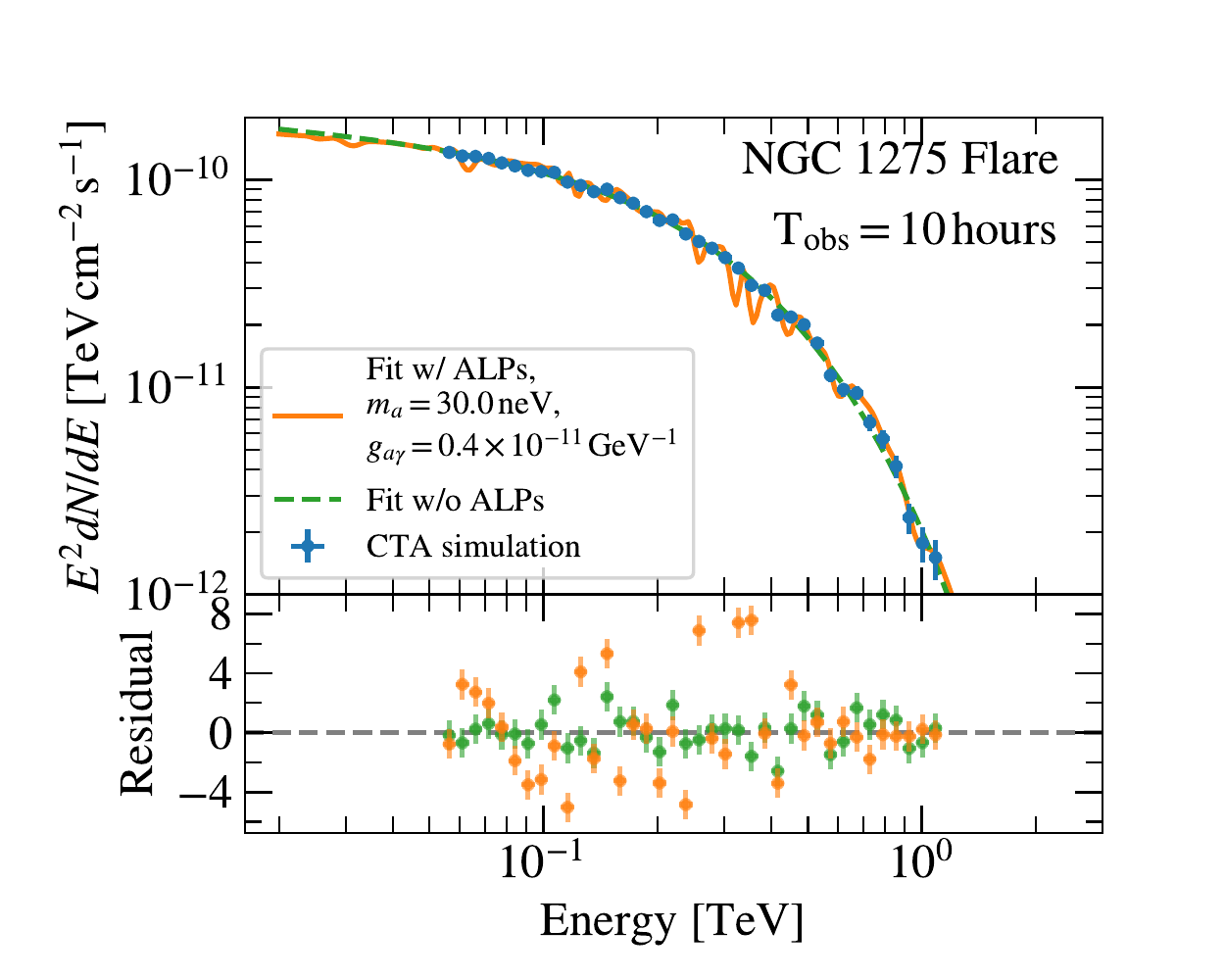}
    \caption{\label{fig:alp-spec} 
Simulated spectra of the radio galaxy NGC\,1275 to assess the sensitivity of CTA to ALP-induced irregularities.
The differential spectrum of NGC\,1275 multiplied by the square of energy is displayed as a function of \gr energy as blue points in the top panels, either in the quiescent (\textit{left}) or flaring state (\textit{right}) of the \gr source.
For illustration, the observation is simulated without an ALP effect and is modeled both without ALPs (green dashed line) and with a fixed set of magnetic-field realization and ALP parameters that are excluded at 95\,\% confidence by the flaring state simulation  (orange line).
Residuals are shown in the lower panel for the best-fit models with and without ALP effect as orange and green points, respectively. The residuals are defined as $(f - \phi(E)) / \sigma_f$, where $f$ is the measured flux at energy $E$ with uncertainty $\sigma_f$ and $\phi(E)$ is the model prediction. 
     }
\end{figure}

We show examples of simulated observations during the average and flaring states with the northern array of CTA in Fig.~\ref{fig:alp-spec}. As for IGMF studies, we adopt an energy threshold of 50\,GeV to limit deviations from the underlying spectrum at lower energies. The input spectrum includes \gr absorption, marginal in this case, but no ALPs. The figure also shows spectral fits both with and without ALPs. The chosen parameters correspond to values for which ALPs could account for all dark matter and one random realization of the Perseus magnetic field is used. The fits are performed with \textsc{Minuit}~\cite{minuit} by maximizing the likelihood in Eq.~\eqref{eq:lnl} summed over all energy bins passing selection for a particular realization. Contrarily to the other science cases presented in this work, we use forty bins instead of ten bins per decade. Such a fine binning is necessary in order to resolve the small-scale oscillations induced by the photon-ALP interactions, even though they are smeared out due to the finite energy resolution. 
The parameters of interest are $\params = (g_{a\gamma}, m_a)$ and the nuisance parameters are $\nuisparams = (N_0,\Gamma, \vec{B}, s, \delta)$. 
For the flaring state, an additional nuisance parameter is the cut-off energy of the spectrum, $E_\mathrm{cut}$.
We assume that the nuisance parameters $s$ and $\delta$ follow Gaussian likelihoods and correspondingly add the term $-2\ln\mathcal{L}_\mathrm{sys} = (s/\sigma_s)^2 + (\delta/\sigma_\delta)^2$ to the likelihood given in Eq.~\eqref{eq:lnl}, 
where both $\sigma_s$ and $\sigma_\delta$ are assumed to be equal to $6\,\%$, in line with Sec.~\ref{sec:cta-sim_subsec_sys_unc}.\footnote{This approach is preferred to the bracketing IRF approach adopted in the other sections. 
As described in App.~\ref{App:ALPcov}, Monte-Carlo simulations are required in order to derive the correct thresholds for the likelihood ratio test. These thresholds are needed to determine a detection significance and exclusion regions in the ALP parameter space. 
When simulating the observations with the nominal IRFs and then reconstructing them with the bracketing ones, the thresholds can change and extensive Monte-Carlo simulations would be needed to determine adapted threshold values.
The bracketing approach can also induce small-scale fluctuations in the count spectrum, which can mimic an ALP signal. 
This signal can be mistaken to be significant if the thresholds of the likelihood ratio test are not adjusted consistently. 
The presented approach incorporates systematic uncertainties in the fit and the threshold values can be derived from simulations self-consistently. 
We have checked that the resulting projected exclusions are more conservative (less constraining) than the simple bracketing approach and do not result in false detections. 
}

The null hypothesis corresponds to the case where ALPs do not couple to photons ($g_{a\gamma} = 0$), {i.e.}\ \Grs are subject to absorption on the EBL only. We perform the integral of Eq.~\eqref{eq:npred} using \gammapy.

For each of the 100 simulated random magnetic-field realizations, we scan a logarithmic $11\times10$ parameter grid with $m_a \in [1,10^3]\,\mathrm{neV}$ and $g_{a\gamma} \in [0.03,10] \times 10^{-11}\,\mathrm{GeV}^{-1}$. For the chosen mass and coupling range, $E_\mathrm{crit}$ falls into the CTA energy range. For lower and higher masses, ALP-induced oscillations occur above and below the CTA energy range, respectively. Couplings $g_{a\gamma} > 6.6\times10^{-11}\,\mathrm{GeV}^{-1}$ are excluded by the CAST experiment~\cite{cast2017} and observations of globular clusters~\cite{ayala2014}. We do not expect the oscillations to be detectable for couplings smaller than the minimum value tested, 
since the median survival probability (median with respect to the magnetic-field realizations) alters the intrinsic spectrum by less than $1\,\%$ below 1\,TeV. 
In order to minimize the computational time to scan the parameter space and magnetic-field configurations, we opt for the use of the Asimov data set~\cite{cowan2011}, in which the observed number of counts is set equal to the number of expected counts. We checked that the results derived from an Asimov data set are representative of the mean results of Monte-Carlo realizations.

For each ALP parameter grid point, we thus end up with 100 values (corresponding to the 100 magnetic-field realizations) of the likelihood of Eq.~\eqref{eq:lnl}. As it is improbable that the $B$~field that maximizes $\mathcal{L}$ is included in the simulated magnetic-field realizations, instead of profiling over the realizations, we sort the likelihoods in each ALP grid point in terms of the magnetic-field realization and use the
likelihood value that corresponds to a certain quantile $Q$ of this distribution (profiling would correspond to $Q=1$).
As noted in Ref.~\cite{ajello2016}, we do not expect that either the $\ts$ or the log-likelihood ratio values, $\lambda$, with respect to the best-fit ALP parameters follow $\chi^2$ distributions (see Sec.~\ref{sect::data_analysis} for a full definition of $\ts$ and $\lambda$). 
Monte-Carlo simulations are therefore necessary to set confidence intervals with appropriate coverage and to fully account for trials factors.
The threshold values adopted for $\lambda$ to exclude a given set of ALP parameters are known to depend on the tested ALP parameters themselves~\cite{ajello2016}.
Thus, a full statistical treatment would entail Monte-Carlo simulations for all tested ALP parameters. 
For 100 pseudo experiments, 100 magnetic field realizations, and 110 tested ALP parameters, a total of $1.1\times10^6$ fits need to be performed per tested injected signal.
Here, we opt to test three sets of illustrative combinations of $g_{a\gamma}$ and $m_a$ and leave the simulation of the full parameter space for future work. 
The chosen parameter values are $(m_a / \mathrm{neV}, g_{a\gamma} / 10^{-11}\,\mathrm{GeV}^{-1}) = (0,0), (30, 0.4), (40, 4)$.
The first set corresponds to the case where no ALP is present. 
The second set corresponds to the case where ALPs could constitute all of dark matter.
The third case produces large amplitude oscillations below the chosen energy threshold and small ones of the order of a few percent over the entire energy range probed with the simulated observations.

For each of the tested parameter sets and 100 pseudo experiments, we run the full analysis over the ALP parameters and magnetic-field realizations and
set $Q = 0.95$, following Ref.~\cite{ajello2016}. Based on the results presented in App.~\ref{App:ALPcov}, we adopt $\lambda$ thresholds of 23.3 and 26.9 for limits at the 95\% and 99\% confidence level, respectively. We checked that re-calculating the limits for 500 magnetic-field realizations instead of 100 has only a negligible effect on our results. 

\begin{figure}[t]
    \centering
    \includegraphics[width = 0.99\linewidth]{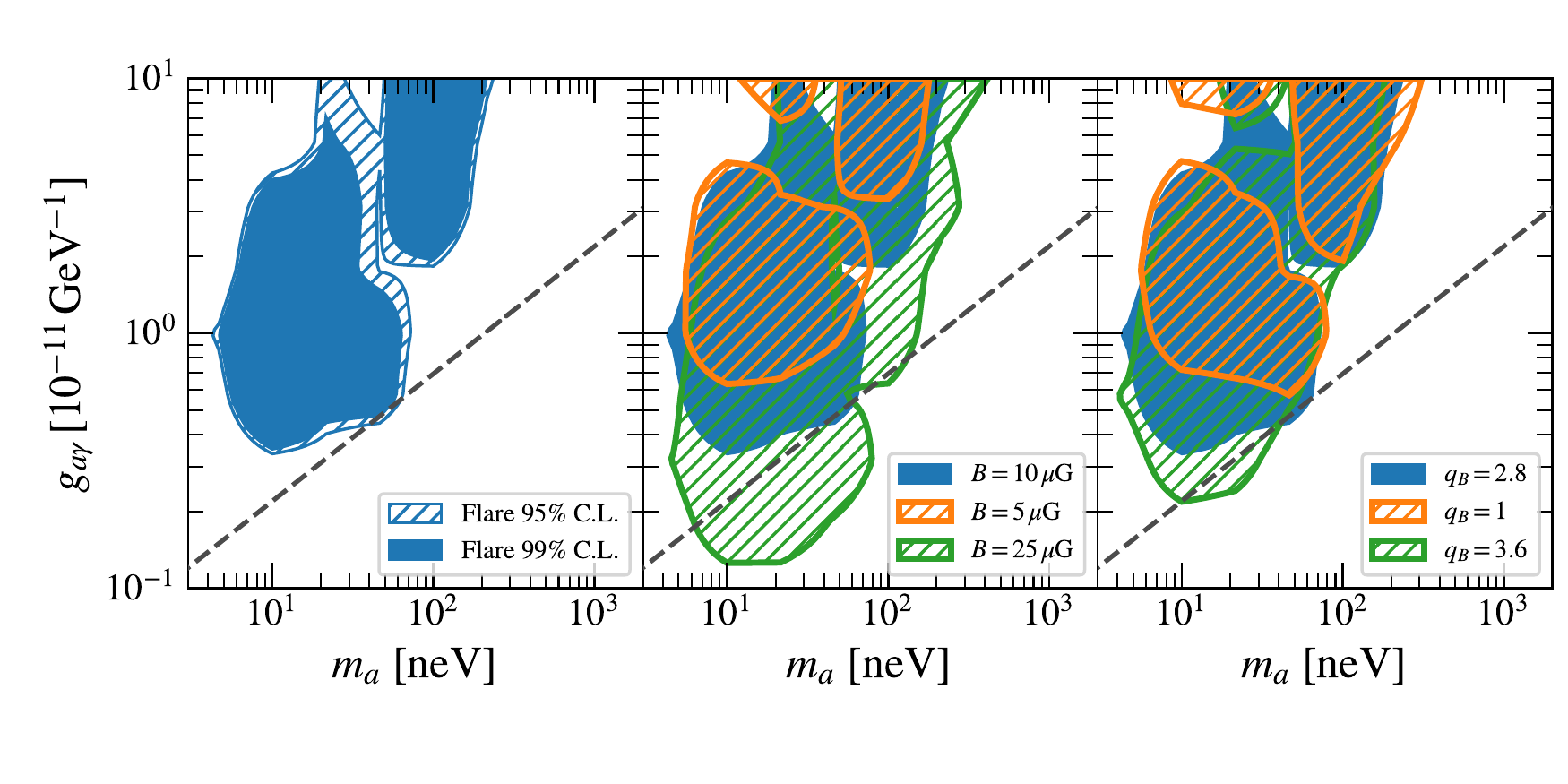}
    \caption{\label{fig:alp-sys} 
Projected CTA constraints on the ALP parameter space (mass, $m_a$, versus coupling to photons, $g_{a\gamma}$) for different assumptions on the intra-cluster magnetic-field properties. 
\textit{Left:} 95\,\% (hashed) and 99\,\% (filled) confidence level exclusion regions for the flaring state of the AGN. 
\textit{Center:} 95\,\% confidence level exclusion regions for the flaring state assuming a central magnetic field strength of 5\,$\mu\,$G (orange), 10\,$\mu\,$G (blue) and 25\,$\mu\,$G (green). 
\textit{Right:} 95\,\% confidence level exclusion regions for the flaring state assuming an index of the magnetic-field turbulence of 1 (orange), 2.8 (blue) and 3.6 (green). 
Below the dark-grey dashed line, ALPs could make up 100\,\% of the dark-matter content of the Universe (above the dashed line, ALPs would contribute sub-dominantly to the overall dark-matter density)~\cite{arias2012}. 
    }
\end{figure}

\subsection{CTA sensitivity to ALP signatures in NGC\,1275 observations}
\label{sec:alp-results}

We show the 95\,\% and 99\,\% confidence limits for the flaring state in the left panel of Fig.~\ref{fig:alp-sys}. 
Interestingly, the quiescent state does not lead to significant exclusions in the ALP parameter space, whereas the flaring state reaches parameters where ALPs could constitute the entirety of dark matter~\cite{arias2012}  (parameters below the dashed line).
For the quiescent state, the maximum likelihood ratio test value is found to be ${\sim}\,12$, which is well below the threshold value for a 95\,\% confidence limit. 
One of the reasons for the lack of constraints from the quiescent state is that a significant number of background events is accumulated for the long 300-hour exposure. The constraints are weakened by the background cut, $\nexcess / \noff > 0.05$ (see Sec.~\ref{sec:cta-sim}), which removes energy bins with central energies ${\lesssim}\,100\,$GeV from the analysis. This selection effect is illustrated in Fig.~\ref{fig:alp-spec}, where it also can be seen that the flaring spectrum extends to lower energies.
When relaxing the background cut from 5\,\% to 1\,\%, the exclusions marginally improve and ALP couplings close to $g_{a\gamma} = 10^{-10}\,\mathrm{GeV}^{-1}$ at $m_a = 100$\,neV can be ruled out. 
This highlights the necessity of observing NGC~1275 in a high flux state in order to set constraints when systematic uncertainties are not sub-dominant. 

The effect of changing the central magnetic-field strength is displayed in the middle panel of Fig.~\ref{fig:alp-sys} for the 95\,\% confidence level exclusion regions derived from the flaring state. If the actual magnetic field of the Perseus cluster is as low as $5\,\mu$G, which is still compatible with lower limits derived from non-observations of \gr emission of the intra-cluster medium~\cite{magic-preseus2016}, the ALP dark-matter parameter space could not be probed. Adopting instead a central field as suggested from Faraday rotation measurements strengthens the CTA sensitivity considerably: ALP dark matter could be probed between 6\,neV and 150\,neV. 

In the right panel of Fig.~\ref{fig:alp-sys}, we show the 95\,\% confidence level exclusion regions derived from the flaring state for different choices of the index of turbulence of the cluster magnetic field, taken from the confidence interval provided in Ref.~\cite{vacca2012}. Higher values of the turbulence index $q_B$ correspond to more power at larger turbulence scales and a larger coherence length of the magnetic field 
($q_B = 11/ 3$ would correspond to Kolmogorov turbulence). Larger coherence lengths lead to deeper oscillation features visible in the spectrum and hence a higher sensitivity to their detection or exclusion, since these large fluctuations are not smeared out by the energy resolution. This trend is also visible in Fig.~\ref{fig:alp-sys}.
On the other hand, for $q_B = 1$, which corresponds to white noise, probing ALP dark matter parameters seems to be difficult with the single observation simulated here. 

We investigate the dependence of the exclusion regions on the scale factor between electron density and magnetic field, $\eta$, and the minimum turbulence scale. The limits weaken slightly for higher values of $\eta$ since the magnetic field decreases more rapidly with increasing distance from the cluster center. The limits weaken as well for a decreasing minimum turbulence scale, as oscillations become faster with energy~\cite{ajello2016}. When convolved with the CTA energy resolution, the oscillations are less likely to be distinguished from Poisson noise. Changing the Galactic magnetic field parameters to values suggested by measurements with \emph{Planck}~\cite{2016A&A...596A.103P} (``Jansson12c'') has a marginal effect on the limits. 

We have also investigated the effect of a different energy binning for the flaring state.
On the one hand, decreasing the number of bins by a factor of two relaxes the exclusions slightly, especially above $g_{a\gamma} \gtrsim 4\times10^{-11}$ and $m_a \lesssim 30\,$neV.
This is to be expected since the oscillations have small amplitudes and high frequencies in this region of the parameter space. 
On the other, when increasing the number of bins by a factor of two, the limits improve slightly and ALPs with $g_{a\gamma} = 3\times10^{-12}\,\mathrm{GeV}^{-1}$ and $m_a = 30\,$neV are additionally excluded compared to the nominal binning.  

\begin{figure}[t]
    \centering
    \includegraphics[width = 0.99\linewidth]{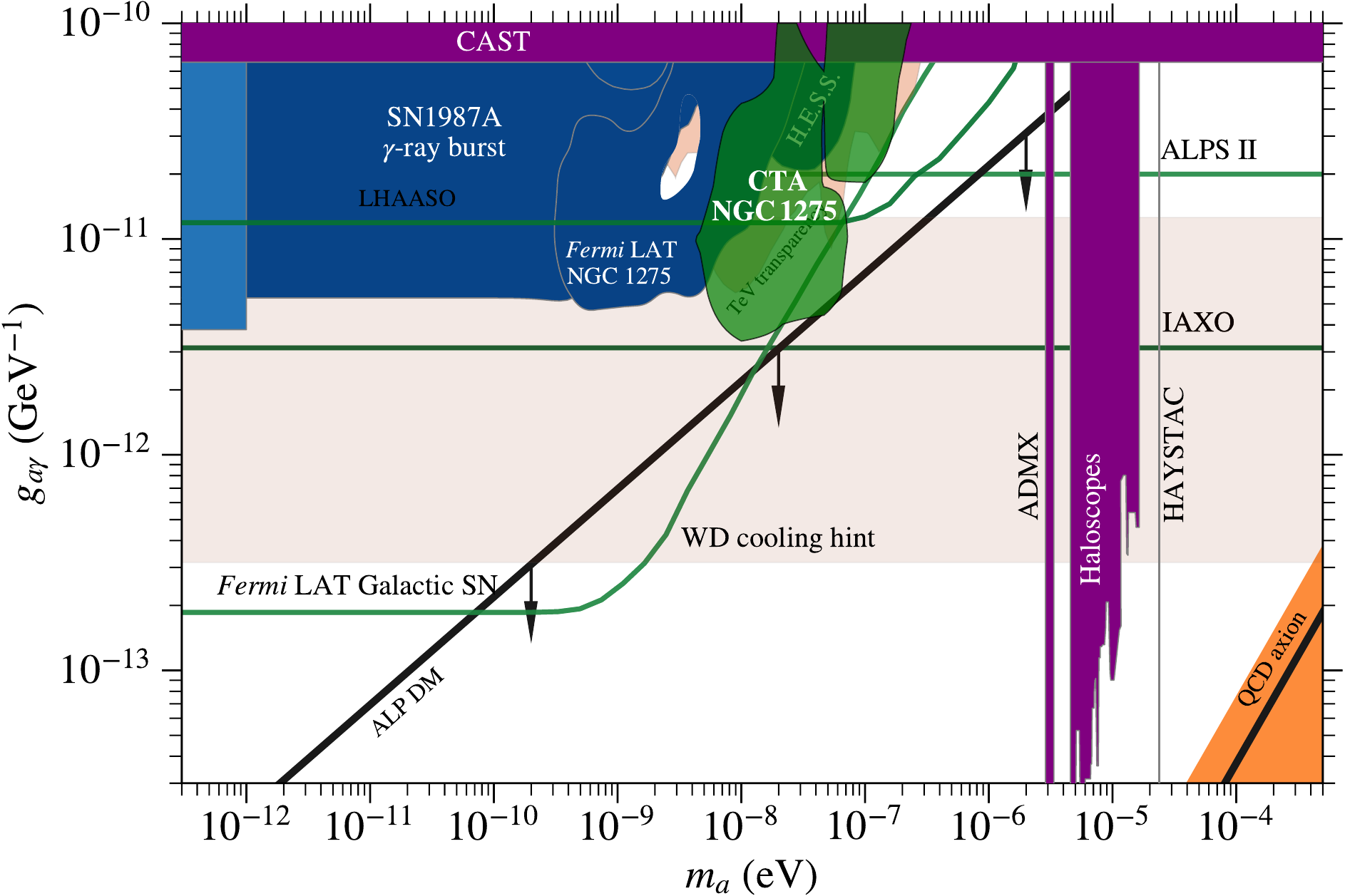}
    \caption{\label{fig:alp-summary} 
Projected CTA constraints on ALPs, as a function of their mass, $m_a$, and coupling to photons, $g_{a\gamma}$. The green filled region illustrates the 95\% confidence level exclusion region obtained from CTA observations of a flaring state of the radio galaxy NGC\,1275. Purple and blue filled regions illustrate exclusion regions from current-generation instruments. Salmon pastel regions correspond to hints for ALPs from an additional cooling of white dwarfs (WD) and an increased transparency of the Universe to TeV \Grs, as labelled in the figure. Other projected sensitivities are shown as green lines. The preferred parameter space for the QCD axion is shown as a dark orange band. See Ref.~\cite{2016arXiv161107784M} and references therein, updated here with the LHAASO sensitivity derived in Ref.~\cite{2017arXiv171201839V}. Figure created with the \textsc{gammaALPsPlot} package, see \url{https://github.com/me-manu/gammaALPsPlot}.    
    }
\end{figure}

However, it should be noted that a different energy binning might also affect the likelihood distributions (see App.~\ref{App:ALPcov}, the same is true for changing, e.g., the cut on $\nexcess / \noff$). 
Thus, different threshold values for deriving confidence intervals from the likelihood ratio test might have to be selected. 
We leave the optimization of the energy binning and the cut values used for the energy bins for future work, when the response of the deployed array is characterized.

The projected 95\,\% upper limits for the fiducial cluster magnetic-field setup and the flaring state of NGC\,1275 are compared to other limits and sensitivities in Fig.~\ref{fig:alp-summary} (green filled region). Notably, CTA observations will improve upon current H.E.S.S. limits on the photon-ALP coupling by almost an order of magnitude. CTA will probe ALP masses an order of magnitude higher than those probed by \emph{Fermi}-LAT observations of NGC\,1275. 
Between ${\sim}\,6$\,neV and ${\sim}\,200$\,neV, CTA could provide the most constraining limits on ALPs to date and could start to probe the parameter space where dark matter could consist entirely of ALPs. 
In the same mass range, CTA observations would also be more sensitive than future searches with LHAASO~\cite{2010ChPhC..34..249C} for an anisotropy in the \gr diffuse emission above several tens of TeV~\cite{2017arXiv171201839V}. In the same energy range, CTA observations could reach a similar sensitivity as future observations with the IAXO experiment~ \cite{2015JPhCS.650a2009F} and the ALPS~II laboratory experiment~\cite{2013JInst...8.9001B}. However, in contrast to dedicated laboratory searches, the possible constraints from CTA will be dominated by systematic uncertainties in the model assumptions, as shown in Fig.~\ref{fig:alp-sys}.

The mass range probed with CTA is too high to test recent claims for evidence of ALPs~\cite{2018JCAP...04..048M,2018PhRvD..97f3003X}. These claims are based on the modeling of \emph{Fermi}-LAT observations of bright pulsars and Galactic supernova remnants, which would be better described by accounting for ALP-induced irregularities than by smooth functions alone. 
The parameters suggested by these analyses are however incompatible with results from CAST and globular cluster observations~\cite{ayala2014,cast2017}.

Observations of several AGN can be combined to further improve the CTA sensitivity. For instance, the \gr source IC\,310 is also located in the Perseus cluster and is detected up to 10\,TeV with the MAGIC telescopes~\cite{magic-ic310}. Such studies will make CTA a prime instrument for searching for dark matter in the form of WISPs, complementarily with searches for more-massive dark matter candidates (see, {e.g.}\ Ref.~\cite{2020arXiv200716129C}). 

\section{Probing {Lorentz invariance up to} the Planck scale}
\label{sec:LIV}

The precise measurements of blazar spectra with CTA also can serve as a test of Lorentz invariance (\LI). Like any other fundamental principle, exploring its limits of validity has been an important motivation for theoretical and experimental research~\cite{Colladay:1998,Kostelecky:2008}. Lorentz invariance violation (\LIV) is motivated as a possible consequence of theories beyond the Standard Model, such as quantum-gravity phenomenology or string theory (see, {e.g.}, Refs.~\cite{Alfaro:2005,Kostelecky:1988,Amelino:2001} and references therein). Although \LIV signatures  are  expected  to produce small effects,  the high energies and  long  distances of  astrophysical \gr sources provide an unprecedented opportunity for observational constraints on LIV. The potential signatures of LIV in the \gr\ channel include, {e.g.}, energy-dependent time delays, photon decay, vacuum Cherenkov radiation, and pair-production threshold shifts \cite{1998Natur.393..763A,1999ApJ...518L..21K,Stecker:2001,Stecker:2003}. Previous studies have indicated that CTA will be especially useful for LIV tests. For instance, Refs.~\cite{2013APh....43..189D,2013APh....43..252I,2015NPPP..265..314D} present prospects for searches for time delays using \gr\ light-curves of GRBs, blazars and pulsars. References~\cite{2017ICRC...35..623G, 2019ICRC...36..739M} alternatively explored the pair-production channel, relying solely on spectral observations of blazars. In the present work, we consider more extensively the CTA potential to test LIV-induced modifications of the pair-production threshold in \gr\ interactions with the EBL. Should the pair-production threshold be affected by LIV, this channel would be a sensitive probe of first- and second-order modifications of dispersion relations.

Phenomenological generalizations of \LIV effects converge on the introduction of a general function of energy, $E$, and momentum, $p$, in the particle energy dispersion relation. The functional form of the correction can change for different field particles and underlying theories but usually leads to similar phenomenology~\cite{2015ApJ...812...60B,1999ApJ...518L..21K,2008PhRvD..78l4010J,Stecker:2001,Stecker:2003,MH_2016azo} (for alternative models, see also~Ref.~\cite{DSR}): an effective \LIV dispersion relation at leading order $n$ can be written as\footnote{Natural units are used, $c=\hbar=1$.}
\begin{equation}\label{eq:GDR}
    E_{a}^{2} -  p_{a}^{2} = m_a^2 \pm |\delta_{a,n}|A_a^{n+2},
\end{equation}
where $a$ stands for the particle type, $A$ stands for either energy or momentum, and $\delta_{a,n}$ is the \LIV parameter. 

In some effective field theories, $\delta_{a,n}=\zeta^{(n)}/M^n$, where $\zeta^{(n)}$ are \LIV coefficients and $M$ is the energy scale of the new physics, such as the Planck energy scale, $E_{\rm Pl}\approx 10^{28}$\,eV, or some energy scale inspired by quantum gravity, $E_{\rm QG}$.  In this work, $A=p\sim E_{\gamma}$ to first order. To compare our results with the limits established by previous \gr tests, \LIV is considered only for photons in subluminal phenomena, corresponding to a negative corrective term in Eq.~\eqref{eq:GDR}. The most recent bounds on superluminal phenomena, such as photon decay resulting in a flux suppression at the highest energies, can be found in Ref.~\cite{2019arXiv191108070H}. It should be noted that subluminal LIV could also result in an apparent flux suppression or reduced sensitivity at the highest energies, as induced by a delayed development of $\gamma$-ray air showers~({e.g.}, Ref.~\cite{2017JCAP...05..049R,Satunin:2019gsl}).  

For simplicity, the \LIV correction will be named $|\delta_{n}| = ( E_{\LIV}^{(n)})^{-n}$ for $n=1,2$. If LIV in the electron sector is also considered, the LIV parameter $1/E_{\LIV}$ becomes a linear combination of the LIV contributions from the different particle species~\cite{MH_2016azo}. In the most common scenario, photons dominate over electrons, and the derived results in this work remain the same. In the second most common scenarios, \LIV is universal for photons and electrons, and a factor of $1/(1-1/2^n)$ should be considered in the final results~\cite{2015ApJ...812...60B,MH_2019ehp}, {i.e.}\ an improvement in the constraints by a factor of two and four thirds for $n=1$ and $n=2$, respectively.

Considering LIV, the threshold energy of pair production for head-on collisions in Eq.~\eqref{Eq:threshold} is modified as {\cite{2008PhRvD..78l4010J}}:
\begin{equation}
    \epsilon_{th}'= \frac{m_e^2}{E_{\gamma}'} + \frac{E_{\gamma}'^{n+1}}{4\left(E_{\LIV}^{(n)}\right)^n}.
\end{equation}
The LI scenario is recovered with $E_{\LIV} \rightarrow \infty$.

If LIV is present, the observed spectrum is modified in a particular way which depends on the value of $E_{\LIV}$~\cite{2015ApJ...812...60B, 1999ApJ...518L..21K, 2008PhRvD..78l4010J,2019ApJ...870...93A}. As a consequence of the increased threshold energy of EBL photons, the number of target photons interacting with the highest-energy \Grs is decreased, effectively increasing the transparency of the Universe to \Grs above tens of TeV. Compton scattering can also lead to LIV signatures in the observed energy spectrum; however its effects are small in comparison with those expected from pair production for \Grs below 1\,PeV \cite{Abdalla:2018sxi}. The most constraining limits on $E_{\LIV}$ exploiting constraints on the \gr optical depth from single-source studies have been obtained in Ref.~\cite{2019ApJ...870...93A}. In this case, limits at the $2\,\sigma$ confidence level have been set as $E_{\LIV}^{(1)}\ge 2.6 \times 10^{28}$\,eV and $E_{\LIV}^{(2)}\ge 7.8 \times 10^{20}$\,eV. Combined constraints from multiple \gr sources enabled Ref.~\cite{2019PhRvD..99d3015L} to constrain LIV up to $E_{\LIV}^{(1)}\ge 0.7 \times 10^{29}$\,eV and $E_{\LIV}^{(2)}\ge 1.6 \times 10^{21}$\,eV, using the same benchmark EBL model as the one used in the present work.

In this study, we probe the potential of CTA to detect or constrain LIV with two blazars, Mrk\,501 and 1ES\,0229+200, which have been proposed as good candidates for such searches (see, {e.g.}, Refs.~\cite{2014JCAP...06..005F, 2016A&A...585A..25T}). A flaring state of Mrk\,501 and a long-term observation of 1ES\,0229+200 are simulated for 10\,hours and 50\,hours, respectively, using the intrinsic parameters provided in App.~\ref{sec:intr-models}. We assume that the intrinsic spectra are power laws with exponential cut-offs, with the comoving energy cut-off set at $E_{\rm cut}' = 10$ and $50$\,TeV. Both values are compatible with observations of these \gr sources during extreme states, with a cut-off as low as 10\,TeV even being disfavored during the high-state of Mrk\,501 observed by HEGRA in 1997~\cite{2001A&A...366...62A}. A cut-off at $50$\,TeV could be considered as an optimistic value based on current models of extreme blazars, an assumption which can hardly be assessed by current observations due to their limited sensitivity.

We start by simulating with \gammapy and \ctools spectral observations of CTA affected by LIV in order to determine at which confidence level CTA can detect LIV signatures. We then simulate LI spectra to determine the exclusion capability of CTA. We implemented the LIV effect in the  {\textsc{ebltable}\xspace} Python module\footnote{\url{https://github.com/me-manu/ebltable}} and checked that the difference in best-fit LIV parameters reconstructed with \gammapy and \ctools is small with respect to statistical uncertainties and systematic uncertainties of instrumental or modeling origin. The minor differences are also sub-dominant with respect to the uncertainties induced by the current knowledge of the EBL.

\begin{figure}[t]
    \centering
    \includegraphics[width=.49\linewidth]{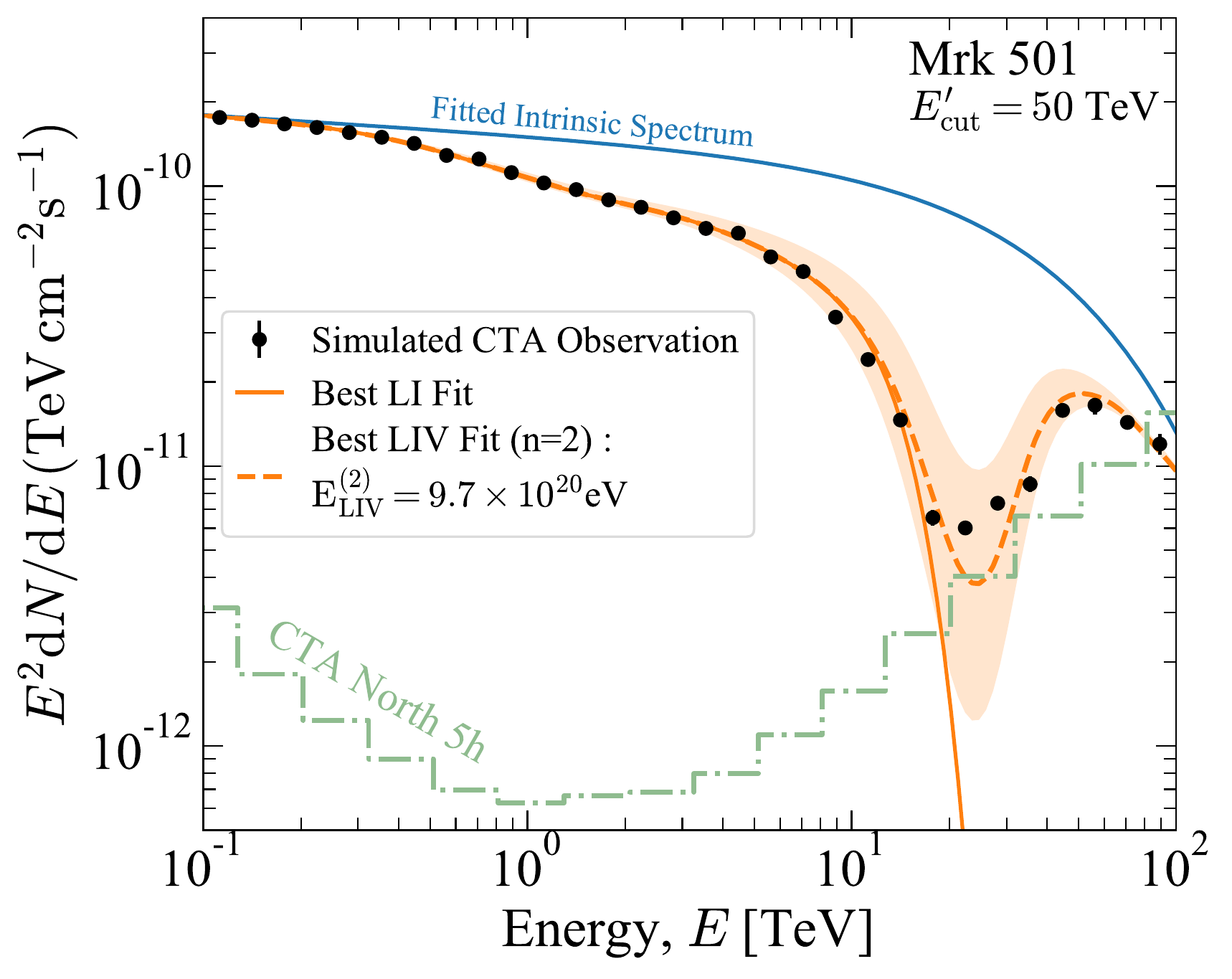}\hfill    \includegraphics[width=0.49\linewidth]{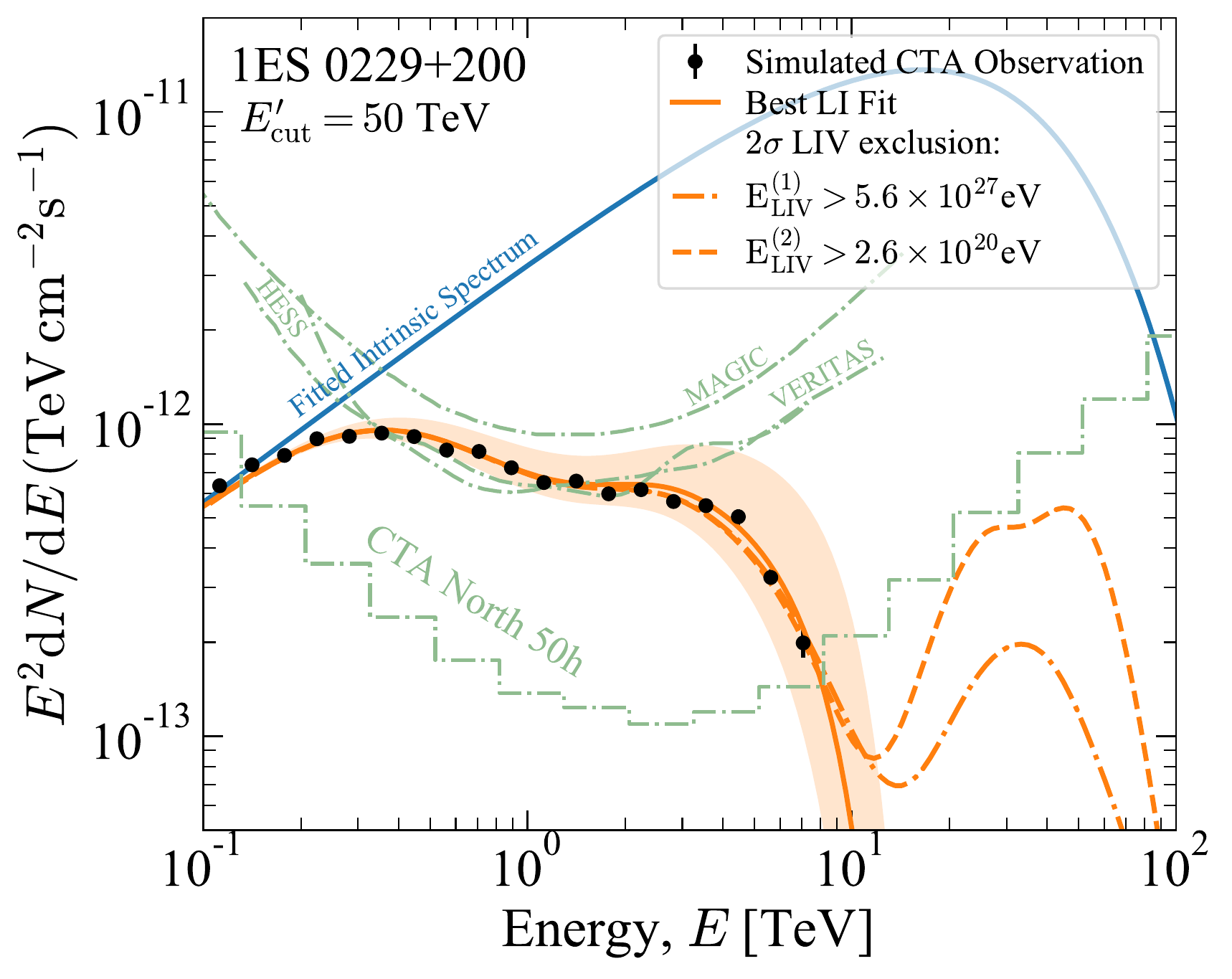}
    \caption{
Simulated spectra of the blazars Mrk\,501 and 1ES\,0229+200 to assess the sensitivity of CTA to LIV-induced modification of \gr transparency.
The differential photon spectra multiplied by the square of energy are displayed as a function of \gr energy as black points.
{\it Left: } Simulation of a 10\,h observation of Mrk\,501 in an elevated state with $E_\mathrm{cut}' = 50$\,TeV assuming LIV with $E_{\LIV}^{(2)}= 10^{21}$\,eV. The best-fit LIV model for $n=2$ is shown as a dashed orange line.
\textit{Right}: Simulation of a 50\,h observation of 1ES\,0229+200  in an average emission state with $E_{\rm cut}' = 50$\,TeV with LI propagation. The LIV models excluded at $2\,\sigma$ for $n = 1$ and 2 are shown as dash-dotted and dotted orange lines, respectively. For comparison, the best-fit LI models are shown as continuous orange lines. The systematic uncertainty induced by the current knowledge of the EBL is shown as a shaded orange band around the injected model. The intrinsic spectrum is shown as a blue line. The 5\,h (left) and 50\,h (right) sensitivities of CTA-North are shown as green dashed line segments. The sensitivities of current-generation instruments, tabulated for a 50\,h exposure, are shown on the right as green lines.    
    }%
    \label{LIV:fig1}%
\end{figure}

\begin{figure}[t]
    \centering
    \includegraphics[width=0.49\linewidth]{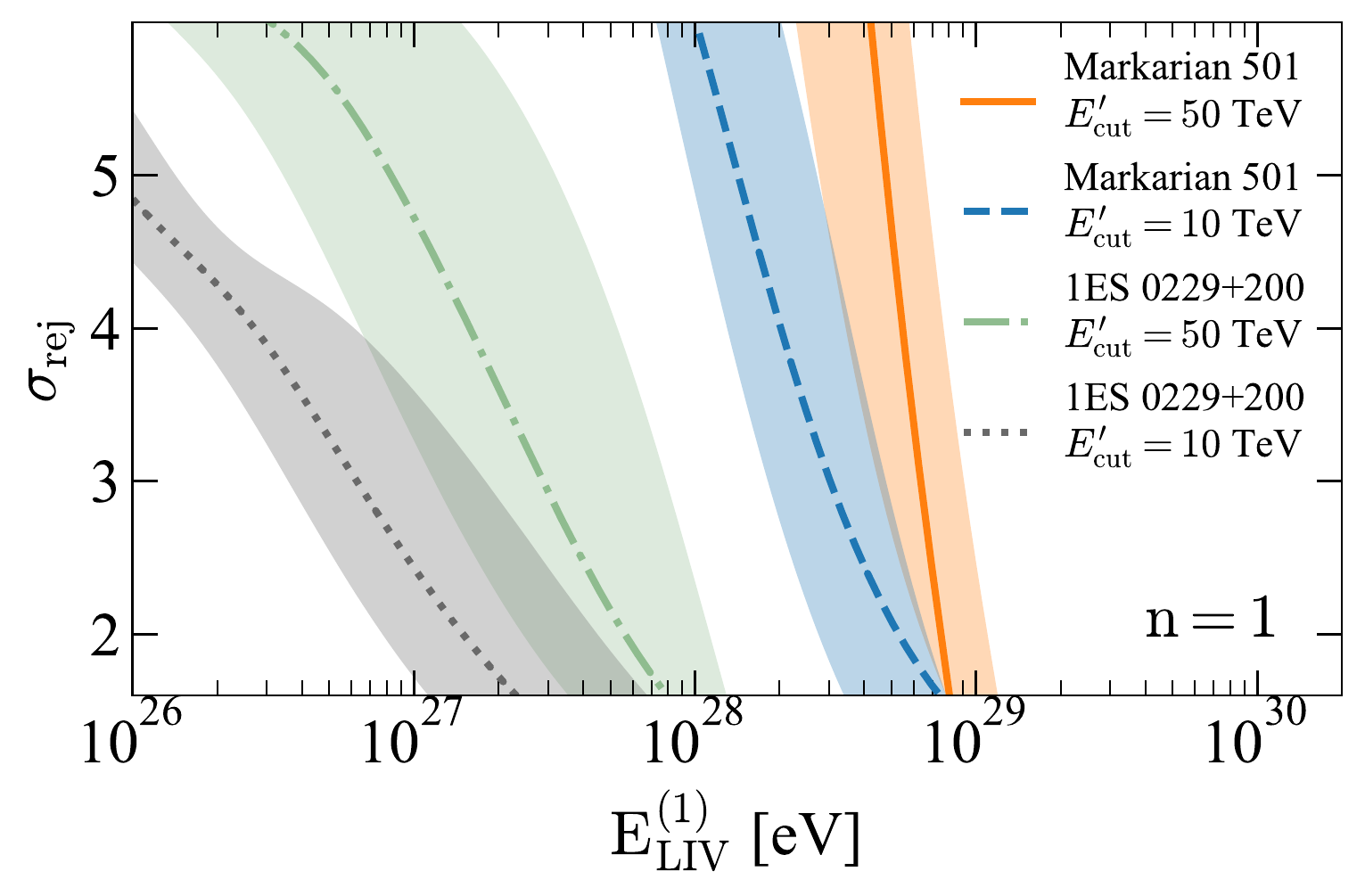}
    \hfill 
    \includegraphics[width=0.49\linewidth]{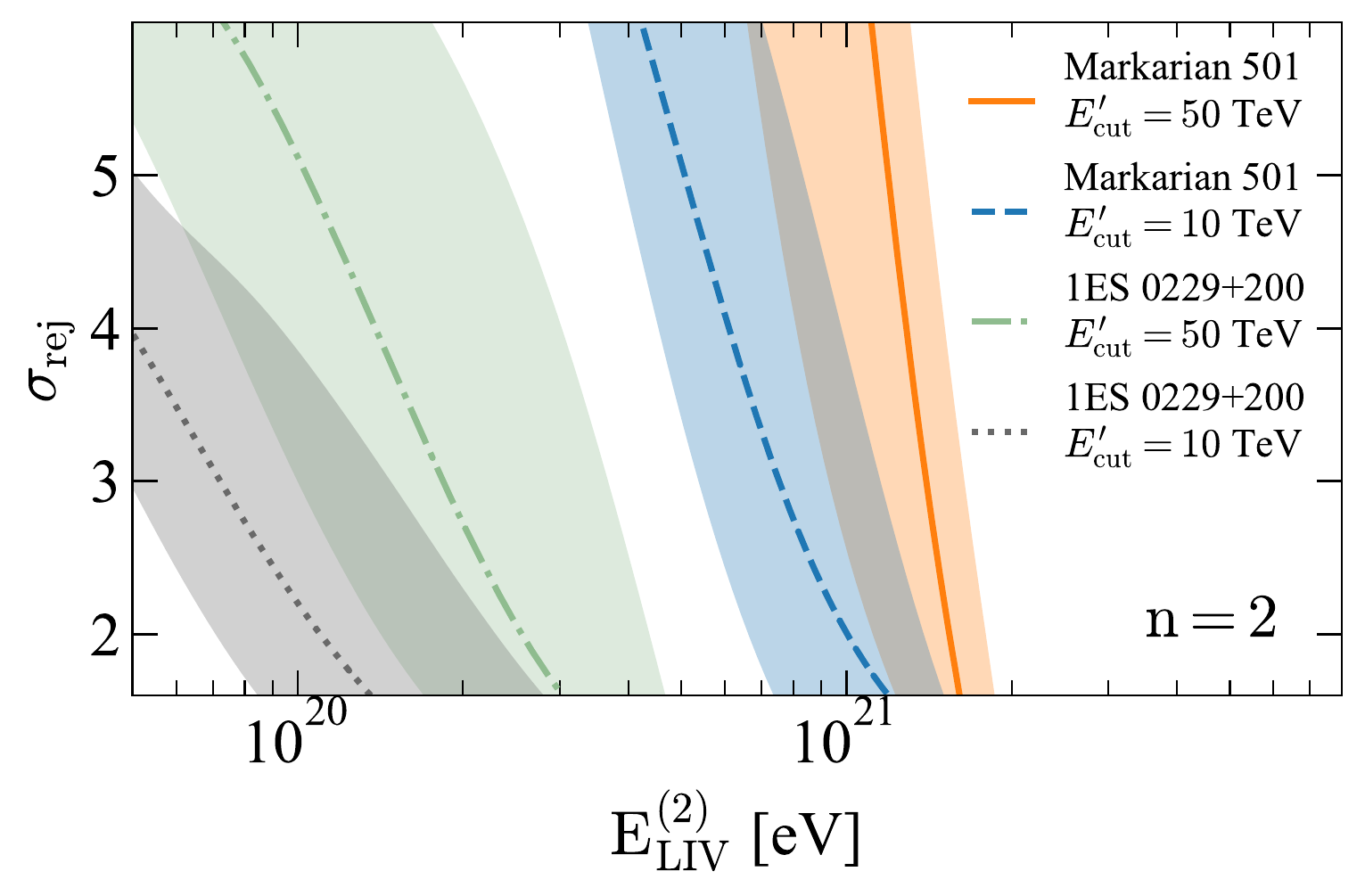}
    \caption{
Exclusion level of \LIV models as a function of \LIV energy scale for CTA observations of Mrk\,501 and 1ES\,0229+200. The lines indicate the rejection level, in significance units, for cut-off energies of 10\,TeV (blue for Mrk\,501, grey for 1ES\,0229+200)  and 50\,TeV (orange for Mrk\,501, green for 1ES\,0229+200), assuming a LIV modification of order $n=1$ (\emph{left}) and $n=2$ (\emph{right}). 
The bands represent the variations induced by the ${\pm}\,1\,\sigma$ uncertainties on the adopted benchmark EBL model~\cite{2011MNRAS.410.2556D}.
    }
    \label{LIV:fig2}%
\end{figure}

\subsection{CTA potential to find a LIV signal}\label{LIV:SubSection1}

With LIV coefficients on the order of current limits,  $E_{\LIV}^{(1)}= 10^{29}$\,eV and $E_{\LIV}^{(2)}= 10^{21}$\,eV, we simulate CTA observations of Mrk\,501 and 1ES\,0229+200 assuming two different cut-off energies (10 and 50\,TeV). 
The free fit parameters to reconstruct the simulated spectra are the intrinsic source parameters, $\boldsymbol{\theta}_\mathrm{int}$, and $\boldsymbol{\pi} = E_\mathrm{LIV}$. To constrain the LIV energy scale, we generate a logarithmic grid with $E_{\LIV}^{(1)}\in[10^{26}\,\unit{eV},10^{30}\,\unit{eV}]$ ($E_{\LIV}^{(2)} \in[5\times10^{19}\,\unit{eV},5\times10^{21}\,\unit{eV}]$) and an equidistant logarithmic spacing of 0.0226 (0.015). For each grid point, we optimize the intrinsic spectral parameters using \textsc{Sherpa} and record the profile likelihood values as discussed in App.~\ref{sect::data_analysis}.

Figure~\ref{LIV:fig1} (left) illustrates the simulation of the spectrum of Mrk\,501 observed with CTA North under the assumption that $E_{\rm cut}' = 50$\,TeV and $E^{(2)}_\mathrm{LIV} = 10^{21}\,\mathrm{eV}$. The intrinsic (observed) spectrum corresponding to the best-fit parameters is shown as a continuous blue (orange) line. The orange band represents the variation of the best-fit parameters when uncertainties on our benchmark EBL model (as derived from modeling of galaxy counts) are taken into account~\cite{2011MNRAS.410.2556D}. For comparison, the sensitivity of 50\,h observations from current ground-based instruments (H.E.S.S., MAGIC and VERITAS) and from CTA North are shown as dash-dotted lines in the right-hand side panel of Fig.~\ref{LIV:fig1}. Most of the KSP live time dedicated to observations of Mrk\,501 is planned to be allocated to CTA North. Nonetheless, observations at high-zenith angle (low elevation) from the southern site could complement the constraints on LIV during an elevated flux state of the blazar. The expected recovery of the flux at the highest energies due to the LIV effect would be easily detected with both arrays of CTA, while being out of the range of current-generation instruments.

The statistical significance of the detection of the LIV signal is given by $\sigma^2 = \lambda_{\LI} - \lambda_{\LIV}$ ($\chi^2$ with one degree of freedom), with $\lambda$ defined in App.~\ref{sect::data_analysis}. All of the reconstructed values are compatible with the simulated LIV scenario within uncertainties. LIV effects at $E_{\LIV}^{(1)}= 10^{29}$\,eV and $E_{\LIV}^{(2)}= 10^{21}$\,eV are not detected for a somewhat pessimistic cut-off value around 10\,TeV. The effect at $E_{\LIV}^{(1)}= 10^{29}$\,eV would be near detection ($4-5\,\sigma$) assuming a cut-off around 50\,TeV in the spectrum of Mrk\,501 and would be clearly discovered ($8-25\,\sigma$) for $E_{\LIV}^{(2)}= 10^{21}$\,eV in both the spectra of Mrk\,501 and 1ES\,0229+200 with high-energy cut-offs.

\subsection{Excluding LIV signal with CTA}

CTA observations of Mrk\,501 and 1ES\,0229+200 are also simulated assuming that the Lorentz symmetry is not broken. We show the example of the latter blazar in Fig.~\ref{LIV:fig1} (right) for a cut-off at 50\,TeV. We use the profile log-likelihood values for the \LI and \LIV hypotheses to exclude the \LIV scenario at a confidence level given by $\sigma_{\rm rej}^2 = \lambda_{\LIV}-\lambda_{\LI}$. The results for the \LIV leading order $n =1, 2$ are presented in Fig.~\ref{LIV:fig2}, in which we show the capability of CTA to exclude the LIV model as a function of $E_{\LIV}^{(1, 2)}$.

\begin{figure}[t]
    \centering
    \includegraphics[width=0.49\linewidth]{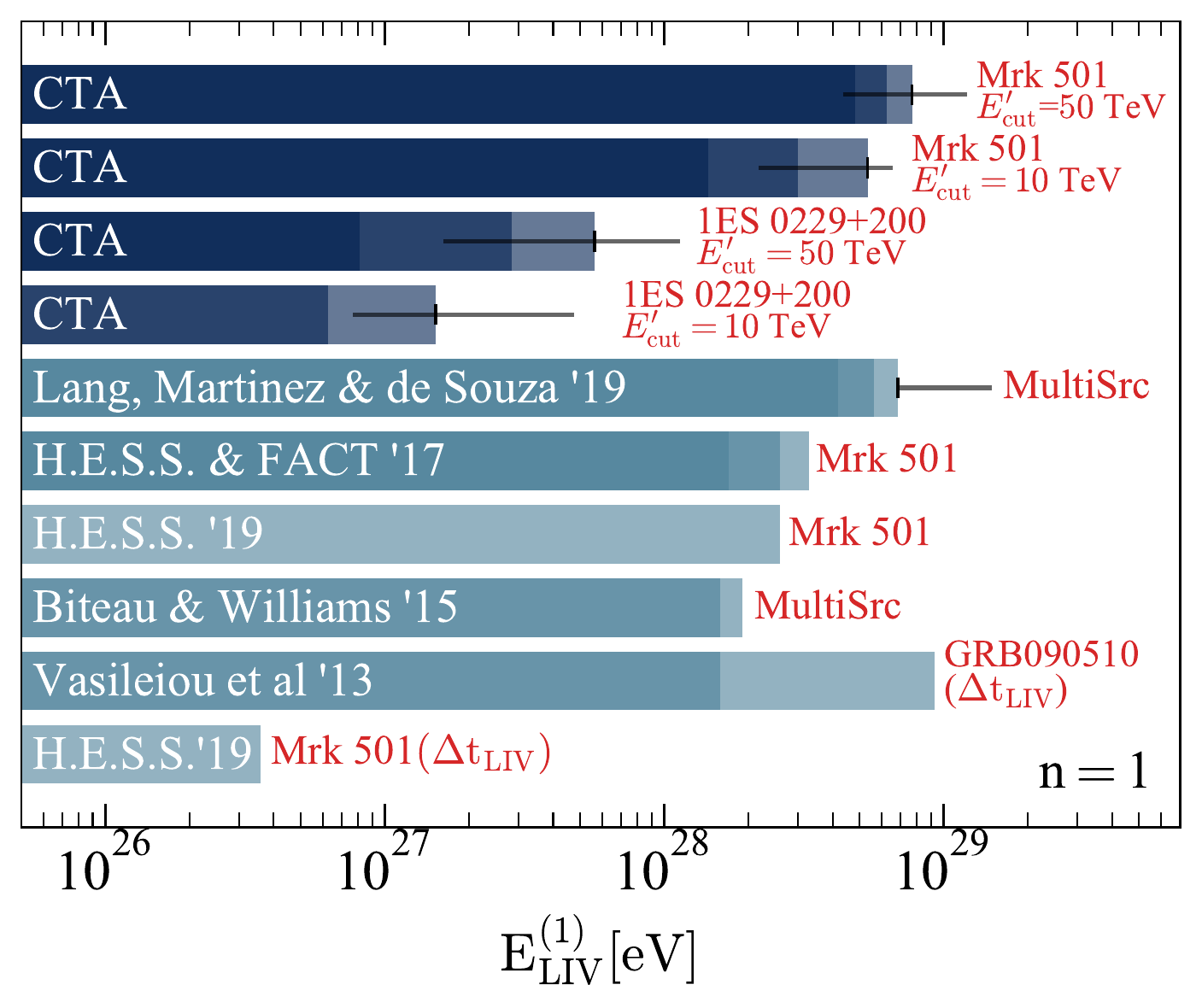}
    \hfill
    \includegraphics[width=0.49\linewidth]{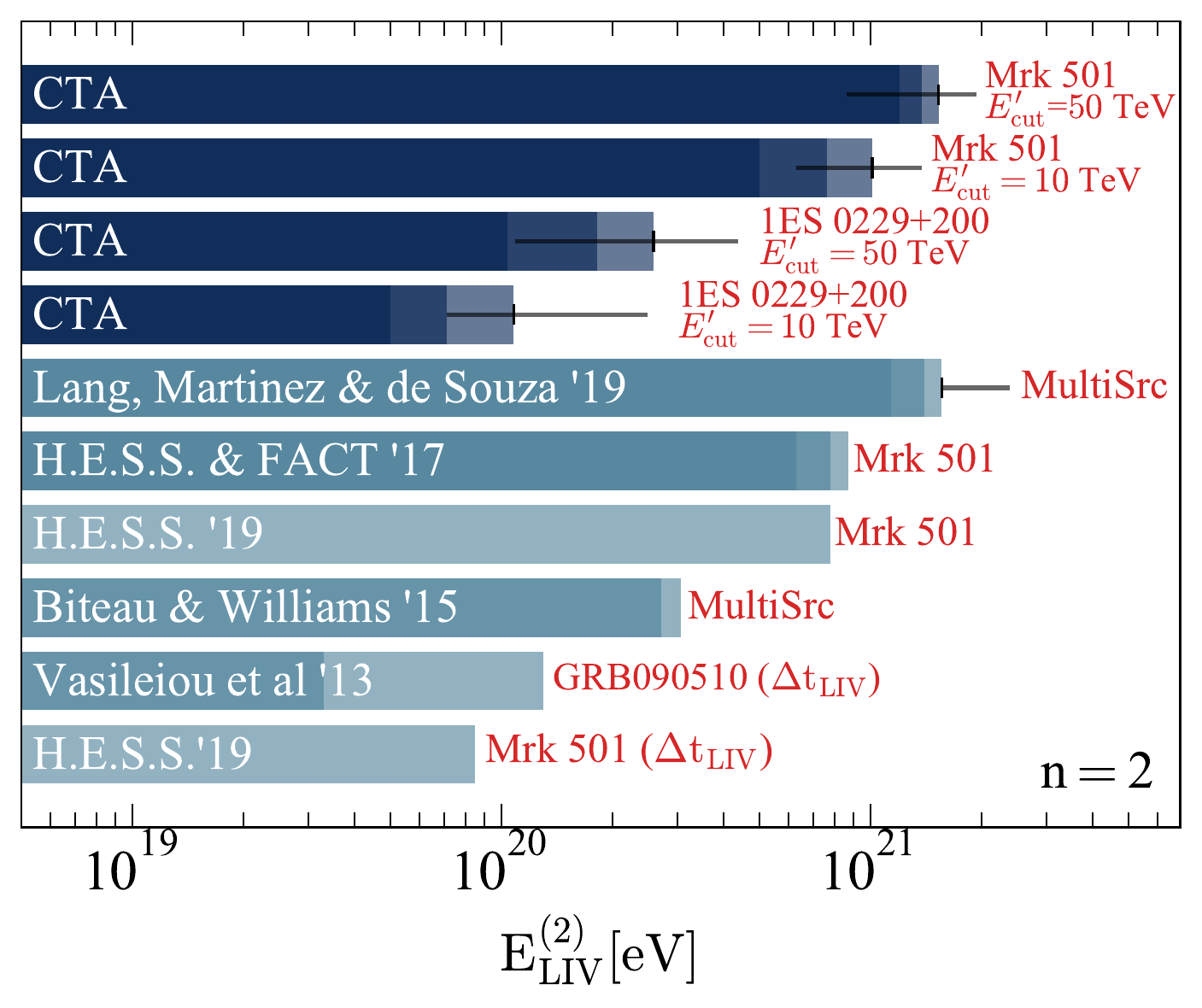}
\caption{
Projected CTA constraints on LIV-induced modifications of \gr transparency. 
CTA statistical limits on the \LIV energy scale from single \gr sources are presented in blue, for both a first- (\textit{left}) and second-order (\textit{right}) modification of the dispersion relation. Exclusion at the 2, 3 and $5\,\sigma$ level are marked with increasingly darker colors. Systematic uncertainties are indicated by black lines for the 2$\,\sigma$ limits. 
Existing subluminal searches in the photon sector, using a similar technique, are shown in cyan for comparison, from top to bottom \cite{2019PhRvD..99d3015L}, \cite{Cologna:2016cws}, \cite{2019ApJ...870...93A}, and \cite{2015ApJ...812...60B}.
Also for comparison, strong limits on energy-dependent time delays from extragalactic \gr sources are shown as annotated with $\Delta t_{\rm LIV}$, from bottom to top \cite{Vasileiou:2013vra,2019ApJ...870...93A}. 
We have translated limits from Ref.~\cite{2015ApJ...812...60B} (``Biteau \& Willams '15'') to the photon sector and the quadratic term.  
``MultiSrc'' stands for limits obtained from the joint analysis of multiple \gr sources.
Uncertainties on the EBL are already accounted for in the constraints from Ref.~\cite{2015ApJ...812...60B} and are shown in black for Ref.~\cite{2019PhRvD..99d3015L} (``Lang, Martinez \& de Souza '19'').
The best limits are around one order of magnitude above the Planck {energy} scale, $E_{\rm Pl} \sim10^{28}$\,eV, for $n=1$. 
}\label{fig:LIV_lim}
\end{figure}

Finally, we compare  in Fig.~\ref{fig:LIV_lim} the LIV energy scales which can be excluded with CTA as well as limits from current-generation instruments using similar analysis techniques. Focusing on constraints from Mrk\,501 during a flaring state (10\,h) and from 1ES\,0229+200 during its quiescent flux state (50\,h), LIV energy scales of $\big(7.7^{+4.4}_{-3.5}\big) \times 10^{28}\,$eV and $\big(5.6^{+5.8}_{-4.0}\big)\times 10^{27}\,$eV could be excluded at the $2\,\sigma$ confidence level for $n=1$. The quoted uncertainties result from the still-limited constraints on the CIB component of the EBL and from the systematic uncertainties on the IRFs. For $n=2$, the limits are $\big(1.5^{+0.4}_{-0.7}\big) \times 10^{21}\,$eV and $\big(2.6^{+1.8}_{-1.5}\big)\times 10^{20}\,$eV, respectively. These CTA limits are {one to two orders of magnitude better than the recent limits based on energy-dependent time delays from H.E.S.S. observations of Mrk\,501 (``{H.E.S.S.} '19'' band)}. The CTA limits are also a factor of two to three more constraining than those obtained by current instruments using the same channel in single-source observations. With similar prospects for multi-source analyses (see ``Lang {\it et al.} '19'' band), the potential to probe LIV-induced modifications of the pair-production threshold, complemented with constraints on time delays from blazars, GRBs and pulsars, will make the CTA observatory an attractive explorer of fundamental symmetries in the photon sector at energy scales beyond the reach of accelerator-based experiments.

\section{Perspectives on CTA constraints of \gr propagation}\label{sec:Ccl}

We have demonstrated that CTA observations will provide unprecedented sensitivity in the use of \gr observations to study cosmology and physics beyond the Standard Model. The assumed live times for the different science goals discussed in this work are summarized in Table~\ref{tab:obs-time-summary}. The observations are all part of the CTA KSPs and should be completed within the first five to ten years of operations of the full observatory. 

\begin{table}[t]
    \centering
    \begin{tabular}{l|cc}
    \hline
    Science case &  Source / Source sample &  Live time (hours) \\
    \hline \hline
    \multirow{4}{*}{EBL}
        & AGN long-term monitoring  & $5\times50$ \\
    {}
        & AGN high quality spectra  & $15\times20$ \\
    {}
        & AGN flares                & $28\times10$ \\
    {} & {} & $\sum = $ 830\\
    \hline
    IGMF & 1ES\,0229+200 & 50 \\
    \hline
    ALPs & NGC\,1275 & 10 \\
    \hline
    \multirow{2}{*}{LIV} 
        & Mrk\,501                  & 10\\
    {} 
        & 1ES\,0229+200             & 50 \\
    \hline
    \end{tabular}
    \caption{
Summary of live times to achieve the necessary sensitivity for the science cases presented in this work. 
The indicated live times correspond to the optimized baseline arrays of CTA. 
For science cases explored in this work with one to two AGN  (IGMF, ALPs, LIV), further observations of blazars from the Northern and Southern arrays of CTA would provide independent and complementary constraints. 
Observations of other extragalactic \gr sources, such as GRBs, could complement the constraints, particularly on the EBL at high redshifts ($z>1$).
    }
    \label{tab:obs-time-summary}
\end{table}

The assumed ${\sim}\,800$\,hours of AGN observations will enable the high-precision probe of the EBL imprint with statistical uncertainties ranging from 5 to 15\% with increasing redshift and systematic uncertainties below 25\,\% at $z < 0.7$, increasing to ${\sim}\,50\%$ up to $z \sim 2$. These results, covering a twice larger redshift range with a two-to-three times better precision with respect to current-generation IACTs, are based on about a quarter of the AGN KSP live time, featuring the spectra of the highest quality, and do not include the foreseen extragalactic survey nor observations of other extragalactic transients  \cite{2019scta.book.....C}. CTA observations of GRBs, which constitute an emerging class of \gr sources detected at VHE \cite{2019Natur.575..455M,2019Natur.575..464A}, would further strengthen the determination of the EBL photon density and evolution \cite{2017ApJ...850...73D}. These results will also enable an estimation of the cosmic star-formation history (see Refs.~\cite{2012MNRAS.426.1097R,2018Sci...362.1031F}) and enable an independent determination of cosmological parameters \cite{2019ApJ...885..137D}. Sensitivity studies dedicated to these more specific topics shall be conducted in the future. 

We have illustrated the capabilities of CTA to detect or constrain the IGMF, ALPs, and LIV with observations of one to two prototypical AGN, observed with the northern CTA array due to the declinations of the \gr sources. These sensitivities for the IGMF, ALPs, and LIV science cases should be seen as the minimum CTA capabilities. The combined sensitivities from multiple \gr sources observed from both the northern and southern sites will significantly enlarge the observed energy range and will enable the probe of a broader parameter space, as demonstrated, {e.g.},  for the IGMF in Ref.~\cite{2018ApJS..237...32A}. Joint \gr measurements from CTA and \emph{Fermi}~LAT would be particularly desirable, since most of the reprocessed cascade emission would end up in the \emph{Fermi}-LAT energy range. Nevertheless, even with the stand-alone observation of a single AGN, CTA will be able to search for \gr halos for IGMF strengths up to $B = 3\times 10^{-13}\,\unit{G}$ (four times better than current IACTs), probe ALP couplings to photons reaching the dark-matter parameter space (improvement by a factor four to five), and constrain LIV up to nearly an order of magnitude {above} the Planck {energy} scale (factor of two to three improvement for single-source observations). 

One of the remaining uncertainties, applicable to all science cases discussed in this work and most importantly to the study of the EBL, lies in the choice of the intrinsic spectral source model~\cite{2019A&A...627A.110B}. Future multi-wavelength observations throughout the electromagnetic spectrum and theoretical modeling of blazar emission will further help to reduce the associated systematic uncertainties and provide a better determination of the intrinsic spectra~({e.g.}, Ref.~\cite[][]{2010ApJ...715L..16M}). 

CTA constraints on the EBL will be complemented by infrared observations from JWST ({e.g.},  Ref.~\cite{2019MNRAS.483.2983Y}) and future mid- to far-infrared satellites, such as SPICA ({e.g.},  Ref.~\cite{2017PASA...34...57S}), enabling high-resolution constraints on the cosmic star-formation history. CTA results on the IGMF will be complementary to future searches for an intergalactic rotation measure at radio frequencies with SKA \cite{2015aska.confE..92J,2018MNRAS.480.3907V}.
Further synergies with SKA include a better characterization of Galactic and intra-cluster magnetic fields \cite{bonafede2015} which will reduce the uncertainties when searching for photon-ALP oscillations. Any hint for photon-ALP interactions found with CTA could be tested by future laboratory experiments like ALPS~II \cite{2013JInst...8.9001B} and IAXO \cite{2015JPhCS.650a2009F}. Searches for LIV could be combined with constraints from cosmic-ray observations with the upcoming upgraded Pierre Auger and Telescope Array observatories \cite{2016arXiv160403637T,Sagawa:2015yrf}. The CTA observatory could even help in identifying extragalactic cosmic-ray accelerators through hadronic spectral imprints, arising either from the source or along the line of sight (see Ref.~\cite{2019scta.book..231Z} and Refs. therein). Last but not least, combined constraints with archival \textit{Fermi}-LAT data as well as joint observations of variable sources with CTA, be they flares from AGN, GRBs or newly discovered transient extragalactic events, may prove to be invaluable to $\gamma$-ray cosmology. Joint observations with current-generation MeV-TeV \gr observatories have provided the community with tremendous scientific return. Continued operation of the \textit{Fermi}-LAT satellite during the early science phase of CTA is highly desirable.

The science cases discussed in the present work illustrate some of the synergies between the CTA KSPs and upcoming multi-wavelength and multi-messenger facilities, which will provide essential means for collective progress in the fields of astrophysics, astroparticles, cosmology and fundamental physics.

\acknowledgments

We gratefully acknowledge financial support from the following agencies and organizations:

\bigskip

State Committee of Science of Armenia, Armenia;
The Australian Research Council, Astronomy Australia Ltd, The University of Adelaide, Australian National University, Monash University, The University of New South Wales, The University of Sydney, Western Sydney University, Australia;
Federal Ministry of Education, Science and Research, and Innsbruck University, Austria;
Conselho Nacional de Desenvolvimento Cient\'{\i}fico e Tecnol\'{o}gico (CNPq), Funda\c{c}\~{a}o de Amparo \`{a} Pesquisa do Estado do Rio de Janeiro (FAPERJ), Funda\c{c}\~{a}o de Amparo \`{a} Pesquisa do Estado de S\~{a}o Paulo (FAPESP), Ministry of Science, Technology, Innovations and Communications (MCTIC), Brasil;
Ministry of Education and Science, National RI Roadmap Project DO1-153/28.08.2018, Bulgaria;
The Natural Sciences and Engineering Research Council of Canada and the Canadian Space Agency, Canada;
CONICYT-Chile grants CATA AFB 170002, ANID PIA/APOYO AFB 180002, ACT 1406, FONDECYT-Chile grants, 1161463, 1170171, 1190886, 1171421, 1170345, 1201582, Gemini-ANID 32180007, Chile;
Croatian Science Foundation, Rudjer Boskovic Institute, University of Osijek, University of Rijeka, University of Split, Faculty of Electrical Engineering, Mechanical Engineering and Naval Architecture, University of Zagreb, Faculty of Electrical Engineering and Computing, Croatia;
Ministry of Education, Youth and Sports, MEYS  LM2015046, LM2018105, LTT17006, EU/MEYS CZ.02.1.01/0.0/0.0/16\_013/0001403, CZ.02.1.01/0.0/0.0/18\_046/0016007 and CZ.02.1.01/0.0/0.0/16\_019/0000754, Czech Republic; 
Academy of Finland (grant nr.317636 and 320045), Finland;
Ministry of Higher Education and Research, CNRS-INSU and CNRS-IN2P3, CEA-Irfu, ANR, Regional Council Ile de France, Labex ENIGMASS, OCEVU, OSUG2020 and P2IO, France;
Max Planck Society, BMBF, DESY, Helmholtz Association, Germany;
Department of Atomic Energy, Department of Science and Technology, India;
Istituto Nazionale di Astrofisica (INAF), Istituto Nazionale di Fisica Nucleare (INFN), MIUR, Istituto Nazionale di Astrofisica (INAF-OABRERA) Grant Fondazione Cariplo/Regione Lombardia ID 2014-1980/RST\_ERC, Italy;
ICRR, University of Tokyo, JSPS, MEXT, Japan;
Netherlands Research School for Astronomy (NOVA), Netherlands Organization for Scientific Research (NWO), Netherlands;
University of Oslo, Norway;
Ministry of Science and Higher Education, DIR/WK/2017/12, the National Centre for Research and Development and the National Science Centre, UMO-2016/22/M/ST9/00583, Poland;
Slovenian Research Agency, grants P1-0031, P1-0385, I0-0033, J1-9146, J1-1700, N1-0111, and the Young Researcher program, Slovenia; 
South African Department of Science and Technology and National Research Foundation through the South African Gamma-Ray Astronomy Programme, South Africa;
The Spanish groups acknowledge the Spanish Ministry of Science and Innovation and the Spanish Research State Agency (AEI) through grants AYA2016-79724-C4-1-P, AYA2016-80889-P, AYA2016-76012-C3-1-P, BES-2016-076342, FPA2017-82729-C6-1-R, FPA2017-82729-C6-2-R, FPA2017-82729-C6-3-R, FPA2017-82729-C6-4-R, FPA2017-82729-C6-5-R, FPA2017-82729-C6-6-R, PGC2018-095161-B-I00, PGC2018-095512-B-I00, PID2019-107988GB-C22; the “Centro de Excelencia Severo Ochoa” program through grants no. SEV-2016-0597, SEV-2016-0588, SEV-2017-0709, CEX2019-000920-S; the “Unidad de Excelencia María de Maeztu” program through grant no. MDM-2015-0509; the “Ramón y Cajal” programme through grants RYC-2013-14511, RYC-2017-22665; and the MultiDark Consolider Network FPA2017-90566-REDC. They also acknowledge the Atracci\'on de Talento contract no. 2016-T1/TIC-1542 granted by the Comunidad de Madrid; the “Postdoctoral Junior Leader Fellowship” programme from La Caixa Banking Foundation, grants no.~LCF/BQ/LI18/11630014 and LCF/BQ/PI18/11630012; the “Programa Operativo” FEDER 2014-2020, Consejer\'ia de Econom\'ia y Conocimiento de la Junta de Andaluc\'ia (Ref. 1257737), PAIDI 2020 (Ref. P18-FR-1580) and Universidad de Ja\'en; “Programa Operativo de Crecimiento Inteligente” FEDER 2014-2020 (Ref.~ESFRI-2017-IAC-12), Ministerio de Ciencia e Innovaci\'on, 15\% co-financed by Consejer\'ia de Econom\'ia, Industria, Comercio y Conocimiento del Gobierno de Canarias; the Spanish AEI EQC2018-005094-P FEDER 2014-2020; the European Union’s “Horizon 2020” research and innovation programme under Marie Skłodowska-Curie grant agreement no. 665919; and the ESCAPE project with grant no. GA:824064;
Swedish Research Council, Royal Physiographic Society of Lund, Royal Swedish Academy of Sciences, The Swedish National Infrastructure for Computing (SNIC) at Lunarc (Lund), Sweden;
State Secretariat for Education, Research and Innovation (SERI) and Swiss National Science Foundation (SNSF), Switzerland;
Durham University, Leverhulme Trust, Liverpool University, University of Leicester, University of Oxford, Royal Society, Science and Technology Facilities Council, UK;
U.S. National Science Foundation, U.S. Department of Energy, Argonne National Laboratory, Barnard College, University of California, University of Chicago, Columbia University, Georgia Institute of Technology, Institute for Nuclear and Particle Astrophysics (INPAC-MRPI program), Iowa State University, the Smithsonian Institution, Washington University McDonnell Center for the Space Sciences, The University of Wisconsin and the Wisconsin Alumni Research Foundation, USA.

\bigskip

The research leading to these results has received funding from the European Union's Seventh Framework Programme (FP7/2007-2013) under grant agreements No~262053 and No~317446.
This project is receiving funding from the European Union's Horizon 2020 research and innovation programs under agreement No~676134.

The research leading to these results has received funding from  the  European  Union's  Horizon 2020 research and innovation program under the Marie Sk{\l}odowska-Curie grant agreement GammaRayCascades No~843800.

\bigskip

The \textit{Fermi} LAT Collaboration acknowledges generous ongoing support
from a number of agencies and institutes that have supported both the
development and the operation of the LAT as well as scientific data analysis.
These include the National Aeronautics and Space Administration and the
Department of Energy in the United States, the Commissariat \`a l'Energie Atomique
and the Centre National de la Recherche Scientifique / Institut National de Physique
Nucl\'eaire et de Physique des Particules in France, the Agenzia Spaziale Italiana
and the Istituto Nazionale di Fisica Nucleare in Italy, the Ministry of Education,
Culture, Sports, Science and Technology (MEXT), High Energy Accelerator Research
Organization (KEK) and Japan Aerospace Exploration Agency (JAXA) in Japan, and
the K.~A.~Wallenberg Foundation, the Swedish Research Council and the
Swedish National Space Board in Sweden.
 
Additional support for science analysis during the operations phase is gratefully
acknowledged from the Istituto Nazionale di Astrofisica in Italy and the Centre
National d'\'Etudes Spatiales in France. This work performed in part under DOE
Contract DE-AC02-76SF00515.

\bibliographystyle{jhep}
\bibliography{references}

\appendix

\newpage 

\section{Data analysis}

\subsection{Statistical treatment}
\label{sect::data_analysis}
The simulated CTA observations are analyzed for each particular science case using the profile likelihood method~\cite{rolke2005}. The logarithm of the Poisson likelihood for each energy bin is given by
\begin{equation}
    \label{eq:lnl}
    \ln\mathcal{L}(\mu, b; \alpha_\mathrm{exp} | \non, \noff)
     = \non \ln(\mu + b) - (\mu + b) + \noff \ln(b / \alpha_\mathrm{exp}) - b / \alpha_\mathrm{exp},
\end{equation}
where $\mu$ is the expected number of counts,\footnote{We have suppressed the subscript $i$ used in Eq.~\eqref{eq:npred}.} $\non$ is the number of counts in the signal region, $\noff$ is the number of counts in the background region and $\alpha_\mathrm{exp}$ is the exposure ratio between these two regions.
The likelihood depends on the parameters $(\params,\nuisparams)$ through $\mu\equiv \mu(\params,\nuisparams)$.

We derive best-fit parameters by maximizing the likelihood above and find their confidence intervals or respective limits using the profile log-likelihood ratio test~\cite{rolke2005}, 
\begin{equation}
        \lambda(\params) = 2\sum\limits_i\ln\left(\frac{\mathcal{L}(\mu_i(\widehat{\params},\widehat{\nuisparams}), \widehat{b}; \alpha_\mathrm{exp} | \non, \noff)}{\mathcal{L}(\mu_i(\params,\widehat{\widehat{\nuisparams}}),\widehat{\widehat{b}}; \alpha_\mathrm{exp} | \non, \noff)} \right),
\end{equation}
where $\widehat{\params}, \widehat{\nuisparams}$, and  $\widehat{b}$ denote the parameters that maximize the unconditional likelihood. The parameters $\widehat{\widehat{b}}$ and $\widehat{\widehat{\nuisparams}}$ are the background and nuisance parameters that maximize the likelihood for fixed parameters, $\params$. If multiple \gr sources are considered, the log-likelihood ratio tests of individual \gr sources can be combined by simply summing the values of $\lambda$ obtained for each \gr source.

The null hypothesis is usually defined here as the configuration of parameters for which the sought-after effect is absent, or equivalently, $\params = \vec{0}$. In this case,  the log-likelihood ratio test defines the test statistic, 
\begin{equation}
    \ts = \lambda(\params = \boldsymbol{0}).
\end{equation}
The $\ts$ values provide a measure of the incompatibility of the data with the null hypothesis compared to the test hypothesis. According to Wilks' theorem~\cite{Wilks1938}, the $\ts$ values should be distributed as a $\chi^2$ with $\nu$ degrees of freedom under the null hypothesis, where $\nu$ is the difference between the number of degrees of freedom of the null and alternative hypotheses. Wilks' theorem applies if the two hypotheses are nested, 
when the parameters are not at their boundaries in the null distribution, and if all nuisance parameters are defined under both the null and alternative hypotheses. 

\subsection{Systematic treatment}\label{sect::systematics}

Systematic uncertainties are estimated using the bracketing method. Constraints on the parameter of interest, $\hat{\pi} \pm \sigma_{\rm stat}$, obtained from bracketing cases that correspond to ${\pm}\, 1\,\sigma$ variations of the corresponding nuisance parameter, $\theta_-$ and $\theta_+$, can be combined to estimate the level of systematic uncertainties induced by the nuisance parameter. Formally, for a probability density function (pdf) of the nuisance parameter, $f(\theta)$, and a pdf at fixed $\theta$ value, $g(\pi |\theta)$, the pdf of the parameter of interest can be obtained through the marginalization $g(\pi) = \int d\theta f(\theta) g(\pi | \theta)$. The variance of the parameter distribution for $\theta = 0$ corresponds to the statistical uncertainty, $\sigma_{\rm stat}^2 = \mathrm{V}(g(\pi |0))$, while the variance of the marginalized distribution is a combination of the statistical and systematic uncertainties, with the latter being defined by the relation $\sigma_{\rm stat}^2 + \sigma_{\rm sys}^2= \mathrm{V}(g(\pi))$. The estimation of the pdf in bracketing cases, with $\tilde{f}(\theta) = \frac{1}{2}\big(\delta(\theta-\theta_-)+\delta(\theta-\theta_+)\big)$, can be used to derive an estimator of the parameter pdf as:
\begin{equation}
    \label{eq:dirac_approx}
    \tilde{g}(\pi) = \frac{1}{2}\big( g(\pi|\theta_-) + g(\pi|\theta_+) \big),    
\end{equation}
where the estimator of the variance induced by statistical and systematic uncertainties is taken as $\tilde{\mathrm{V}}_{\rm stat+sys} = \int d\pi (\pi-\hat{\pi})^2 \tilde{g}(\pi)$. This estimator can be shown to be unbiased in the simple case of normal distributions. Let us assume that the parameter of interest is distributed as $g(\pi|\theta) = \mathcal{N}(\pi | p(\theta), \sigma_{\rm stat})$, where $\mathcal{N}(x | \hat{x}, \sigma)$ is a normal distribution of median $\hat{x}$ and standard deviation $\sigma$. The function $p$ links variations of the nuisance parameter to variations of the parameter of interest, such as $p(\theta_\pm) = \pi_\pm$. Then it is straightforward to show that $\int d\pi \left(\pi - \hat{\pi}\right)^2 g(\pi|\theta_-) = \sigma_{\rm stat}^2 + (\pi_{-} -\hat{\pi})^2 $, where $\pi_{-} - \hat{\pi}$ is equivalent to $\sigma_{\rm sys}$ in this test case. It should be noted that we assume for the sake of simplicity that the statistical uncertainty does not depend on the nuisance parameter and that the function $p$ scales linearly with the nuisance parameter. Under such assumptions, 
\begin{align}
    \tilde{\mathrm{V}}_{\rm stat+sys} 
     &= \frac{1}{2}\left( \int d\pi \left(\pi - \hat{\pi}\right)^2 \tilde{g}(\pi|\theta_-) + \int d\pi \left(\pi - \hat{\pi}\right)^2 \tilde{g}(\pi|\theta_+) \right) \nonumber \\
     &=  \sigma_{\rm stat}^2 + \sigma_{\rm sys}^2.
\end{align}

We extend this approach by estimating the lower and upper systematic uncertainties induced by the nuisance parameter, $\theta$, as
\begin{align}
    \sigma_{\rm sys, -}^2 &= \frac{\int_{-\infty}^{\hat{\pi}}d\pi (\pi - \hat{\pi})^2 \tilde{g}(\pi)}{\int_{-\infty}^{\hat{\pi}}d\pi\tilde{g}(\pi)} - \sigma_{\rm stat}^2, \nonumber \\
     \sigma_{\rm sys, +}^2 &= \frac{\int_{\hat{\pi}}^{\infty}d\pi (\pi - \hat{\pi})^2 \tilde{g}(\pi)}{\int_{\hat{\pi}}^{\infty}d\pi\tilde{g}(\pi)} - \sigma_{\rm stat}^2, 
\end{align}
where $\tilde{g}(\pi)$ is obtained from the parameter distribution for nuisance parameters $\theta_\pm$, as indicated in Eq.~\eqref{eq:dirac_approx}. The systematic uncertainties induced by multiple nuisance parameters, such as uncertainties on the effective area and on the energy scale, are combined quadratically, effectively treating these parameters as independent variables.

\section{Intrinsic AGN models}\label{sec:intr-models}

In this Appendix, we list the spectral models used for the different AGN samples described in Sec.~\ref{subsect::ebl_ksp_sel}, which are used to derive the sensitivity of CTA for the detection of the EBL attenuation. We provide the detection significances of the simulated AGN above $E(\tau=1)$ for different assumed exponential cut-offs at $0.1/(1+z)\,$TeV, $1/(1+z)\,$TeV and $10/(1+z)\,$TeV. The spectral models and detection significances of the long-term monitoring sample (Sec.~\ref{sec:longterm}) are listed in Table~\ref{table:longterm}. The high-quality spectral sample (Sec.~\ref{sec:hq}) is detailed in Table~\ref{table:hq1-10}. Parameters of the GeV- and TeV-flare samples (Sec.~\ref{sec:flare}) are provided in Tables~\ref{table:gevflare} and~\ref{table:tevflare}.

\begin{table}[h]
    \caption{
Properties of AGN in the long-term monitoring sample. 
The spectral model parameters are adapted from Ref.~\cite{2015ApJ...812...60B}. 
A live time of 50\,hours is used to assess the detectability of the AGN with CTA.
The columns display from left to right: 
(1) the 3FGL source name;
(2) the counterpart name; 
(3; 4) the AGN type and SED class from 4LAC (or 4FGL$^*$, or 3LAC$^\dagger$, or VizieR$^\mathsection$, or Ref.~\cite{2020NatAs...4..124B}$^\ddagger$);
(5) the redshift;
(6) the redshift reference;
(7) the reference energy, $E_0$, for the source spectral model;
(8) the corresponding flux level at $E_0$;
(9) the power-law photon index, $\Gamma$, or corresponding parameter for log-parabola models;
(10) the curvature parameter, $\beta$, for log-parabola models;
(11--13) $N_{\sigma}^{0.1}$, $N_{\sigma}^{1}$, $N_{\sigma}^{10}$ the significance of detection obtained for $E\geqslant E(\tau=1)$ considering an exponential cut-off at $0.1/(1+z)$\,TeV, $1/(1+z)$\,TeV and $10/(1+z)$\,TeV, respectively. 
    }
    \tiny
    \centering
    \addtolength{\tabcolsep}{-2.5pt} 
    \begin{tabular}{llllSlccSSSSS}
    \hline
      3FGL Name & Counterpart & Type & Class & $z$ & $z_{\rm ref}$ & $E_0$ & Normalization & $\Gamma$ & $\beta$ & $N_{\sigma}^{0.1}$ & $N_{\sigma}^{1}$ & $N_{\sigma}^{10}$\\
                &             &      &            &&               & [TeV]           & [$/$cm$^{2}\,$s\,TeV]    &          &         &                  &              &       \\
    \hline\hline
    J1104.4+3812 & Mrk\,421 & BLL & HSP & 0.031 & \cite{ulrich75} & 3.34 & $6.52\times 10^{-12}$ & 2.71 & 0.20 & 0.1 & -0.3 & 18.7 \\
    J1653.9+3945 & Mrk\,501 & BLL & HSP & 0.034 & \cite{ulrich75} & 1.42 & $8.27\times 10^{-12}$ & 2.19 & & -0.2 & 0.0 & 19.9 \\
    J1517.6$-$2422 & AP\,Librae & BLL & LSP & 0.049 & \cite{dis74} & 0.42 & $7.86\times 10^{-12}$ & 2.17 & & -0.1 & -0.0 & 9.1 \\
    J2202.7+4217 & BL\,Lac & BLL & LSP & 0.069 & \cite{oke74,ver95} & 0.28 & $4.64\times 10^{-11}$ & 2.91 & & 0.0 & 0.4 & 5.8 \\
    J1221.4+2814 & W\,Comae & BLL & ISP & 0.102 & \cite{wei85} & 0.36 & $6.24\times 10^{-11}$ & 2.77 & & 0.2 & 10.8 & 33.0 \\
    J2158.8$-$3013 & PKS\,2155$-$304 & BLL & HSP & 0.116 & \cite{fal93} & 0.33 & $1.42\times 10^{-10}$ & 2.75 & & 0.2 & 27.6 & 68.4 \\
    J1428.5+4240 & H\,1426+428 & BLL & HSP & 0.129 & \cite{rem89} & 2.30 & $1.49\times 10^{-12}$ & 0.81 & & -0.1 & 12.9 & 44.8 \\
    J0232.8+2016 & 1ES\,0229+200 & BLL & EHSP$^\ddagger$ & 0.140 & \cite{sch93} & 1.64 & $1.71\times 10^{-12}$ & 1.26 & & 0.1 & 14.7 & 40.8 \\
    J1103.5$-$2329 & 1ES\,1101$-$232 & BLL & EHSP$^\ddagger$ & 0.186 & \cite{rem89} & 0.59 & $1.23\times 10^{-11}$ & 1.41 & & 0.6 & 24.0 & 56.3 \\
    J1015.0+4925 & 1ES\,1011+496 & BLL & HSP & 0.212 & \cite{alb07} & 0.18 & $5.16\times 10^{-10}$ & 3.24 & & 0.3 & 19.8 & 32.2 \\
    J0222.6+4301 & 3C\,66A & BLL & ISP & 0.340 & \cite{fur13a,tor18} & 0.14 & $5.45\times 10^{-10}$ & 2.44 & & 0.5 & 17.3 & 26.9 \\
    \hline
    \end{tabular}
\label{table:longterm}
\end{table}

\begin{table}[h]
    \caption{
Properties of AGN in the high-quality spectra sample.
The spectral model parameters are adapted from Ref.~\cite{2015ApJ...812...60B}. 
A live time of 20\,hours is used to assess the detectability of the AGN with CTA.
See Table~\ref{table:longterm} for column details. 
    }
    \tiny
    \centering
    \addtolength{\tabcolsep}{-2.5pt} 
    \begin{tabular}{llllSlccSSSSS}
    \hline
      3FGL Name & Counterpart & Type & Class & $z$ & $z_{\rm ref}$ & $E_0$ & Normalization & $\Gamma$ & $\beta$ & $N_{\sigma}^{0.1}$ & $N_{\sigma}^{1}$ & $N_{\sigma}^{10}$\\
                &             &      &            &&               & [GeV]           & [$/$cm$^{2}\,$s\,TeV]    &          &         &                  &              &       \\
    \hline\hline
    J0214.4+5143 & TXS\,0210+515 & BLL$^\dagger$ & EHSP$^\ddagger$ & 0.049 & \cite{mar96} & 49.7 & $2.10\times 10^{-10}$ & 1.55 & & 0.2 & 0.0 & 11.4 \\
    J0550.6$-$3217 & PKS\,0548$-$322 & BLL & EHSP$^\ddagger$ & 0.069 & \cite{fos76} & 34.7 & $2.75\times 10^{-10}$ & 1.80 & & -0.2 & 0.4 & 10.7 \\
    J2250.1+3825 & B3\,2247+381 & BLL & HSP & 0.119 & \cite{lau98} & 40.9 & $8.49\times 10^{-10}$ & 1.65 & & -0.4 & 9.9 & 29.2 \\
    J1917.7$-$1921 & 1H\,1914$-$194 & BLL & HSP & 0.137 & \cite{car03} & 26.2 & $5.85\times 10^{-9}$ & 2.05 & & -0.3 & 11.9 & 30.8 \\
    J1010.2$-$3120 & 1RXS\,J101015.9$-$3119 & BLL & HSP & 0.143 & \cite{pir07} & 41.7 & $7.05\times 10^{-10}$ & 1.65 & & -0.3 & 11.6 & 33.1 \\
    J0648.8+1516 & RX\,J0648.7+1516 & BLL$^\dagger$ & HSP$^\dagger$ & 0.179 & \cite{ali11} & 33.1 & $2.55\times 10^{-9}$ & 1.83 & & 0.3 & 14.2 & 27.4 \\
    J1221.3+3010 & PG\,1218+304 & BLL & EHSP$^\ddagger$ & 0.184 & \cite{ahn12} & 22.2 & $1.24\times 10^{-8}$ & 1.47 & 0.22 & 0.1 & 7.3 & 14.1 \\
    J0349.2$-$1158 & 1ES\,0347$-$121 & BLL & EHSP$^\ddagger$ & 0.188 & \cite{woo05} & 44.6 & $3.37\times 10^{-10}$ & 1.65 & & -0.2 & 5.8 & 15.3 \\
    J1224.5+2436 & MS\,1221.8+2452 & BLL & HSP & 0.219 & \cite{sto91} & 29.0 & $1.89\times 10^{-9}$ & 1.94 & & -0.2 & 6.4 & 12.0 \\
    J0303.4$-$2407 & PKS\,0301$-$243 & BLL & HSP & 0.266 & \cite{fal00} & 23.0 & $1.23\times 10^{-8}$ & 2.20 & & -0.1 & 13.8 & 22.1 \\
    J0543.9$-$5531 & 1RXS\,J054357.3$-$5532 & BLL & HSP & 0.273 & \cite{pit14} & 25.1 & $4.63\times 10^{-9}$ & 2.08 & & -0.3 & 7.9 & 13.2 \\
    J0416.8+0104 & 1ES\,0414+009 & BLL & ETEV$^\ddagger$ & 0.287 & \cite{hal91} & 28.0 & $1.34\times 10^{-9}$ & 1.98 & & -0.1 & 3.7 & 6.3 \\
    J1931.1+0937 & RX\,J1931.1+0937 & BLL$^\dagger$ & HSP$^\dagger$ & >0.476 & \cite{sha13} & 21.8 & $9.84\times 10^{-9}$ & 2.31 & & 0.5 & 6.1 & 8.3 \\
    J0033.6$-$1921 & KUV\,00311$-$1938 & BLL & HSP & >0.506 & \cite{pit14} & 23.1 & $8.36\times 10^{-9}$ & 2.21 & & 0.0 & 7.4 & 9.5 \\
    J1427.0+2347 & PKS\,1424+240 & BLL & HSP & >0.604 & \cite{fur13b} & 17.2 & $6.35\times 10^{-8}$ & 2.00 & 0.18 & -0.1 & 7.4 & 10.9 \\
    \hline
    \end{tabular}
\label{table:hq1-10}
\end{table}

\begin{tiny}
\addtolength{\tabcolsep}{-3pt} 
\begin{longtable}{llllSlccSSSSS}
    \caption{
Properties of AGN in the GeV-flare sample.
The spectral model parameters derived from the LAT analysis described in Sec.~\ref{sec:flare}. 
A live time of 10\,hours is used to assess the detectability of the AGN with CTA.
See Table~\ref{table:longterm} for column details. 
}\\
    \hline
      3FGL Name & Counterpart & Type & Class & $z$ & $z_{\rm ref}$ & $E_0$ & Normalization & $\Gamma$ & $\beta$ & $N_{\sigma}^{0.1}$ & $N_{\sigma}^{1}$ & $N_{\sigma}^{10}$\\
                &             &      &            &&               & [GeV]           & [$/$cm$^{2}\,$s\,TeV]    &          &         &                  &              &       \\
    \hline\hline     
    \endfirsthead
    \caption{ (continued) Properties of AGN in the GeV-flare sample.} \\
    \hline
      3FGL Name & Counterpart & Type & Class & $z$ & $z_{\rm ref}$ & $E_0$ & Normalization & $\Gamma$ & $\beta$ & $N_{\sigma}^{0.1}$ & $N_{\sigma}^{1}$ & $N_{\sigma}^{10}$\\
                &             &      &            &&               & [GeV]           & [$/$cm$^{2}\,$s\,TeV]    &          &         &                  &              &       \\
    \hline\hline     
    \endhead
    J0522.9$-$3628 & PKS\,0521$-$36 & AGN & LSP & 0.055 & \cite{dan79} & 0.32 & $1.14\times 10^{-2}$ & 2.89 & & -0.2 & 0.1 & 1.3 \\
    J0739.4+0137 & PKS\,0736+01 & FSRQ & LSP & 0.191 & \cite{hew93} & 2.00 & $3.74\times 10^{-5}$ & 2.33 & & 0.2 & 28.1 & 47.3 \\
    J0017.6$-$0512 & PMN\,J0017$-$0512 & FSRQ & LSP & 0.226 & \cite{sha12} & 0.86 & $9.94\times 10^{-5}$ & 2.05 & & 1.0 & 65.9 & 101.8 \\
    J0303.4$-$2407 & PKS\,0301$-$24 & BLL & HSP & 0.260 & \cite{pit14} & 2.00 & $3.51\times 10^{-5}$ & 2.08 & & 1.1 & 75.0 & 114.7 \\
    J1700.1+6829 & GB6\,J1700+6830 & FSRQ & LSP & 0.301 & \cite{hen97} & 2.00 & $3.51\times 10^{-5}$ & 2.12 & & 1.1 & 59.6 & 83.6 \\
    J0854.8+2006 & OJ\,287 & BLL & LSP & 0.306 & \cite{mil78,nil10} & 2.00 & $4.71\times 10^{-5}$ & 2.14 & & 2.4 & 74.8 & 102.5 \\
    J1751.5+0939 & PKS\,1749+096 & BLL & LSP & 0.322 & \cite{sti88} & 2.00 & $1.51\times 10^{-4}$ & 2.26 & & 3.5 & 98.2 & 131.8 \\
    J1153.4+4932 & PKS\,1150+497 & FSRQ & LSP & 0.334 & \cite{hew93} & 0.52 & $6.32\times 10^{-4}$ & 2.53 & & 0.4 & 6.7 & 9.9 \\
    J1349.6$-$1133 & PKS\,1346$-$112 & FSRQ & ISP & 0.340 & \cite{era03} & 0.77 & $3.50\times 10^{-4}$ & 2.79 & & 0.0 & 2.9 & 4.0 \\
    J1512.8$-$0906 & PKS\,1510$-$089 & FSRQ & LSP & 0.361 & \cite{tho90} & 2.00 & $4.06\times 10^{-4}$ & 2.13 & & 35.4 & 344.3 & 429.5 \\
    J0958.6+6534 & S4\,0954+65 & BLL & LSP & 0.367 & \cite{law96} & 2.00 & $3.98\times 10^{-5}$ & 2.36 & & 1.1 & 29.6 & 42.5 \\
    J0510.0+1802 & PKS\,0507+17 & FSRQ & LSP & 0.416 & \cite{per98} & 2.00 & $1.47\times 10^{-4}$ & 2.18 & & 7.9 & 125.4 & 158.9 \\
    J1642.9+3950 & B3\,1641+399 & FSRQ & LSP & 0.594 & \cite{hew93} & 2.00 & $3.02\times 10^{-5}$ & 2.09 & & 3.3 & 46.4 & 61.3 \\
    J2035.3+1055 & PKS\,2032+107 & FSRQ & LSP & 0.601 & \cite{ant87} & 0.80 & $4.84\times 10^{-4}$ & 2.68 & & 0.3 & 6.8 & 8.2 \\
    J0112.8+3207 & 4C\,31.03 & FSRQ & LSP & 0.603 & \cite{wil76} & 2.00 & $1.99\times 10^{-5}$ & 2.70 & & 0.1 & 2.0 & 3.1 \\
    J2236.5$-$1432 & PKS\,2233$-$148 & BLL & LSP & >0.609 & \cite{sha13} & 2.00 & $1.22\times 10^{-4}$ & 2.51 & & 2.0 & 34.0 & 45.0 \\
    J2345.2$-$1554 & PMN\,J2345$-$1555 & FSRQ & LSP & 0.621 & \cite{sha12} & 2.00 & $7.54\times 10^{-5}$ & 2.26 & & 10.3 & 90.5 & 110.5 \\
    J1958.0$-$3847 & PKS\,1954$-$388 & FSRQ & LSP & 0.630 & \cite{bro75} & 2.00 & $3.12\times 10^{-5}$ & 2.03 & & 11.0 & 94.6 & 116.5 \\
    J1849.2+6705 & 4C\,+66.20 & FSRQ & LSP & 0.657 & \cite{sti93} & 2.00 & $7.94\times 10^{-5}$ & 2.04 & & 21.1 & 134.1 & 185.8 \\
    J1801.5+4403 & S4\,1800+44 & FSRQ & LSP & 0.663 & \cite{hew93} & 2.00 & $6.87\times 10^{-5}$ & 2.44 & & 2.7 & 31.9 & 42.1 \\
    J1506.1+3728 & B2\,1504+37 & FSRQ & LSP & 0.674 & \cite{sti94} & 0.65 & $2.80\times 10^{-4}$ & 2.13 & & 4.5 & 45.0 & 58.0 \\
    J1800.5+7827 & S5\,1803+78 & BLL & LSP & 0.680 & \cite{law96} & 2.00 & $2.18\times 10^{-5}$ & 2.35 & & 1.5 & 14.1 & 23.6 \\
    J1159.5+2914 & Ton\,599 & FSRQ & LSP & 0.725 & \cite{whi00} & 2.00 & $8.98\times 10^{-5}$ & 1.89 & & 27.2 & 193.0 & 236.8 \\
    J1748.6+7005 & S4\,1749+70 & BLL & ISP & 0.770 & \cite{law96} & 1.14 & $1.49\times 10^{-4}$ & 2.36 & & 2.2 & 16.4 & 28.1 \\
    J1848.4+3216 & GB6\,B1846+3215 & FSRQ & LSP & 0.798 & \cite{sha09} & 2.00 & $4.66\times 10^{-5}$ & 2.44 & & 2.9 & 24.9 & 33.3 \\
    J0442.6$-$0017 & NRAO\,190 & FSRQ & LSP & 0.845 & \cite{sha12} & 2.00 & $5.33\times 10^{-5}$ & 2.06 & & 18.2 & 118.4 & 146.2 \\
    J0339.5$-$0146 & PKS\,0336$-$01 & FSRQ & LSP & 0.852 & \cite{wil78} & 2.00 & $6.21\times 10^{-5}$ & 2.20 & & 12.4 & 85.4 & 106.4 \\
    J2254.0+1608 & 3C\,454.3 & FSRQ & LSP & 0.859 & \cite{hew93} & 2.00 & $1.01\times 10^{-3}$ & 2.55 & & 34.7 & 183.0 & 218.8 \\
    J0538.8$-$4405 & PKS\,0537$-$441 & BLL & LSP & 0.892 & \cite{bec02} & 2.00 & $3.98\times 10^{-5}$ & 2.05 & & 11.6 & 85.6 & 109.5 \\
    J1733.0$-$1305 & PKS\,1730$-$130 & FSRQ & LSP & 0.902 & \cite{jun84} & 2.00 & $6.38\times 10^{-5}$ & 2.33 & & 6.3 & 49.8 & 64.5 \\
    J0505.3+0459 & PKS\,0502+049 & FSRQ & LSP & 0.954 & \cite{dri97} & 2.00 & $1.29\times 10^{-4}$ & 2.30 & & 10.9 & 81.4 & 101.8 \\
    J0238.6+1636 & A\,0235+164 & BLL & LSP & 0.954 & \cite{coh87} & 2.00 & $5.42\times 10^{-5}$ & 2.44 & & 2.4 & 25.8 & 32.7 \\
    J1310.6+3222 & OP\,313 & FSRQ & LSP & 0.998 & \cite{sha12} & 2.00 & $2.83\times 10^{-5}$ & 2.34 & & 1.4 & 14.7 & 20.8 \\
    J0457.0$-$2324 & PKS\,0454$-$234 & FSRQ & LSP & 1.003 & \cite{sti89} & 2.00 & $5.21\times 10^{-5}$ & 2.33 & & 3.7 & 33.9 & 45.7 \\
    J0909.1+0121 & PKS\,B0906+015 & FSRQ & LSP & 1.023 & \cite{hew93} & 2.00 & $2.83\times 10^{-5}$ & 2.85 & & 0.5 & 2.2 & 3.4 \\
    J2232.5+1143 & CTA\,102 & FSRQ & LSP & 1.037 & \cite{hew93} & 2.00 & $7.00\times 10^{-4}$ & 2.26 & & 41.3 & 227.3 & 276.4 \\
    J0725.2+1425 & 4C\,14.23 & FSRQ & LSP & 1.038 & \cite{sha12} & 2.00 & $6.56\times 10^{-5}$ & 2.98 & & 0.4 & 2.6 & 3.8 \\
    J0428.6$-$3756 & PKS\,0426$-$380 & BLL & LSP & 1.110 & \cite{sba05} & 2.00 & $3.09\times 10^{-5}$ & 2.13 & & 9.0 & 52.2 & 72.6 \\
    J2147.2+0929 & PKS\,2144+092 & FSRQ & LSP & 1.113 & \cite{whi88} & 2.00 & $4.04\times 10^{-5}$ & 2.85 & & 0.8 & 5.0 & 7.8 \\
    J2147.3$-$7536 & PKS\,2142$-$75 & FSRQ & LSP & 1.139 & \cite{jau78} & 2.00 & $2.72\times 10^{-5}$ & 2.20 & & 4.2 & 29.6 & 39.8 \\
    J1717.4$-$5157 & PMN\,J1717$-$5156 & FSRQ$^\dagger$ & LSP$^\dagger$ & 1.158 & \cite{cho13} & 2.00 & $3.86\times 10^{-5}$ & 2.36 & & 4.7 & 26.5 & 39.0 \\
    J0237.9+2848 & 4C\,+28.07 & FSRQ & LSP & 1.206 & \cite{sha12} & 2.00 & $4.62\times 10^{-5}$ & 2.10 & & 11.9 & 73.3 & 95.6 \\
    J0532.7+0732 & OG\,050 & FSRQ & LSP & 1.254 & \cite{sha12} & 2.00 & $3.38\times 10^{-5}$ & 2.30 & & 3.6 & 26.2 & 36.1 \\
    J2025.6$-$0736 & PKS\,2023$-$077 & FSRQ & LSP & 1.388 & \cite{dri97} & 0.50 & $2.15\times 10^{-3}$ & 1.68 & 0.08 & 22.3 & 109.3 & 157.1 \\
    J1033.8+6051 & S4\,1030+61 & FSRQ & LSP & 1.401 & \cite{hew10} & 2.00 & $3.20\times 10^{-5}$ & 2.83 & & 0.2 & 1.7 & 2.2 \\
    J2323.5$-$0315 & PKS\,2320$-$035 & FSRQ & LSP & 1.411 & \cite{bro75} & 2.00 & $3.34\times 10^{-5}$ & 1.95 & & 20.9 & 102.9 & 132.9 \\
    J0403.9$-$3604 & PKS\,0402$-$362 & FSRQ & LSP & 1.423 & \cite{hew93} & 2.00 & $9.11\times 10^{-5}$ & 2.81 & & 2.3 & 10.9 & 16.4 \\
    J1457.4$-$3539 & PKS\,1454$-$354 & FSRQ & LSP & 1.424 & \cite{hoo03} & 2.00 & $8.40\times 10^{-5}$ & 2.28 & & 13.0 & 67.6 & 90.8 \\
    J0921.8+6215 & OK\,+630 & FSRQ & LSP & 1.453 & \cite{sti93} & 0.60 & $4.41\times 10^{-4}$ & 2.17 & & 4.3 & 25.9 & 33.2 \\
    J1522.1+3144 & B2\,1520+31 & FSRQ & LSP & 1.487 & \cite{sow05} & 2.00 & $5.88\times 10^{-5}$ & 2.34 & & 5.6 & 30.2 & 41.8 \\
    J1427.9$-$4206 & PKS\,1424$-$41 & FSRQ & LSP & 1.522 & \cite{whi88} & 2.00 & $8.46\times 10^{-5}$ & 2.39 & & 7.5 & 41.1 & 58.1 \\
    N/A & PKS\,1824$-$582 & FSRQ & LSP$^\mathsection$ & 1.530 & \cite{hea08} & 2.00 & $7.00\times 10^{-5}$ & 2.66 & & 1.2 & 10.4 & 14.2 \\
    J0730.2$-$1141 & PKS\,0727$-$11 & FSRQ$^\dagger$ & LSP$^\dagger$ & 1.591 & \cite{zen02} & 2.00 & $4.09\times 10^{-5}$ & 2.52 & & 2.4 & 13.6 & 21.0 \\
    J1635.2+3809 & PKS\,1633+382 & FSRQ & LSP & 1.813 & \cite{wil76} & 2.00 & $2.38\times 10^{-5}$ & 3.23 & & -0.1 & 0.0 & 0.5 \\
    J2311.0+3425 & B2\,2308+34 & FSRQ & LSP & 1.817 & \cite{wil76} & 2.00 & $3.04\times 10^{-5}$ & 1.90 & & 7.4 & 49.1 & 69.1 \\
    J0808.2$-$0751 & PKS\,0805$-$07 & FSRQ & LSP & 1.837 & \cite{whi88} & 2.00 & $5.55\times 10^{-5}$ & 2.21 & & 6.5 & 41.0 & 59.8 \\
    J2201.7+5047 & NRAO\,676 & FSRQ$^\dagger$ & LSP$^\dagger$ & 1.899 & \cite{sow05} & 0.43 & $4.35\times 10^{-3}$ & 2.80 & & 0.3 & 3.2 & 4.9 \\
    N/A & B2\,0748+33 & FSRQ & LSP$^\mathsection$ & 1.935 & \cite{hew93} & 2.00 & $4.43\times 10^{-5}$ & 3.08 & & 0.5 & 0.7 & 1.3 \\
    J0108.7+0134 & 4C\,+01.02 & FSRQ & LSP & 2.099 & \cite{cha91} & 2.00 & $8.46\times 10^{-5}$ & 2.54 & & 4.5 & 25.1 & 34.6 \\
    J1332.0$-$0508 & PKS\,1329$-$049 & FSRQ & LSP & 2.150 & \cite{tho90} & 2.00 & $6.12\times 10^{-5}$ & 2.79 & & 1.1 & 7.2 & 10.6 \\
    N/A & TXS\,0552+398 & FSRQ$^*$ & LSP$^\mathsection$ & 2.363 & \cite{tor12} & 2.00 & $4.18\times 10^{-5}$ & 2.33 & & 2.0 & 14.4 & 20.2 \\
    J1833.6$-$2103 & PKS\,1830$-$211 & FSRQ$^\dagger$ & LSP$^\dagger$ & 2.507 & \cite{Lid99} & 2.00 & $7.28\times 10^{-5}$ & 2.45 & & 3.3 & 22.1 & 31.1 \\
    J1345.6+4453 & B3\,1343+451 & FSRQ & LSP & 2.534 & \cite{sha12} & 2.00 & $1.92\times 10^{-5}$ & 3.96 & & -0.1 & 0.0 & -0.1 \\
    \hline\noalign{\smallskip}
    \label{table:gevflare}
    \end{longtable}
\end{tiny}

\begin{table}[h]
    \caption{
Properties of AGN in the TeV-flare sample. 
The spectral model parameters are adapted from Ref.~\cite{2015ApJ...812...60B}. 
A live time of 10\,hours is used to assess the detectability of the AGN with CTA.
See Table~\ref{table:longterm} for column details. 
    }
    \tiny
    \centering
    \addtolength{\tabcolsep}{-2.5pt} 
    \begin{tabular}{llllSlccSSSSS}
    \hline
      3FGL Name & Counterpart & Type & Class & $z$ & $z_{\rm ref}$ & $E_0$ & Normalization & $\Gamma$ & $\beta$ & $N_{\sigma}^{0.1}$ & $N_{\sigma}^{1}$ & $N_{\sigma}^{10}$\\
                &             &      &            &&               & [TeV]           & [$/$cm$^{2}\,$s\,TeV]    &          &         &                  &              &       \\
    \hline\hline
    J0316.6+4119 & IC\,310 & RDG & ETEV$^\ddagger$ & 0.019 & \cite{owe95} & 0.52 & $7.68\times 10^{-11}$ & 1.66 & & 0.1 & 0.3 & 4.4 \\
    J1104.4+3812 & Mrk\,421 & BLL & HSP & 0.031 & \cite{ulrich75} & 1.74 & $8.98\times 10^{-11}$ & 2.17 & & 0.2 & 0.2 & 27.6 \\
    J1653.9+3945 & Mrk\,501 & BLL & EHSP$^\ddagger$ & 0.034 & \cite{ulrich75} & 5.01 & $5.50\times 10^{-12}$ & 2.07 & & 0.7 & 0.4 & 33.7 \\
    J2347.0+5142 & 1ES\,2344+514 & BLL$^\dagger$ & HSP$^\dagger$ & 0.044 & \cite{per96} & 0.85 & $5.68\times 10^{-11}$ & 1.68 & & -0.1 & 0.4 & 32.0 \\
    J2202.7+4217 & BL\,Lac & BLL & LSP & 0.069 & \cite{oke74,ver95} & 0.29 & $1.34\times 10^{-9}$ & 2.84 & & -0.0 & 5.4 & 33.6 \\
    J1221.4+2814 & W\,Comae & BLL & ISP & 0.102 & \cite{wei85} & 0.45 & $1.20\times 10^{-10}$ & 2.52 & & -0.2 & 14.9 & 38.5 \\
    J2158.8$-$3013 & PKS\,2155$-$304 & BLL & HSP & 0.116 & \cite{fal93} & 0.36 & $2.41\times 10^{-9}$ & 2.41 & 0.22 & 0.1 & 94.4 & 182.9 \\
    J0809.8+5218 & 1ES\,0806+524 & BLL & HSP & 0.137 & \cite{bad98} & 0.32 & $1.02\times 10^{-9}$ & 3.72 & & -0.4 & 26.0 & 46.1 \\
    J0721.9+7120 & S5\,0716+714 & BLL & ISP & >0.232 & \cite{dan13} & 0.31 & $4.71\times 10^{-10}$ & 1.58 & & 0.4 & 51.7 & 81.7 \\
    J0222.6+4301 & 3C\,66A & BLL & ISP & 0.340 & \cite{fur13a,tor18} & 0.14 & $5.45\times 10^{-10}$ & 2.45 & & 0.5 & 7.6 & 11.9 \\
    J1224.9+2122 & PKS\,1222+216 & FSRQ & LSP & 0.432 & \cite{hew93} & 0.15 & $6.75\times 10^{-9}$ & 2.20 & & 3.8 & 81.2 & 103.9 \\
    J1256.1-0547 & 3C\,279 & FSRQ & LSP & 0.536 & \cite{hew93} & 0.12 & $8.63\times 10^{-9}$ & 3.63 & & 3.1 & 38.7 & 50.9 \\
    J1443.9+2502 & PKS\,1441+25 & FSRQ & LSP & 0.940 & \cite{sha12} & 0.09 & $5.99\times 10^{-9}$ & 2.76 & & 2.7 & 25.2 & 33.6 \\
    J0221.1+3556 & S3\,0218+35 & FSRQ & LSP & 0.944 & \cite{coh03} & 0.11 & $7.87\times 10^{-9}$ & 1.76 & & 6.2 & 51.6 & 68.0 \\
    \hline\hline
    \end{tabular}
\label{table:tevflare}
\end{table}

\newpage

\section{Blazars with uncertain redshifts}
\label{uncertain_z}
Nine blazars are excluded from the GeV-flare sample due to their uncertain redshifts. For six of them, the redshift is ambiguous as it is based on the detection of only one line. For the remaining three, two conflicting results have been published and the true values remain uncertain. Details are given in Table~\ref{table:uncertain_z}.  

\begin{table}[h]
  \small
\centering
\caption{
Properties of blazars excluded from the GeV-flare sample due to an uncertain redshift. 
The columns display from left to right:
    (1) the source name;
    (2) the name of the associated counterpart;
    (3) the AGN type;
    (4) the possible redshift value;
    (5) reference(s) for the possible redshift value;
    (6) reason for the exclusion: either conflicting results (``Conflict'') or detection of only one line (``Unique Line''). 
} 
    \label{table:uncertain_z}
\begin{center}
   \begin{tabular}{lllccl} 
      \hline
    3FGL Name                  &   Counterpart           &    Type     &     Tentative $z$     & $z_{\rm ref}$     & Comment \\
      \hline
      \hline
    J1230.3+2519     & ON\,246                  &   BLL        & 0.135?     &  \cite{pai17,nas96}     &   Conflict   \\ 
    J1312.7+4828     & GB6\,B1310+484    &  FSRQ      & 0.501?    &  \cite{sha12,mun03}  & Conflict    \\ 
    J2329.3$-$4955     &  PKS\,2326$-$502       &  FSRQ     & 0.518        &   \cite{jau84}                        & Unique Line  \\  
    J2250.7$-$2806     & PMN\,J2250$-$2806    & FSRQ      & 0.525       &  \cite{sha12}                        &  Unique Line  \\      
    J0904.8$-$5734     &   PKS\,0903$-$57        &  FSRQ      & 0.695      &   \cite{tho90}                       & Unique Line    \\
    J1625.7$-$2527     & PKS\,1622$-$253        & FSRQ      & 0.786       & \cite{dise94}                        &  Unique Line \\   
    J0210.7$-$5101    & QSO\,B0208$-$512                &  BLL         &  0.999?  & \cite{wis00,jon09}     &   Conflict    \\ 
    J1048.4+7144     & S5\,1044+71            & FSRQ     & 1.15         &   \cite{pol95}                        &  Unique Line \\
    J2244.1+4057    & TXS\,2241+406       & FSRQ      & 1.171       &   \cite{sha12}                        & Unique Line   \\     
      \hline
      \hline
\end{tabular}
\end{center}
\end{table}

\newpage
\section{EBL constraints for various energy cutoffs}\label{sec:ebl-extra-results}

\begin{figure}[h]
  \center
  \includegraphics[width=\columnwidth]{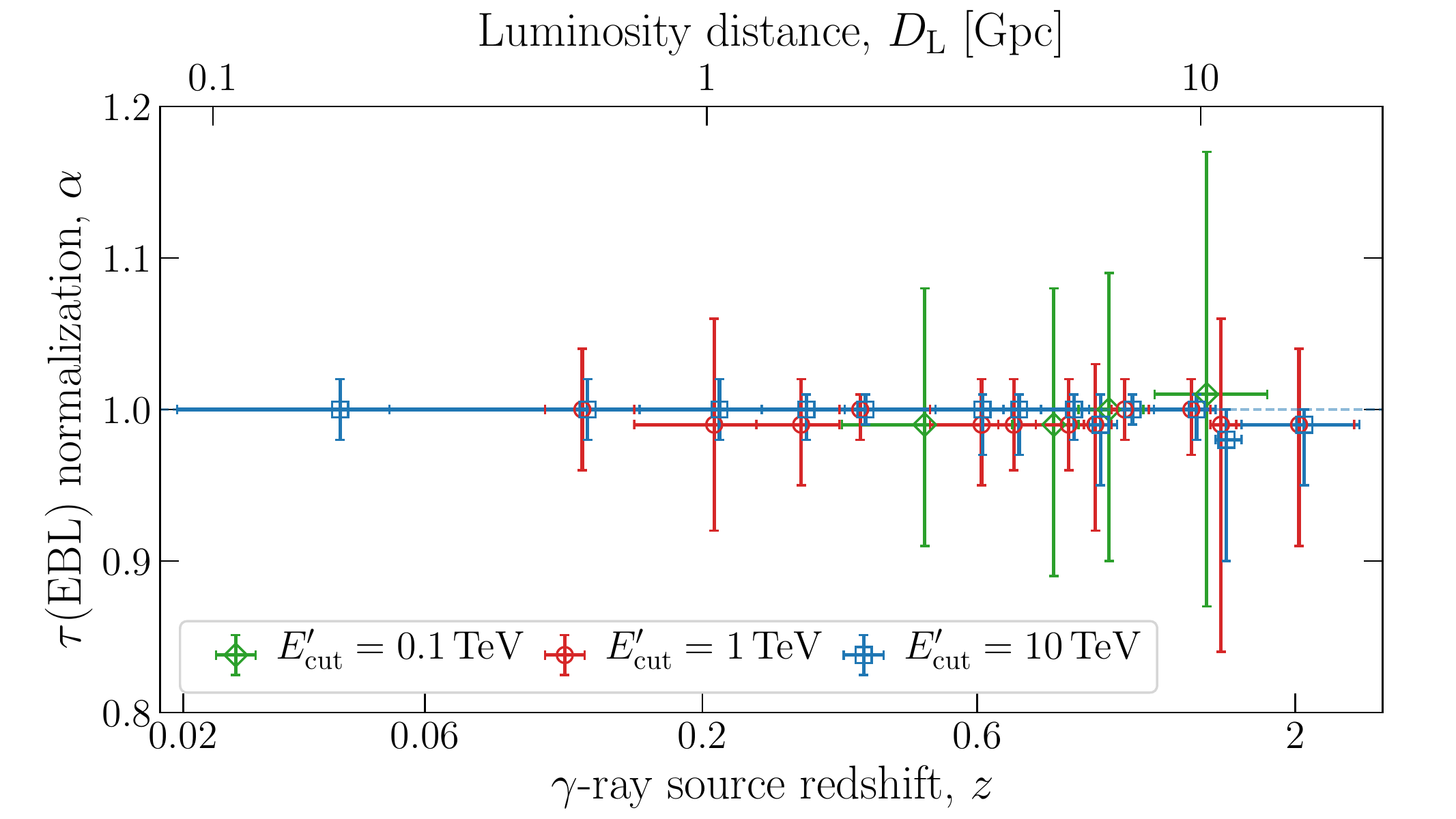}
  \caption{Median reconstructed EBL scale factor as a function of redshifts of the \gr sources, considering for each AGN a comoving exponential cut-off at $E_{\rm cut}' = 0.1$ (green), 1 (red) and 10\,TeV (blue). Only statistical uncertainties are indicated for the sake of comparison.}
  \label{fig::ebl_constrains10}
\end{figure}

We compare in  Fig.~\ref{fig::ebl_constrains10} the reconstructed optical-depth normalization, $\alpha$, with respect to the benchmark EBL model of Ref.~\cite{2011MNRAS.410.2556D}, for intrinsic exponential cut-offs at $0.1/(1+z)\,$TeV, $1/(1+z)\,$TeV and $10/(1+z)\,$TeV. 


\section{Coverage study for limits on the ALP parameter space}
\label{App:ALPcov}

In this Appendix, we present results on the coverage study for the ALPs science case, which we derive from 100 Monte-Carlo simulations each for three sets of injected ALP parameters (see Sec.~\ref{Sec:ALPsim}). 

\begin{figure}[t]
    \centering
    \includegraphics[width = \linewidth]{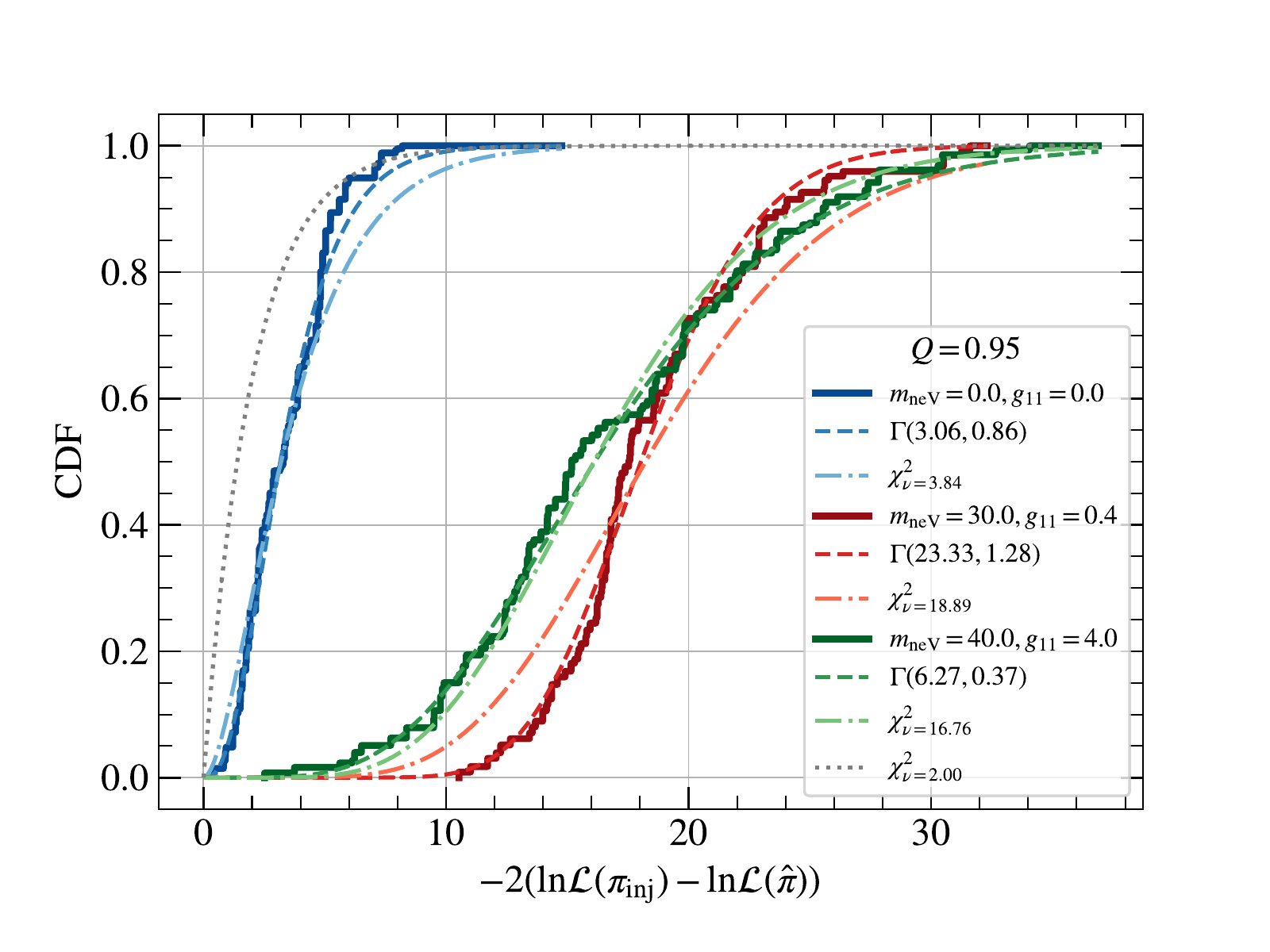}
    \caption{\label{fig:alp-nulldists} 
Cumulative distribution functions (CDFs) of the log-likelihood ratio tests for different injected ALP parameters. Solid lines show the simulated distributions whereas the dashed (dash-dotted) lines correspond to $\Gamma$ ($\chi^2$) distributions with the best-fit parameters reported in the legend. We use the short-hand notation $m_\mathrm{neV} = m_a / \mathrm{neV}$ and $g_{11} = g_{a\gamma} / 10^{-11}\,\mathrm{GeV}^{-1}$. For reference, a $\chi^2$ with $\nu=2$ degrees of freedom is shown as a grey dotted line. The likelihood ratios are derived for magnetic-field quantiles $Q=0.95$.
    }
\end{figure}

The cumulative distribution functions (CDF) of the likelihood ratio test between the hypothesis of the injected signal (denoted with $\boldsymbol{\pi}_\mathrm{inj}$) versus the best-fit hypothesis (with parameters $\widehat{\boldsymbol{\pi}}$) are derived from the pseudo experiments described in Sec.~\ref{Sec:ALPsim}.
All CDFs shown in Fig.~\ref{fig:alp-nulldists} are different from the naive expectation of a $\chi^2$ distribution with $\nu=2$ degrees of freedom (grey dotted line).
The fit of $\chi^2$ distributions with $\nu$ as a free parameter (dash-dotted lines) fail to represent the tails of the simulated distributions at high values of the likelihood ratio test. Instead, modified $\Gamma$ distributions\footnote{
We use the definition $\Gamma(\alpha, \beta) = \beta^\alpha x^{\alpha-1} \exp(-\beta x)/\Gamma(\alpha)$, where $\Gamma(\alpha)$ is the gamma function.} appear to provide a reasonable analytic approximation to the simulated distribution.
As anticipated, the best-fit parameters of the $\Gamma$ distribution obtained for different ALP parameter sets are different (see legend in Fig.~\ref{fig:alp-nulldists} for the corresponding values). 
Since we do not produce pseudo experiments for all tested ALP parameter values, we opt for the choice of using the best-fit $\Gamma$ distribution found for $m_a = 30$\,neV and $g_{a\gamma} = 4\times10^{-12}\,\mathrm{GeV}^{-1}$ (red dashed line) when setting limits. 
For 95\,\% (99\,\%) upper limits this corresponds to threshold values of $\lambda = 23.3$ ($26.9$). 
By construction, for this particular set of parameters, the adopted approach results in the correct coverage of the limits.
For the other tested parameter set with non-zero coupling, this choice results in a slight under-coverage, which we however deem to be negligible as the threshold values are only marginally different (26.1 for a 95\,\% confidence limit, see the green dashed line).
We leave it for future work to check this assumption with more comprehensive simulations that investigate the evolution of the threshold values over a larger portion of the ALP parameter space.  

\end{document}